\documentclass[acmsmall]{acmart}

\AtBeginDocument{%
	\providecommand\BibTeX{{%
			\normalfont B\kern-0.5em{\scshape i\kern-0.25em b}\kern-0.8em\TeX}}}

\setcopyright{acmcopyright}
\copyrightyear{20XX}
\acmYear{20XX}
\acmDOI{10.1145/1122445.1122456}

\acmJournal{JACM}
\acmVolume{1}
\acmNumber{1}
\acmArticle{1}
\acmMonth{1}

\usepackage{graphicx}
\usepackage{color}
\usepackage{amsmath}
\usepackage{booktabs}

\usepackage{amssymb}
\usepackage{graphicx,graphics}
\usepackage{latexsym}
\usepackage{threeparttable}
\usepackage{colortbl}

\usepackage{bm}
\usepackage{subfigure}
\usepackage{multirow}
\usepackage{algorithm}
\usepackage{algpseudocode}

\newtheorem{definition}{Definition}[section]
\newtheorem{lemma}{Lemma}[section]
\newtheorem{Proof}{Proof}[section]
\newcommand{\tabincell}[2]{\begin{tabular}{@{}#1@{}}#2\end{tabular}}

\graphicspath{{./figures/}}

\begin{document}
	
	\title{Indexing Metric Spaces for Exact Similarity Search}
	
	\author{Lu Chen}
	\affiliation{
		\institution{College of Computer Science, Zhejiang University}
		\streetaddress{38 Zheda Road}
		\city{Hangzhou}
		\country{China}
		\postcode{310027}
	}
	\email{luchen@zju.edu.cn}
	
		\author{Yunjun Gao}
	\affiliation{
		\institution{College of Computer Science, Zhejiang University}
		\country{China}
	}
	\email{gaoyj@zju.edu.cn}
	
	\author{Xuan Song}
	\affiliation{
		\institution{College of Computer Science, Zhejiang University}
		\country{China}
	}
	\email{xsong@zju.edu.cn}
	\author{Zheng Li}
	\affiliation{
		\institution{College of Computer Science, Zhejiang University}
		\country{China}
	}
	\email{zhengli3401@zju.edu.cn}
	
	\author{Yifan Zhu}
	\affiliation{
		\institution{College of Computer Science, Zhejiang University}
		\country{China}
	}
	\email{xtf\_z@zju.edu.cn}
	
	\author{Xiaoye Miao}
	\affiliation{
		\institution{Center for Data Science, Zhejiang University}
		\streetaddress{86 Yuhangtang Road}
		\city{Hangzhou}
		\country{China}
		\postcode{310058}
	}
	\email{miaoxy@zju.edu.cn}
	
	\author{Christian S. Jensen}
	\affiliation{
		\institution{Department of Computer Science, Aalborg University}
		\state{Aalborg}
		\country{Denmark}
	}
	\email{csj@cs.aau.dk}
	
	\begin{abstract}
	With the continued digitization of societal processes, we are seeing an explosion in available data. This is referred to as big data. In a research setting, three aspects of the data are often viewed as the main sources of challenges when attempting to enable value creation from big data: volume, velocity, and variety. Many studies address volume or velocity, while fewer studies concern the variety. Metric spaces are ideal for addressing variety because they can accommodate any data as long as it can be equipped with a distance notion that satisfies the triangle inequality. To accelerate search in metric spaces, a collection of indexing techniques for metric data have been proposed. However, existing surveys offer limited coverage, and {a comprehensive empirical study exists has yet to be reported}. We offer {a comprehensive survey of existing metric indexes} that support exact similarity search:  {we summarize existing partitioning, pruning, and validation techniques} used by metric indexes to {support exact similarity search}; we provide the time and space complexity analyses of index construction; and we offer {an} empirical comparison of their query processing performance. Empirical studies are important when evaluating metric indexing performance, because performance can depend highly on the effectiveness of available pruning and validation as well as on the data distribution, which means that complexity analyses often offer limited insights. This article aims at revealing strengths and weaknesses of different indexing techniques to offer guidance on selecting an appropriate indexing technique for a given setting, and to provide directions for future research on metric indexing.
	\end{abstract}

	
	
	\maketitle

	\section{Introduction}
\label{sec:Introduction}
\begin{figure}[H]
		\vspace{-5mm}
		\centering
		\includegraphics[width=0.8\linewidth]{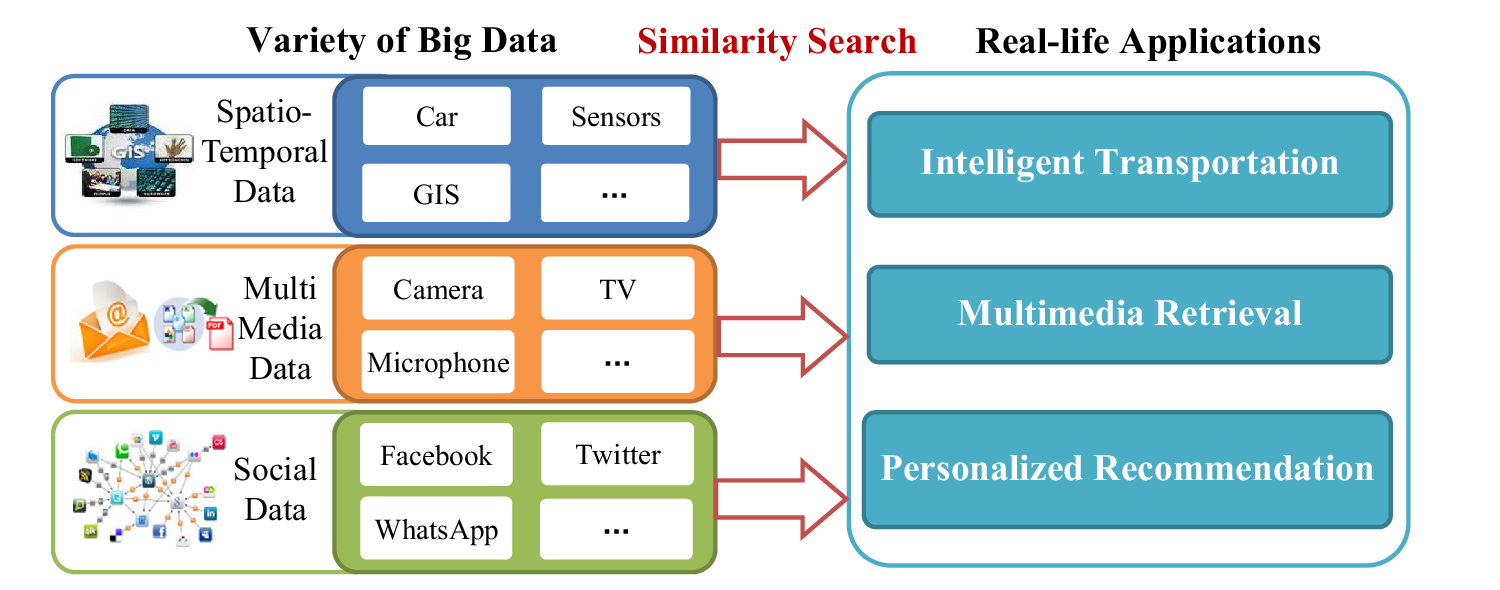}
		\setlength{\abovecaptionskip}{0.2cm}
		\caption{{Applications of Similarity Search on Variety of Big Data}}
		\label{fig:Applications}
		\vspace{-0.5cm}
	\end{figure}

Massive and increasing volumes of data are being generated. For example, as suggested in Fig.~\ref{fig:Applications}, global navigation satellite systems, e.g., GPS, and communication-network based positioning technologies enable the generation of massive spatio-temporal data by in-vehicle devices and smartphones. Massive multimedia data are being generated by surveillance cameras, smart-assistant microphones, and smartphones. Further, massive volumes of social media data is being generated by services such as Facebook, Twitter, and {WhatsApp}. This proliferation of data offers opportunities for value creation, benefiting businesses as well as society. On the flipside, this state of affairs also presents difficult challenges due to the sheer volume, velocity, and variety of the data. Here, volume refers to massive data, velocity refers to the phenomenon that the data arrives at a rapid rate, and variety refers to the characteristic that the data stems from a wide range of sources and is diverse in terms of its structure and data type. {Many studies and products address volume or velocity, while variety is receiving less attention. More specifically, many platforms (e.g., MapReduce~\cite{refn-1}, Hadoop\footnote{Apache Hadoop. http://hadoop.apache.org/ (2008)}, Spark\footnote{Apache Spark. http://spark.apache.org/ (2014)}, Flink\footnote{Apache Flink. http://flink.apache.org/ (2014)}, and Storm\footnote{Apache Storm. http://storm.apache.org/ (2014)}) address volume and velocity, while only few (notably Azure Cosmos DB\footnote{Microsoft. https://azure.microsoft.com/services/cosmos-db/ (2017)} and GeminiDB\footnote{Huawei. https://www.huaweicloud.com/intl/product/geminidb.html (2019)}) focus on the variety aspect by supporting a range of models (such as tables, graphs, and documents).} The use of metric spaces enables addressing variety because any type of data that can be associated with {a distance notion} that satisfies the triangle inequality can be treated as metric data. Hence, by using metric spaces, unified solutions can be developed that enable value creation from diverse data.
	
{Due to the generality of metric spaces, search in metric spaces plays an important role in many real-life applications, with similarity search taking center stage~\cite{ref32, ref99, ref124}.} Given a query object, similarity search finds similar objects according to the definition of similarity. In intelligent transportation, similarity queries can be used to find the nearest restaurant for a user. In multimedia retrieval, similarity queries can be utilized to identify images similar to a specified image. In recommender systems, similarity queries can be employed to generate personalized recommendations for users. In addition, similarity queries can be used to accelerate data mining tasks. For example, similarity search can be used as the first step in {clustering~\cite{refn-3} or outlier detection~\cite{refn-2}}. In the above applications, metric spaces can accommodate a wealth of data types (e.g., locations, images, and strings), and can be used to support a wide range of distance metrics that quantify the similarity of objects, including the shortest path distance for locations, the $L_p$-norm  and {earth mover's distance~\cite{refn-14}} for images, and the edit distance for strings. Metric spaces require no assumptions on the representations of objects, and any {distance notation} satisfying the triangle inequality can be accommodated.

A number of indexing techniques exist that aim to accelerate search in metric spaces. The main goal of this article is to present a comprehensive survey that {describes and analyzes existing metric indexes}. Although several studies~\cite{ref32, ref60, ref99, ref124} offer valuable surveys of metric indexing techniques, no study has yet offered comprehensive theoretical and experimental analysis. This study extends a previous study~\cite{ref37} by the authors that covers on pivot-based metric indexing. We extend that study as follows: (i) we cover {existing metric indexes} that support exact similarity search; (ii) we summarize {the partitioning methods}, and improve the coverage of the pruning and validation techniques used for {exact similarity search}; (iii) we cover the time and space complexity related to index construction; and (iv) we provide empirical comparisons of the similarity search performance achieved by {metric indexes}.

{A key strength of metric space solutions is that they are applicable to a broad range of data types found in real-world applications. While we can design specific indexes for different data types (R-trees~\cite{ref13} for low-dimensional vectors, q-gram inverted indexes for strings~\cite{refn-4}, HI-tree~\cite{MKPM21} for documents, etc.), metric indexing offers a wholesale solution that spans a broad range of data, the key requirement simply being that a distance function is provided that satisfies the triangle inequality. Since the early 1980’s, major DBMS vendors have followed a “one-size-fits-all” approach, due in part to the low maintenance cost and ease of use by customers. However, in today’s setting, the “one-size-fits-all” approach is increasingly being abandoned~\cite{refn-5,refn-6} due to a number of reasons such as the availability of independent DB engines and better optimization techniques. Further, some database vendors are developing multi-model database systems (notably Azure Cosmos DB developed by Microsoft in 2017 and GeminiDB developed by Huawei in 2019) that aim to support a range of models (e.g., tables, graphs, and documents) within a single system. Such systems easily become very complex. Metric indexing offers an approach to reduce this complexity.} 
	
The rest of the article is organized as below. Section \ref{sec:BASIC CONCEPTS} presents the basic concepts of metric indexes. Section \ref{sec:DETAILED CATEGORIZATION OF METRIC INDEXES} provides detailed categorizations of metric indexes. Section \ref{sec:TECHNIQUES FOR METRIC INDEXING AND QUERYING} summarizes the techniques used by metric indexes.  Section \ref{sec:METRIC INDEXES FOR EXACT METRIC SIMILARITY SEARCH} describes each metric index in turn. Experimental findings are reported in Section \ref{sec:EXPERIMENTAL COMPARISION AMONG METRIC INDEXES}. Finally, Section \ref{sec:SUMMARY AND RESEARCH DIRECTIONS} concludes and provides future research directions.

\section{BASIC CONCEPTS}\label{sec:BASIC CONCEPTS}
We provide the basic definitions of metric spaces and {similarity search}, and we introduce and categorize the metric indexes. Table \ref{tab:Notations used Throughout the Paper} summarizes notations used throughout this article.

\subsection{Metric Space}\label{subsec:Metric Space}
A metric space is a two-tuple $(M, d)$, in which $M$ is an object domain and $d$ is a distance function that measures the "similarity" between objects in $M$. In particular, the distance function $d$ has four properties: (i) \emph{Symmetry}: $d(q, o) = d(o, q)$; (ii) \emph{non-negativity}: $d(q, o) \geq 0$; (iii) \emph{identity}: $ d(q, o) =  0$ iff $ q = o $; and (iv) \emph{triangle inequality}: $d(q, o) \leq d(q, p) + d(p, o)$.

\begin{table}
	\setlength{\abovecaptionskip}{0.1cm}
	\caption{Notations used Throughout this Article}
	\label{tab:Notations used Throughout the Paper}
	\resizebox*{\textwidth}{!}{
		\begin{tabular}{ll}
			\toprule
			\textbf{Symbol} & \textbf{Description}\\
			\midrule
			\textit{O, C, P}      &  The sets of data, centers, and pivots \\
			\textit{S}          &  The sample set of $O$ \\
			\textit{o, c, p}      &  An object, center, and pivot \\
			{\textit{M}}          & {An object domain}\\
			\textit{$d(o, p)$}  &  The distance between objects $o$ and $p$\\			
			\textit{n, $n_s$}    &  The cardinality of dataset $O$ and the cardinality of the sample set $S$\\		
			\textit{s}          &  The storage size of an object \\
			\textit{m}          &  The tree arity or the capacity of a tree node \\
			\textit{l, $l_c$}    &  The number of pivots and the number of candidate pivots \\
			\textit{g}          &  The number of pivots in a pivot group \\
			\textit{$n_d$}      &  The number of values for a discrete $ d() $ or the maximum distance of a continuous $d()$ \\
			\textit{MRQ(q, r)}   &  The metric range query with query object \textit{q} and search radius \textit{r} \\
			\textit{M\textit{k}NNQ(q, k)} & The \textit{k} nearest neighbor query with query object \textit{q} and result cardinality \textit{k} \\
			{$\textit{ND}_k $} & {The distance from $q$ to its nearest neighbor}\\
			{$N_i$, $AN(o)$} & {A tree node, the accessory nodes of $o$} \\
		   {MBB} & {The minimum bounding box of a node} \\
			{$\phi(o)$} &  {The mapped vector  $\left\langle d(o, p_1), d(o, p_2), \cdots, d(o, p_l)\right\rangle$ using a pivot set $P = \{p_1, \cdots, p_l\}$}\\
			{$SR(q)$} & \ {The mapped search region using $P$}\\
			{$R_i$, $A(p_i)$} &  {A particular partition obtained via different partitioning techniques, a partition corresponds to pivot $p_i$ } \\
			{$d_{med}$} &  {The medium distance used to obtain two partitions of equal size} \\
			{$\delta$, $\rho$, $\alpha$} & {Distance thresholds} \\
			{$\mu, \sigma^2$} & {The mean and variance of distances between data objects}\\ 	
			{$ \mu_{p_i} $} & {The expected value of $d(o, p_i) (o \in A(p_i))$}\\ 	
			{$ N(c)$} & {The so-called nearest neighbors for a node $c$ defined in SAT}\\ 			
			\bottomrule
	\end{tabular}}
	\vspace{-0.5cm}
\end{table}

Any type of data combined with a distance function that satisfies the above four properties constitutes a metric space. Hence, the notation of metric space is very general. A typical example of a metric space is a vector space associated with the $L_p$-norm $(p \geq 1)$. Another example of a metric space is a set of strings along with the edit distance. {Note that vectors are easy and efficient to process. Thus, vector spaces (including Euclidean spaces) often find use in real-world applications. Vector spaces are also used by the different solutions for metric space. For example, pivot mapping  is used to represent objects as vectors of their distances to pivots (cf. Section~\ref{sec:TECHNIQUES FOR METRIC INDEXING AND QUERYING}.2). Unlike embedding techniques (e.g., word2vec~\cite{refn-7}) that map other data types to vectors, pivot-based solutions can ensure accurate (and thus explainable) results when given a specific {metric}. This may be important in applications where liability is an issue. For example, although genes can be embedded as vectors~\cite{refn-8}, embedding does not enable accurate and explainable results, which are important in discovering the evolutionary relationship between species~\cite{refn-9}, genome databases, and so on.}

Various indexes exist that support particular metric spaces, e.g., the R-tree~\cite{ref13}, KD-tree~\cite{ref14}, and TV-tree~\cite{ref71} for low-dimensional vector spaces, where properties (e.g., the dimension information of vector spaces) of the particular metric space are utilized to accelerate the search. However, such specific properties are not available in general metric spaces, and thus, they cannot be used for search space pruning in general metric spaces. For example, the specific properties of vectors cannot be applied to strings. By using metric space properties, indexing solutions can be developed for processing a wide variety of data. Nonetheless, due to the generality of metric spaces, we can only use the four distance properties discussed above for pruning and hence accelerate search. In Section~\ref{sec:TECHNIQUES FOR METRIC INDEXING AND QUERYING}, we detail {the techniques} that can be used in general metric spaces.

{\bfseries Intrinsic Dimensionality of Metric Space.} The dimensionality is an important aspect of vector space data. The higher the dimensionality, the worse the search performance will generally be. However, metric spaces are not limited to vector spaces, but also include other data types such as strings and sets. Instead of dimensionality, we can use the more general notion of intrinsic dimensionality that applies to any data type in metric spaces. {As discussed in previous studies~\cite{ref15, ref32, ref43, ref75, ref97, ref105, ref121, ref124}, intrinsic dimensionality affects metric space search performance just like dimensionality affects vector space search performance. Many methods exist to compute intrinsic dimensionality. They can be classified into local and global methods. The local methods~\cite{ACHKRT19, H13, H17a, H17b, H20} provide different dimensionality estimates for each data object or each subset of a dataset, while the global methods~\cite{NPRB17} provide a dimensionality estimate for an entire dataset. A simple global estimate is introduced in a previous survey~\cite{ref32}, i.e., intrinsic dimensionality is calculated as $\mu^{2}/2\sigma^{2} $, where $\mu$ and $ \sigma^2$ are the mean and variance of the distances between data objects. This simple notion of intrinsic dimensionality is found to be verified effective in an experimental study~\cite{NPRB17}. Hence, we use this simple but effective definition of intrinsic dimensionality of datasets in our experiments.}
	
\subsection{Similarity Queries in Metric Spaces}\label{subsec:Metric Similarity Queries}
We proceed to define {similarity search in metric spaces}, including metric range query and metric \textit{k} nearest neighbor query. We focus on {similarity queries} because of their importance and widespread use, and because they are suitable for studying efficiency and effectiveness of the designed indexes.

\begin{definition}\label{defn:metricRangeQuery}
{\bf (METRIC RANGE QUERY).} Given an object set $O$, a query object $q$, and a search radius $r$ in a metric space, a metric range query returns the objects in $O$ that are within distance $r$ of $q$, i.e., $ MRQ(q, r) = \left\{ o| o \in O \land d(q, o) \leq r \right\}$.
\end{definition}	

\begin{definition}\label{defn:metricKNN}
{\bf (METRIC ${\bm K}$ NEAREST NEIGHBOR QUERY)}. Given an object set $O$, a query object $q$, and an integer $k$ in a metric space, a metric $k$ nearest neighbor query finds $k$ objects in $O$ that are most similar to q, i.e., $ M\textit{k}NNQ(q, k) = \left\{S|S\subseteq O \land |S|=k \land \forall{s} \in S, \forall{o} \in O -  S(d(q, s) \leq d(q, o))\right\}$.
\end{definition}
	
	Consider the DNA set $O$ = \{"ATAGCTTCA", "AATCTGA", "AATCTGT", "AAAACGG", "CATCTGT"\}, where edit distance is employed. An example metric range query finds the DNAs from \textit{O} within edit distances no larger than 2 of the query DNA "CAATCTGT", i.e., MRQ("CAATCTGT", 2) = \{"AATCTGA", "AATCTGT", "CATCTGT"\}. An example metric \textit{k} (= 2) nearest neighbor query finds the 2 DNAs from \textit{O} closest to the query DNA "CAATCTGT", i.e., M\textit{k}NNQ("CAATCTGT", 2) = \{"AATCTGT", "CATCTGT"\}.
	
	
		\begin{figure} [H]
		\centering
		\includegraphics[width=0.5\linewidth]{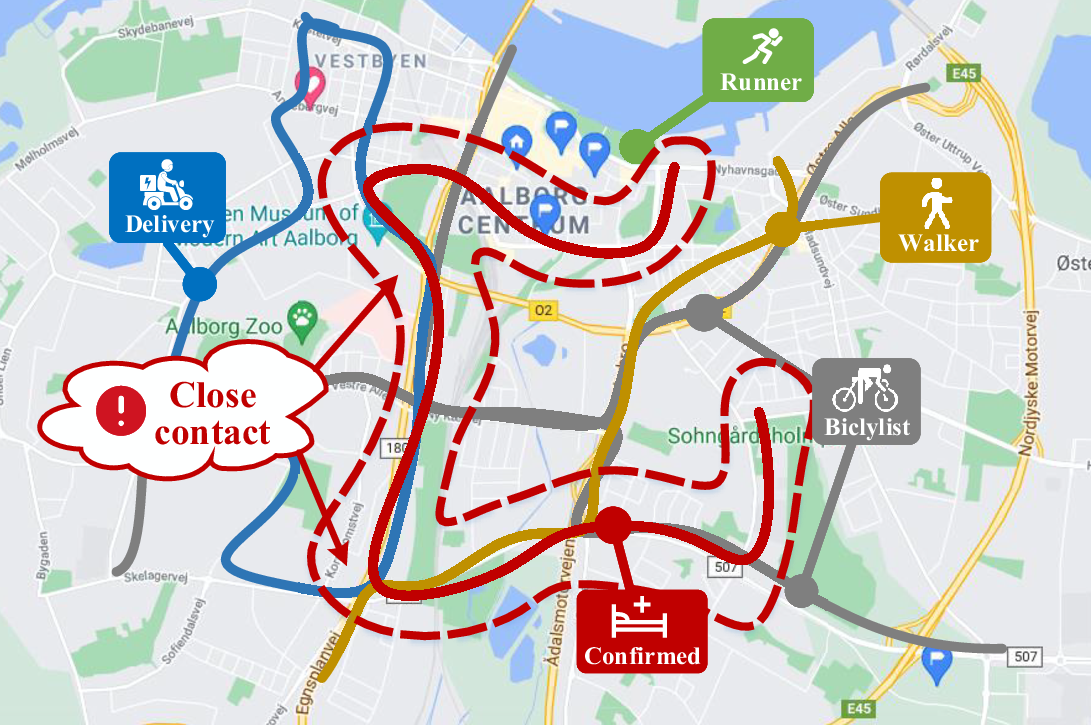}
		\setlength{\abovecaptionskip}{0.1cm}
		\caption{{Illustration of Exact Similarity Search for COVID-19 Infected Person Detection}}
		\label{fig:Covid19}
		\vspace{-0.5cm}
	\end{figure}

	{There are two kinds of “approximate” in similarity search. First, the user may request similar results, i.e., results that are approximately equal to a query object. Second, to achieve higher efficiency, a solution may compute a result that approximates the requested result. The survey covers the computation of approximate queries, but it does not cover solutions that compute results that approximate requested results. Hence, exact similarity search denotes the exact computation of both exact results and approximate results.}
	\begin{enumerate}
	\item[(i)] {In terms of exact results, we provide a motivating example in public safety applications, where accurate (and explainable) results are required.} For example, Fig.~\ref{fig:Covid19} shows a set of trajectories of moving objects (bicyclists, pedestrians, and runners) and a query trajectory of a COVID-19 infected person (the red curve). In this example, obtaining exact results is desirable to stop the spread of COVID-19. {The example uses the $L_2$-norm. This similarity notion – which is also called the $L_2$ metric – measures the distance between two points in two-dimensional Euclidean space and is used widely in many contexts~\cite{ref124}}.
		
	\item[(ii)] {In terms of approximate results, exact similarity search is important in a wide range of applications such as data cleaning~\cite{AGK06}, near-duplicate detection~\cite{KL12,QWLXL11}, and bioinformatics~\cite{QWLXL11}. }
	\end{enumerate}

	\textbf{Strategies for Answering M\textit{k}NNQ using MRQ. }An M\textit{k}NNQ can be answered by an MRQ, if the distance from \textit{q} to its $k^{th}$ nearest neighbor, denoted as $ND_k$, is known. However, $ND_k$ is not known when a query is issued. Three typical methods exist for computing M\textit{k}NNQ~\cite{ref19, ref61}.
	
	\begin{enumerate}
		\item[(i)] \textbf{Strategy 1:} MRQs with incremental growing search radius. Specifically, an MRQ with a small radius is performed first, and then the radius is increased gradually until $k$ nearest neighbors are found. Although this strategy tries to avoid visiting objects already verified, it traverses the metric index multiple times, resulting in high cost.
		\item[(ii)] \textbf{Strategy 2:} Setting the search radius to infinity, and verifying the objects in order while tightening the radius using verification. To be described in Sections~\ref{sec:DETAILED CATEGORIZATION OF METRIC INDEXES} and \ref{sec:METRIC INDEXES FOR EXACT METRIC SIMILARITY SEARCH}, M\textit{k}NNQ processing usually adopts this second strategy due to its better performance.
		\item[(iii)] \textbf{Strategy 3:} Using the candidates to calculate the current $ND_k$, and then performing an MRQ using the current $ND_k$ as the radius to refine the result. The performance of this strategy relies on the quality of the initial candidates.
	\end{enumerate}

	\textbf{Performance Metrics for {Similarity Search in Metric Spaces}.} When evaluating the performance of {a similarity query}, we use running time (i.e., response time) as the performance metric.
{As metric spaces are general and support a broad range of data types, the associated distance functions also range from being simple (e.g., the $L_p$-norm for low-dimensional vectors and the hamming distance) to being complex (e.g., edit distance, the $L_p$-norm for high-dimensional vectors, shortest-path distance, and earth mover's distance~\cite{refn-10}). Existing metric indexes and search algorithms aim to reduce the number of distance computations to improve efficiency. Acknowledging that the distance function may be simple in some applications, indicating that distance computation is not the dominant cost, the survey provides different recommendations for simple versus complex distance functions (cf. Section~\ref{sec:SUMMARY AND RESEARCH DIRECTIONS}).} Since external indexes are stored on disk, the IO time can be estimated as the number of disk page accesses performed  during search. In contrast, metric indexes stored in main memory incur no IO time. To summarize, in order to evaluate the performance of {similarity search in metric spaces}, we utilize three metrics: (i) the running time, (ii) the number of distance computations (\textit{compdists}), and (iii) the number of page accesses (\emph{PA}).

	\begin{table}
		\centering
		\setlength{\abovecaptionskip}{0.1cm}
		\caption{Metric Indexes for Exact Similarity Search {\small(Construction Cost includes the costs of distance computations and other operations, and refers to the following parameters: the dataset cardinality $n$, the tree arity or capacity $m$ of a tree node, the number $l$ of pivots, the cardinality $n_s$ of a sample set, the number $l_c$ of candidate pivots, and the number $k$ of nearest neighbors)}}
		\label{tab:Metric Indexes for Exact Similarity Search}
		\resizebox{\textwidth}{!}{
			\begin{tabular}{llll}
				\hline
				\rowcolor[HTML]{C6D9F1}\textbf{Index}&\textbf{Category}&\textbf{Space Cost}&\textbf{Construction Cost} \\
				\hline
				\rowcolor[HTML]{EFF3FF}BST~\cite{ref69}, MBST~\cite{ref92}             & CP-Index & $ O(ns)$                     & $ \Omega(n\log_{2}n) $ \\
				\rowcolor[HTML]{EFF3FF}VT~\cite{ref46,ref47}                  & CP-Index & $ O(ns)$                     & $ \Omega(n\log_{3}n) $ \\
				\rowcolor[HTML]{EFF3FF}BU-Tree~\cite{ref72}                  & CP-Index & $ O(ns+n^2)$                     & $ O(n^3) $ \\
				\rowcolor[HTML]{C6D9F1}GHT~\cite{ref115}                     & CP-Index & $ O(ns)$                     & $ \Omega(n\log_{2}{n})$ \\
				\rowcolor[HTML]{C6D9F1}GNAT~\cite{ref21}, EGNAT~\cite{ref76,ref89}       & Hybrid   & $ O(ns + nm)$                & $ \Omega(nm\log_{m}n) $ \\
				\rowcolor[HTML]{EFF3FF}SAT~\cite{ref30,ref84,ref85}               & CP-Index & $ O(ns)$                     & $ O(n\log^2{n}/\log{\log{n}}) $  \\
				\rowcolor[HTML]{EFF3FF}DSAT~\cite{ref86,ref87,ref88,ref90,ref91}      & CP-Index & $ O(ns)$                     & $ O(mn\log_{m}{n}) $  \\
				\rowcolor[HTML]{EFF3FF}DSACLT~\cite{ref12,ref22}                & CP-Index & $ O(ns)$                     & $ O(mn\log_{m}{n⁄k}) $  \\
				\rowcolor[HTML]{EFF3FF} \textit{k}NNG~\cite{ref96}    & CP-Index & $ O(ns + nk)$                     & $ O(n^2) $ \\
				\rowcolor[HTML]{C6D9F1}M-tree~\cite{ref38,ref42,ref110}          & CP-Index & $ O(ns + ns/m) $             & $ O(n(m..{m^2})\log_{m}{n}) $ \\
				\rowcolor[HTML]{C6D9F1}PM-tree~\cite{ref112}                  & Hybrid   & $ O(n(s+l) + n(s+l)/m + ls)$ & $ O(n(m..{m^2})\log_{m}{n}+nl\log_{m}{n}) $\\
				\rowcolor[HTML]{EFF3FF}LC~\cite{ref28,ref31}, DLC~\cite{ref91}           & CP-Index & $ O(ns)$                     & $ O(n^2/m) $\\
				\rowcolor[HTML]{EFF3FF}HC~\cite{ref56,ref57}                    & CP-Index & $ O(ns)$                     & $ O(n\log_{2}{n⁄m})$ \\
				\rowcolor[HTML]{C6D9F1}D-index~\cite{ref48,ref49,ref123}          & Hybrid   & $ O(ns + nl + ls)$           & $ O(nl) $ \\
				\rowcolor[HTML]{C6D9F1}$\rm MB^+$-Tree~\cite{ref63}              & CP-Index & $ O((n+n/m)(s+\log_{2}{n⁄m}+\log_{2}{n_d})) $ &$ O(n\log_{2}{n⁄m}+ nm\log_{m}{n}) $\\
				\rowcolor[HTML]{EFF3FF}AESA~\cite{ref102}, ROAESA~\cite{ref118}, iAESA~\cite{ref51,ref52}               & P-Index  & $ O(ns + n^2)$               & $ O(n^2) $ \\
				\rowcolor[HTML]{EFF3FF}LAESA~\cite{ref79}                     & P-Index  & $ O(ns + ls + nl) $          & $ O(nl) $ \\
				\rowcolor[HTML]{EFF3FF}TLAESA~\cite{ref80,ref113}               & Hybrid   & $ O(ns + ls + nl) $          & $ \Omega(n\log_{2}{n} + nl) $\\
				\rowcolor[HTML]{C6D9F1}EPT~\cite{ref103}                      & P-Index  & $ O(ns + lgs + nl)$          & $ O(nlg)$ \\
				\rowcolor[HTML]{C6D9F1}$\rm EPT^*$~\cite{ref37}                   & P-Index  & $ O(ns + l_cs + nl) $        & $ O(nll_{c}n_{s}) $ \\
				\rowcolor[HTML]{EFF3FF}CPT~\cite{ref82}                       & P-Index  & $ O(ns + ns/m + ls +  nl)$   & $ O(n(m..{m^2})\log_{m}n + nl) $ \\
				\rowcolor[HTML]{C6D9F1}BKT~\cite{ref23}, FQT~\cite{ref10}                      & P-Index  & $ O(ns + ln_d)$              & $ O(nl) $\\
				\rowcolor[HTML]{EFF3FF} FHQT~\cite{ref11}, FQA~\cite{ref32}    & P-Index  & $ O(ns + nl)$                & $ O(nl) $ \\
				\rowcolor[HTML]{C6D9F1}VPT~\cite{ref114,ref115,ref122}, DVPT~\cite{ref58}  & P-Index  & $ O(ns)$                     & $ O(n\log_{2}n) $ \\
				\rowcolor[HTML]{C6D9F1}MVPT~\cite{ref18,ref19}                  & P-Index  & $ O(ns)$                     & $ O(n\log_{m}n) $\\
				\rowcolor[HTML]{EFF3FF}Omni-family~\cite{ref20,ref66}           & P-Index  & $ O(ns + nl + nl/m + ls)$    & $ O(nml\log_{m}n) $ \\
				\rowcolor[HTML]{C6D9F1}SPB-tree~\cite{ref33,ref34}              & P-Index  & $ O(ns + n + n/m + ls)$      & $ O(n({l^2..l^3}) + n(m+l)\log_{m}n) $\\
				\rowcolor[HTML]{EFF3FF}M-index~\cite{ref93}                   & Hybrid   & $ O(ns + nl + n + n/m + ls)$ & $ \Omega(nl\log_{l}{n⁄m})  + O(nm\log_{m}n) $ \\
				\rowcolor[HTML]{EFF3FF}M-$\rm index^*$~\cite{ref37}               & Hybrid   & $ O(ns+nl+n+n/m+nl/m+ls)$    & $ \Omega(nl\log_{l}{n⁄m})+O(nm\log_{m}n) $ \\
				\hline
			\end{tabular}
		}
		\vspace{-0.3cm}
	\end{table}

\section{DETAILED CATEGORIZATION OF METRIC INDEXES}
\label{sec:DETAILED CATEGORIZATION OF METRIC INDEXES}

{We first provide the categorization of metric indexes for exact {similarity search}, and then cover the detailed  categorizations for compact-partitioning based methods and pivot-based methods.}
	
		\subsection{Categorization of Metric Indexes for Exact Metric Similarity Queries}
	\label{subsec:Metric Indexes for Exact Metric Similarity Queries}
	A rich set of indexes have been proposed that aim to support efficient metric similarity queries. Table \ref{tab:Metric Indexes for Exact Similarity Search} provides an overview of {existing metric indexes} that support exact {similarity search}. They can be classified into two board categories: compact-partitioning techniques (termed CP-Indexes) and pivot-based techniques (termed P-Indexes). In addition, hybrid indexes combine compact-partitioning and pivot-based methods (termed Hybrid). In the table, we provide the space and the time cost for index construction. {The space and time complexities of most in-memory metric indexes and some secondary-memory metric indexes (such as the M-tree and LC) are already provided in previous surveys or in the original papers, but the time and space complexity results for other in-memory indexes (such as the BU-tree, HC, and EPT$^{(*)}$) and most secondary-memory metric indexes are not provided in the literature.} Some indexes (e.g., the BST family, the GHT family, TLAESA, and the M-index) are unbalanced trees. For these, we assume that the tree structure is balanced to obtain the optimal (i.e., lower bound) construction cost $\Omega(\cdot)$. We omit the {similarity search} complexity for the indexes, because it depends on the pruning ability that depends on the data distribution. As stated in Section \ref{subsec:Metric Similarity Queries}, we instead report on empirical studies that use three performance metrics to quantify the {similarity search} performance. {In Table~\ref{tab:Metric Indexes for Exact Similarity Search}, different background colors are used to distinguish the families of indexes. For instance, BST, MBST, VT, and the BU-tree belong to the BST family, while GHT, GANT, and EGNAT belong to the GHT family.}
	
	\textbf{Compact-partitioning based Methods.} Methods in this category divide the space as compactly as possible, and try to prune unqualified partitions during search. The Bisector Tree (BST)~\cite{ref69} is a binary tree that uses a center with a covering radius to represent a partition. Many variants of the BST, including the Monotonous BST (MBST)~\cite{ref92}, the Voronoi Tree (VT)~\cite{ref46, ref47}, and the Bottom-Up Tree (BU-Tree)~\cite{ref72}, are developed to improve the efficiency of the BST. The Generalized Hyperplane Tree (GHT)~\cite{ref115} is similar to the BST, but does not store covering radius. The Spatial Approximation Tree (SAT)~\cite{ref30, ref84, ref85} is based on Voronoi diagrams, and attempts to approximate the structure of a Delaunay graph. Dynamic and secondary memory extensions of SAT includes DSAT~\cite{ref86, ref87,ref88,ref90,ref91} and DSACLT~\cite{ref12,ref22}. In addition,
	\textit{k}-Nearest-Neighbor Graph (\textit{kNNG})~\cite{ref96} is another popular graph structure. Next, the M-tree~\cite{ref38, ref42,ref110} is a height-balanced tree optimized for secondary memory. It is the first dynamic index that supports insertion and deletion. Several variants of M-trees have been presented, including the MM-tree~\cite{ref98}, Slim-tree~\cite{ref68, ref111}, $\rm M^{+} $-tree~\cite{ref127}, $\rm BM^{+}$-tree~\cite{ref128}, and CM-tree~\cite{ref7}, which use different split functions to reduce the overlap among nodes; the $\rm M^X $-tree~\cite{ref65} and Onion-tree~\cite{refn-16,ref27} that aim to reduce the tree construction cost; the DBM-tree~\cite{ref119, ref120} that allows a controlled imbalance to better accommodate the dataset density variations; the Antipole Tree~\cite{ref26} that aims to minimize the number of clusters; and the BP-tree~\cite{ref1},  $\rm M^{*}$-tree~\cite{ref108}, DF-tree~\cite{ref67}, PM-tree~\cite{ref112}, and $\rm PM^{*} $-tree~\cite{ref108} that combine multiple local or global pivots to further improve the pruning ability of the M-tree. More recently, variants $ \rm M^{\#}$-tree and $\rm PM^{\#}$-tree~\cite{ref100} are designed to avoid duplications of data. The List of Clusters (LC) index~\cite{ref28, ref31} employs a list of clusters to trade construction time for query time. Its dynamic version is called Dynamic LC (DLC)~\cite{ref91}. The construction efficiency of LC can be improved by constructing multiple layers (resulting in Hierarchy of Clusters HC~\cite{ref56, ref57}) or by using cluster reduction~\cite{ref6}. Next, the Metric $\rm B^{+}$-tree ($\rm MB^{+}$-tree)~\cite{ref63} uses relaxed generalized partitioning or hash partitioning to recursively partition the dataset and build a binary tree, while each leaf node denotes a cluster. In particular, each object can be represented as a fixed-length bit string after partitioning and then can be indexed by a $\rm B^{+}$-tree.
	
	\textbf{Pivot-based Methods.} Methods in this category store pre-computed distances from every object in the database to a set of so-called pivots and then utilize these distances to prune unqualified objects during search. The Approximating Eliminating Search Algorithm (AESA)~\cite{ref102} utilizes a pivot table to preserve the distances from each object to other objects. To improve search efficiency, the Reduced-Overhead AESA (ROAESA)~\cite{ref118} and iAESA~\cite{ref51,ref52} adopt the same data structure as AESA, but they sort the pre-computed distances during the search. To reduce the storage of AESA, the Linear AESA (LAESA)~\cite{ref79} only keeps the distances from every object to selected pivots. Unlike LAESA that uses a single set of pivots, Extreme Pivot Table $\rm EPT^{(*)}$~\cite{ref37, ref103} employs several sets of pivots. Next, the Clustered Pivot-table (CPT)~\cite{ref82} clusters the pre-computed distances to improve query efficiency. The Burkhard-Keller Tree (BKT)~\cite{ref23} is designed for discrete distance functions. In contrast to BKT, where pivots at individual levels are different, the Fixed Queries Tree (FQT)~\cite{ref10}, Fixed Height FQT (FHQT)~\cite{ref11}, and Fixed Queries Array (FQA)~\cite{ref32} use the same pivot for all nodes at the same level of the tree. The Vantage-Point Tree (VPT)~\cite{ref114,ref115,ref122} is designed for continuous distance functions and has been extended to a dynamic structure DVPT~\cite{ref58}, and generalized to an \textit{m}-ary tree yielding the MVPT~\cite{ref18,ref19}. The Omni-family~\cite{ref20, ref66} employs selected pivots together with existing structures (e.g., the R-tree) to index pre-computed distances. The Space-filling curve and Pivot-based $\rm B^{+}$-tree (SPB-tree)~\cite{ref33,ref34} utilizes a space-filling curve to map pre-computed distances to integers, which are then indexed by the $\rm B^{+}$-tree. Note that, although SFC used in the SPB-tree can well cluster the data, SPB-tree is not classified to hybrid methods in this survey, as it does not use the partitioning techniques summarized in Section~\ref{sec:TECHNIQUES FOR METRIC INDEXING AND QUERYING}.1.
	
	\textbf{Hybrid Methods.} These methods combine compact partitioning with the use of pivots. The Geometric Near-Neighbor Access Tree (GNAT)~\cite{ref21} is an \textit{m}-way generalization of GHT that utilizes the generalized hyperplane partition method to partition the dataset and also uses cut-regions~\cite{ref73} defined by pivots to accelerate similarity search. A dynamic variant, the Evolutionary Geometric Near-neighbor Access Tree (EGNAT), has also been proposed~\cite{ref76, ref89}. By combining generalized partitioning and ball partitioning, the Tree LAESA (TLAESA)~\cite{ref80,ref113} extends LAESA and organizes the data in a tree. The D-index~\cite{ref48,ref49,ref123} combines hash partitioning and the pivot mapping. It is a multilevel structure which hashes objects to buckets that are search-separable. The PM-tree~\cite{ref112} also uses cut-regions defined by pivots to accelerate similarity queries on the M-tree. The M-$\rm index^{(*)}$~\cite{ref37,ref93} generalizes the iDistance~\cite{ref64} technique for general metric spaces, which compacts the objects by using pre-computed distances to their closest pivots.
	
	\textbf{Other Metric Indexes.}
	This survey cover metric indexes that support exact {similarity search}; thus, indexes that support other metric search are omitted. For example, to answer similarity joins, the eD-index~\cite{ref50} which extends the D-index is designed. To answer probabilistic range query, the UP-Index~\cite{ref4} and UPB-tree/UPB-forest~\cite{ref35,ref36} are developed to index uncertain metric data. To support indexing multiple metric spaces, the $\rm M^2$-tree~\cite{ref39}, $\rm M^3$-tree~\cite{ref25}, and Reference $\rm R^*/RR^*$-tree~\cite{ref55} are proposed.	
	{When the intrinsic dimensionality is high, exact similarity search rarely outperforms sequential scan~\cite{ref97,WSB98}. In such cases, approximate similarity search (i.e., where high-quality approximate answers instead of exact answers are returned) may be attractive. In particular, approximate similarity search is able to offer increased efficiency over exact similarity search, which is beneficial in cases where fast search is needed and where the metric is too expensive, the dataset size is extremely large, etc.}  Many metric indexes are proposed that aim to support approximate {similarity search}, including approximate M-tree variants~\cite{ref40,ref70,ref109,ref125,ref126}, hash based methods~\cite{ref8,ref9,ref78}, permutation-based indexes ~\cite{ref3,ref29,ref53,ref81,ref83,ref116}, and \textit{k}NN graph based methods~\cite{ref74,ref95,ref106,ref107}, to name but a few. The P-Shere tree~\cite{ref59} is built using a sample of query objects. The DAHC-tree~\cite{ref2} is optimized according to the global data distribution for high-dimension space.	
	In addition, metric indexes can use other techniques (e.g., short term memories~\cite{ref101}, bit operations~\cite{ref45}, regrouping~\cite{ref104}, parallel computing~\cite{ref94}, and cost-model-based distance distribution~\cite{ref41}) to improve query efficiency. Different from an individual metric index, an index framework that combines different metric indexes is also available~\cite{ref77}.
	
	\subsection{Categorization of Compact-Partitioning based Metric Indexes}
	\label{subsec:Compact-Partitioning based Metric Indexes}
	Compact partitioning methods can be divided into four categories according to the partitioning technique used: \textit{generalized hyperplane partitioning based indexes, ball partitioning based indexes, hash partitioning based indexes, and hybrid partitioning based indexes} (see Table \ref{tab:Compact-partitioning based Metric Indexes}). Note that, hybrid methods that use partitioning techniques are also discussed here.
	\begin{table}[H]
		\centering
		\setlength{\abovecaptionskip}{0.2cm}
		\caption{Compact-partitioning based Metric Indexes}
		\label{tab:Compact-partitioning based Metric Indexes}
		\small
		\begin{tabular}{llll}
			\hline
			\rowcolor[HTML]{C6D9F1}\textbf{Partitioning Technique \qquad} & \textbf{Index} & \textbf{Storage} & \textbf{Scalability \qquad} \\
			\hline
			\rowcolor[HTML]{EFF3FF}~ & GHT, GNAT & Main-memory & Static \\
			\rowcolor[HTML]{EFF3FF}~ &  EGNAT & Secondary-memory & Dynamic\\
			\rowcolor[HTML]{EFF3FF}~ &  \textit{k}NNG & Main-memory & Dynamic\\
			\rowcolor[HTML]{EFF3FF}\multirow{-4}{*}{\tabincell{l}{Generalized hyperplane \\ partitioning}} & M-index$^{(*)}$ & Secondary-memory & Dynamic\\
			\rowcolor[HTML]{C6D9F1}~ & M-tree and PM-tree & Secondary-memory \qquad \qquad & Dynamic\\
			\rowcolor[HTML]{C6D9F1}\multirow{-2}{*}{Ball partitioning} & LC, DLC, HC & Secondary-memory & Dynamic \\
			\rowcolor[HTML]{EFF3FF}~ & D-index & Secondary-memory & Dynamic\\
			\rowcolor[HTML]{EFF3FF}\multirow{-2}{*}{Hash partitioning } & $\rm MB^+$-tree & Secondary-memory  & Dynamic\\
			\rowcolor[HTML]{C6D9F1}~ & BST, MBST, VT, BU-tree \qquad & Main-memory & Static \\
			\rowcolor[HTML]{C6D9F1}~ & SAT & Main-memory & Static \\
			\rowcolor[HTML]{C6D9F1}~ & DSAT, DSACLT & Secondary-memory & Dynamic\\
			\rowcolor[HTML]{C6D9F1}\multirow{-5}{*}{Hybrid partitioning} & TLAESA & Main-memory & Static \\
			\hline
		\end{tabular}
		\vspace{-0.2cm}
	\end{table}

Indexes in the first category use the generalized hyperplane partitioning (as stated in Definition \ref{defn:GENERALIZED HYPERPLANE PARTITIONING}) to partition the data. Indexes in the second category use ball partitioning (as defined in Definition \ref{defn:BALL PARTITIONING}) to partition the data. Indexes in the third category utilize the hash partitioning (as defined in Definition \ref{defn:HASH PARTITIONING}) to organize the data. Indexes in the fourth category combine ball partitioning and hyperplane partitioning. In particular, they employ the hyperplane partitioning to partition the data, but use ball partitioning for representing each partition.

	\subsection{Categorization Pivot-based Metric Indexes}\label{subsec:Pivot-based Metric Indexes}
	Pivot-based methods can be classified into three categories, i.e., \textit{pivot-based tables, pivot-based trees, and {pivot-based secondary-memory indexes}}, depending on the structure they use for storing pre-computed distances (see Table \ref{tab:Pivot-based Metric Indexes}). Note that, hybrid methods that use pivot mapping techniques are also discussed in this subsection.
	
		\begin{table} [H]
		\centering
		\setlength{\abovecaptionskip}{0.2cm}
		\caption{Pivot-based Metric Indexes}
		\label{tab:Pivot-based Metric Indexes}
		\small
		\begin{tabular}{llll}
			\hline
			\rowcolor[HTML]{C6D9F1}\textbf{Category} & \textbf{Index} & \textbf{Storage} & \textbf{Scalability} \qquad \qquad \\ \hline
			\rowcolor[HTML]{EFF3FF}~ & AESA, ROAESA, iAESA \qquad & Main-memory & Dynamic \\
			\rowcolor[HTML]{EFF3FF}~ & LAESA & Main-memory & Dynamic \\
			\rowcolor[HTML]{EFF3FF}~ & EPT & Main-memory & Dynamic \\
			\rowcolor[HTML]{EFF3FF}\multirow{-4}{*}{Pivot-based tables} & CPT & Secondary-memory \quad & Dynamic \\
			\rowcolor[HTML]{C6D9F1}~ & BKT & Main-memory & Dynamic \\
			\rowcolor[HTML]{C6D9F1}~ & FQT, FHQT, FQA & Main-memory & Dynamic \\
			\rowcolor[HTML]{C6D9F1}~ & TLAESA & Main-memory & static \\
			\rowcolor[HTML]{C6D9F1}~ & GNAT & Main-memory & Static \\
			\rowcolor[HTML]{C6D9F1}~ & VPT, MVPT & Main-memory & Static \\
			\rowcolor[HTML]{C6D9F1}\multirow{-6}{*}{Pivot-based trees} & DVPT & Main-memory & Dynamic \\
			\rowcolor[HTML]{EFF3FF}~ & PM-tree & Secondary-memory & Dynamic \\
			\rowcolor[HTML]{EFF3FF}~ & EGNAT & Secondary-memory & Dynamic \\
			\rowcolor[HTML]{EFF3FF}~ & D-index & Secondary-memory & Dynamic \\
			\rowcolor[HTML]{EFF3FF}~ & Omni-family & Secondary-memory & Dynamic \\
			\rowcolor[HTML]{EFF3FF}~ & M-index$^{(*)}$ & Secondary-memory \qquad & Dynamic \\
			\rowcolor[HTML]{EFF3FF}\multirow{-6}{*}{{\tabincell{l}{Pivot-based \\secondary-memory indexes}} \qquad} & SPB-tree & Secondary-memory & Dynamic \\
			\hline
		\end{tabular}
		\vspace{-0.3cm}
	\end{table}

Indexes in the first category utilize tables to store pre-computed distances. Indexes in the second category use tree structures to store pre-computed distances. 	Indexes in the third category utilize an {secondary-memory index} (e.g., an R-tree or a $\rm B^+$-tree) to store pre-computed distances.	
According to Table \ref{tab:Pivot-based Metric Indexes}, although many of the pivot-based methods are dynamic, all the indexes need to be re-built by re-computing all the stored distances when pivots are updated. However, the pivots do not need to be real objects in the dataset; thus, pivots need not necessarily be updated when inserting or deleting data objects. Among all the metric indexes, only two (BKT and FQT) are designed for discrete distance functions that return a finite range of values. Nevertheless, they can be extended to support continuous distance functions.

\section{TECHNIQUES FOR METRIC INDEXING AND QUERYING}
\label{sec:TECHNIQUES FOR METRIC INDEXING AND QUERYING}
We summarize the partitioning methods for the compact-partitioning based metric indexes as well as the pivot-based filtering and validation techniques for all the metric indexes.

\subsection{Partitioning Methods}\label{subsec:Partitioning Methods}


\begin{figure}[t]
	\centering
	\includegraphics[width=0.9\linewidth]{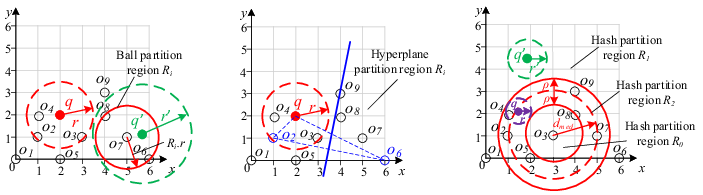} \\
	{\footnotesize  (a) Ball partitioning \quad   \quad  \quad  \quad  \quad (b) Hyperplane partitioning  \quad  \quad   \quad \quad  (c) Hash partitioning}
	\setlength{\abovecaptionskip}{0.2cm}
	\caption{Illustration of Partitioning Methods}
	\label{fig:Partitioning Methods}
	\vspace{-0.5cm}
\end{figure}

Three types of partitioning methods exist, i.e., ball partitioning, generalized hyperplane partitioning, and hash partitioning, as described below.

\begin{definition}\label{defn:BALL PARTITIONING}
	{\bf (BALL PARTITIONING)}. Given a center $c$ and a radius $r$, the set of objects $o$ in ball partition $R_i$ is defined as $\left\{o | o \in O \land d(o, c) \leq r\right\}$.
\end{definition}

In the example illustrated in Fig.~\ref{fig:Partitioning Methods}(a), given the center $o_7$ and radius $d(o_7, o_6)$, we can obtain the ball partition $R_i = \left\{o_6, o_7, o_8\right\}$. {The ball radius can be set to $2^i$ $(-\infty \le i \le +\infty)$ at each tree level $i$~\cite{ref16, ref62}, to contend with the curse of high intrinsic dimensionality or to contend with the cases when distance functions do not satisfy the triangle inequality.}

\begin{definition}\label{defn:GENERALIZED HYPERPLANE PARTITIONING}
	{\bf (GENERALIZED HYPERPLANE PARTITIONING).} Given a set $C$ of centers, let $c_i \in C$ be the center of partition region $R_i$. The set of objects $o$ in generalized hyperplane partition $R_i$ is defined as $\left\{o | o \in O \land \forall{c_j} \ne c_i(d(o, c_i) \leq d(o, c_j))\right\}$.
\end{definition}

In Fig.~\ref{fig:Partitioning Methods}(b), given the two centers $o_2$ and $o_6$, we can get two hyperplane partitions $R_1 = \{o_1, o_2, o_3,\\ o_4, o_5\}$ and $R_2 = \{o_6, o_7, o_8, o_9\}$. The generalized hyperplane partitioning can be relaxed by introducing a threshold $\delta$~\cite{ref44, ref63}, such that partition $R_i$, obtained by relaxed generalized hyperplane partitioning, is defined as $\left\{o | o \in O \land \forall{c_j} \ne c_i(d(o, c_i) \leq d(o, c_j) + \delta)\right\}$. This relaxed method can be used in hyperplane partitioning based indexes (e.g., BST and GNAT).

\begin{definition}\label{defn:HASH PARTITIONING}
	{\bf (HASH PARTITIONING)}. Given a hash function $h$, the set of objects $o$ in hash partition $R_i$, obtained by the hash partitioning, is defined as $ \left\{o | o \in O \land h(o) = i\right\} $.
\end{definition}

A particular hash function, i.e., the $\rho$-split function $bps^{\rho}(c, o)$ uses a center $c$ and the medium distance $d_{med}$ to partition the data into three subsets as defined below. The median distance $d_{med}$ is relative to $c$ and is defined so that the number of objects with distances smaller than $d_{med}$ is the same as the number of objects with distances larger than $d_{med}$.	
\begin{equation}
	bps^{\rho}(c, o) =
	\begin{cases}
		0   & \text{if $d(c, o) \leq  d_{med} - \rho$}\\
		1   & \text{if $d(c, o) >  d_{med} + \rho$}\\
		-   & \text{otherwise}\\
	\end{cases}
\end{equation}

Here, the result "$-$" denotes the last partition, the exclusion partition. In Fig.~\ref{fig:Partitioning Methods}(c), given a center $o_3$, we get three hash partitions: $R_0 = \left\{o_3\right\}$, $R_1 = \left\{o_1, o_6\right\}$, and $R_2 = \left\{o_2, o_4, o_5, o_7, o_8, o_9\right\}$. Here, $R_2$ is the exclusion partition. The $\rho$-split function can be generalized to a set of centers. Given a set of \textit{m} centers \textit{C}, the objects can be divided into $2^m + 1$ partitions. Specifically, $bps^{\rho}(C, o) = bps^{\rho}(c_i, o) \times \sum_{x=1}^i 2^{i-1} (c_i \in C, 1 \leq i \leq m)$ if $bps^{\rho}(c_i, o) = 0$ or $1$; otherwise, $bps^{\rho}(C, o) = 2^m$.

\begin{figure}[t]
	\centering
	\includegraphics[width=0.85\linewidth]{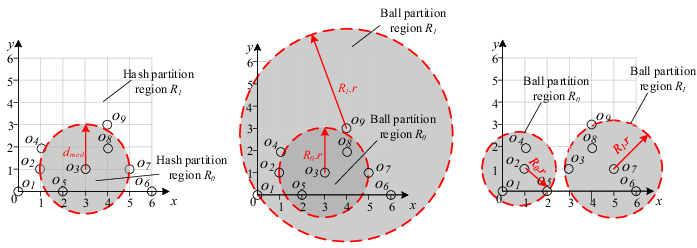} \\
	{\footnotesize \quad \quad \quad (a) Hash partitioning (center $o_3$) \quad (b) Ball partitioning (centers $o_3$ and $o_9$)  \quad (c) Ball partitioning (centers $o_2$ and $o_7$)}
	\setlength{\abovecaptionskip}{0.2cm}
	\caption{Comparison of Ball Partitioning and Hash Partitioning}
	\label{fig:comp}
	\vspace{-0.2cm}
\end{figure}

{The hash partitioning used in metric spaces is mainly extended from ball partitioning (e.g., \cite{ref49}) and hyperplane partitioning (e.g., \cite{ref78}). This is natural because  ball and hyperplane partitioning make it easy to ensure the correctness of {similarity search}, and thus, to support exact {similarity search}. However, hash functions can also be designed specifically for approximate {similarity search} (e.g., \cite{STTV14}), which is a promising future direction for designing metric indexes.}

{Given $m$ centers, ball (or hyperplane) partitioning divides the data space into $m$ subspaces, while $\rho$-split hash partitioning divides the data space into $2^m+1$ subspaces. Although $\rho$-split hash partitioning can be regarded as ball partitioning, it is not a good split for ball partitioning. Fig.~\ref{fig:comp} illustrates ball and hash partitioning. In Fig.~\ref{fig:comp}(a), the hash function $bps^0(o_3, o)$ ($\rho$ = 0) divides the space into two disjoint subspaces, including hash partition region $R_0$ (the grey circular region) and hash partition region $R_1$ (the white region). The hash partitioning can be regarded as a ball partitioning with centers $o_3$ and $o_9$, as shown in Fig.~\ref{fig:comp}(b). However, the split is not good, as the overlap between ball regions $R_0$ and $R_1$ is large, resulting in poor pruning capabilities. If we instead use centers $o_2$ and $o_7$ for ball partitioning, as depicted in Fig.~\ref{fig:comp}(c), we can obtain a better split.  Fig.~\ref{fig:comp} illustrates two differences between hash and ball partitioning: (i) Their representations are different. Metric indexes that use ball partitioning to represent each region as a ball, while metric indexes that use $\rho$-split hash partitioning to only represent some of regions as balls. (ii) How to select the centers is different. Existing studies \cite{ref24, ref75, refn-11, refn-12, refn-13} focus mostly on how to select high-quality pivots for pivot-based metric indexes, and only few studies~\cite{ref30, ref42} consider how to select high-quality centers for compact-partitioning based metric indexes.}

\subsection{Pivot Mapping}\label{subsec:Pivot Mapping}
By using a set of pivots, the objects in a metric space can be mapped to data points in a vector space. In particular, given a pivot set $ P = \left\{p_1, p_2, \cdots, p_l\right\}$, a metric space $(M, d)$ can be mapped to a vector space $(R^l, L_{\infty})$. Specifically, an object $q$ in the metric space is mapped to a point $\phi(q) = \left\langle d(q, p_1), d(q, p_2), \cdots, d(q, p_l)\right\rangle$ in the vector space.

\begin{figure} [H]
	\centering
	\subfigtopskip=0cm
	\subfigbottomskip=0cm
	\subfigcapskip=0cm
	\subfigure[Original metric space]{
		\label{Original metric space}
		\includegraphics[]{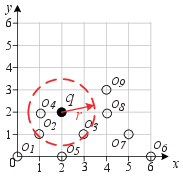}
	}
	\includegraphics[]{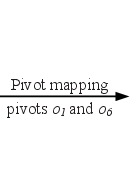}
	\subfigure[Mapped vector space]{
		\label{Mapped vector space}
		\includegraphics[]{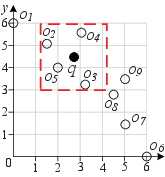}
	}
	\setlength{\abovecaptionskip}{0.2cm}
	\caption{Pivot Mapping}
	\label{fig:PivotMapping}
	\vspace{-0.5cm}
\end{figure}

Consider the example in Fig.~\ref{fig:PivotMapping}, where the $L_2$-norm is used as the distance function. If $P = \left\{o_1, o_6\right\}$, the object set in the original metric space can be mapped to the data points in a two-dimensional vector space, in which the \textit{x}-axis denotes $d(o_i, o_1)$ and the \textit{y}-axis represents $d(o_i, o_6)$ for any object $o_i$. For example, object $o_5$ is mapped to point $\left\langle 2, 4 \right\rangle$.

\subsection{Pivot-based Filtering and Validation}
\label{subsec:Pivot-based Filtering and Validating}
The triangle inequality is the only property that can be used to reduce the search space in general metric spaces. Below, we summarize seven filtering and validation lemmas. In the case of both compact-partitioning and pivot-based metric indexes, centers and pivots are combined with the triangle inequality for pruning and validation. A center used for compact partitioning methods can be regarded as a pivot of pivot-based methods. First, pivot-based filtering~\cite{ref33} based on the pivot mapping is utilized to avoid unnecessary distance computations.

\begin{lemma}\label{lemma:PIVOT FILTERING}
	{\bf (PIVOT FILTERING)}. Given a set $ P = \left\{p_1, p_2, \cdots, p_l\right\} $ of pivots, a query object $q$, and a search radius $r$, let $ SR(q) = \langle[d(q, p_1) - r, d(q, p_1) + r], \cdots, [d(q, p_l) - r, d(q, p_l) + r]\rangle$ be a mapped search region. If a mapped $\phi(o) = \left\langle d(o, p_1), d(o, p_2), \cdots, d(o, p_l)\right\rangle $ locates outside $SR(q)$, the original object $o$ can be pruned safely.
\end{lemma}

\begin{Proof}\rm
	Assume, to the contrary, that an object $o$ exists that satisfies $d(q, o) \leq r $, but $\phi(o) \notin SR(q) $ (i.e., $\exists p_i \in P(d(o, p_i) > d(q, p_i) + r \lor d(o, p_i) < d(q, p_i) - r))$. By the triangle inequality, $d(q, o) \geq |d(q, p_i) - d(o, p_i)| > r $, which contradicts our assumption, and completes the proof.
\end{Proof}

Since the pre-computed distances in $\phi(o)$ are stored together with object $o$ in a metric index, we can avoid distance computations involving object $o$ if $\phi(o) \notin SR(q) $, based on Lemma \ref{lemma:PIVOT FILTERING}. Consider the example in Fig.~\ref{fig:PivotMapping} where the dashed rectangle represents search region $SR(q)$. Here, object $o_1$ can be pruned as $\phi(o_1) \notin SR(q) $. Also, Lemma \ref{lemma:PIVOT FILTERING} can be utilized to prune an entire region (i.e., a minimum bounding box that contains multiple $\phi(o)$s) if it does not intersect $ SR(q) $. Next, based on ball partitioning, we present a range-pivot filtering technique~\cite{ref124}.

\begin{lemma}\label{lemma:RANGE-PIVOT FILTERING}
	{\bf (RANGE-PIVOT FILTERING)}. Given a ball partition region $R_i$ with the center $R_i.p$ and radius $R_i.r$, a query object $q$, and a search radius $r$, if $d(q, R_i.p) > R_i.r + r$, $R_i$ can be pruned.
\end{lemma}

\begin{Proof}\rm
	For any object $o$ in $R_i$, if $ d(q, R_i.p) > R_i.r + r $, then $ d(q, o) \geq d(q, R_i.p) - d(o, R_i.p) >  R_i.r + r - d(o, R_i.p) $ due to the triangle inequality. As $ d(o, R_i.p) \leq R_i.r $ by Definition \ref{defn:BALL PARTITIONING}, we can derive that $d(q, o) > r $. Thus, no object $o$ in $R_i$ can be in the result set, and $R_i$ can be pruned safely.
\end{Proof}

Consider Fig.~\ref{fig:Partitioning Methods}(a), where the red dashed line denotes the search region, and the solid red circle represents the ball region $R_i = \left\{o_6, o_7, o_8\right\}$ with center $R_i.p = o_7$ and radius $R_i.r = d(o_7, o_6)$. As $ d(q, R_i.p) > R_i.r + r $, $R_i$ can be pruned due to Lemma \ref{lemma:RANGE-PIVOT FILTERING}. Note that, for an object $o$ inside ball region $R_i$, if we record its distance $d(o, R_i.p)$ to the partition center $R_i.p$, Lemma \ref{lemma:RANGE-PIVOT FILTERING} can be applied to prune this object by replacing $R_i.r$ with $d(o, R_i.p)$. Hence, Lemma \ref{lemma:RANGE-PIVOT FILTERING} can also be used for pruning single objects. Next, based on the generalized hyperplane partitioning, a double-pivot filtering technique~\cite{ref124} exists.

\begin{lemma}\label{lemma:DOUBLE-PIVOT FILTERING}
	{\bf (DOUBLE-PIVOT FILTERING)}. Given pivots $p_i$ and $p_j$, a query object $q$, and a search radius $r$, if $ d(q, p_i) - d(q, p_j) > {2r} $, $R_i$ can be pruned, as $p_i$ is the corresponding pivot for $R_i$.
\end{lemma}

\begin{Proof}\rm
	For every $o$ in $R_i$, according to the definition of $R_i$, $d(o, p_i) \leq d(o, p_j)$. Based on the triangle inequality, we have $ d(q, p_i) \leq d(o, p_i) + d(q, o) $ and $ d(q, p_j) \geq d(o, p_j) - d(q, o)$. Thus, we can derive that $d(q, p_i) - d(q, p_j) \leq d(o, p_i) + d(q, o) - d(o, p_j) + d(q, o) \leq 2 d(q, o)$ as $d(o, p_i) \leq d(o, p_j) $. If $ d(q, p_i) - d(q, p_j) > {2r}$, then $d(q, o) > r$. Therefore, no object $o (\in R_i)$ can be a result object, and $R_i$ can be pruned.
\end{Proof}
Consider Fig.~\ref{fig:Partitioning Methods}(b), where $o_2$ and $o_6$ are pivots. Since $d(q, o_6) - d(q, o_2) > {2r}$, $R_i = \left\{o_6, o_7, o_8, o_9\right\}$ can be discarded safely according to Lemma \ref{lemma:DOUBLE-PIVOT FILTERING}.

\begin{lemma}\label{lemma:EXCLUSIVE FILTERING}
	{\bf (EXCLUSIVE FILTERING)}. Given a $\rho$-split function $bps^{\rho}(c, o)$, a query object $q$, and a search radius $r$, if $r \leq \rho $ and $bps^{\rho-r}(c, q) =$ `$-$', objects in partitions $R_0$ and $R_1$ can be pruned safely; if $ bps^{r-\rho}(c, q) = 0$, objects in partition $R_1$ can be pruned; and if $ bps^{r-\rho}(c, q) = 1 $, objects in partition $R_0$ can be pruned.
\end{lemma}

\begin{Proof}\rm
	If $r \leq \rho$ and $bps^{\rho - r}(c, q) =$ `$-$', we can get that $d_{med} - \rho + r < d(q, c) \leq d_{med} + \rho - r$. For objects $o$ in $ R_0 $ (i.e., $d(c, o) \leq d_{med} - \rho), d(q, o) \geq d(q, c) - d(c, o) > d_{med} - \rho + r - d_{med} + \rho = r $, and thus, objects in $R_0$ can be pruned. For objects $o$ in $R_1 (i.e., d(c, o) > d_{med} + \rho), d(q, o) \geq d(c, o) - d(c, q) > d_{med} + \rho - d_{med} - \rho  + r = r$, and hence, objects in $R_1$ can be pruned.
	
	If $bps^{r-\rho}(c, q) = 0$, we can get $d(q, c) \leq d_{med} - r + \rho$. For objects in $R_1$ (i.e., $ d(c, o) > d_{med} + \rho), d(q, o) \geq d(c, o) - d(q, c) > d_{med} + \rho - d_{med} + r - \rho > r$. Therefore, objects in $R_1$ can be pruned.
	
	If $bps^{r - \rho}(c, q) = 1$, we can get $d(q, c) > d_{med} + r - \rho$. For objects $o$ in $R_0$ (i.e., $d(c, o) \leq d_{med} - \rho), d(q, o) \geq d(q, c) - d(c, o) > d_{med} + r - \rho - d_{med} + \rho > r $. Thus, objects in $R_0$ can be pruned.
\end{Proof}

Consider Fig.~\ref{fig:Partitioning Methods}(c) and assume that the search region $SR(q, r)$ is the dashed purple circle. Partitions $R_0$ and $R_1$ can be pruned due to $ r \leq \rho $ and $ bps^{\rho - r}(c, q) =$ `$-$'. Lemma \ref{lemma:EXCLUSIVE FILTERING} uses one center for the $\rho$-split function to illustrate the exclusive filtering. However, Lemma \ref{lemma:EXCLUSIVE FILTERING} can also be extended to multiple centers.

Lemmas \ref{lemma:PIVOT FILTERING} through \ref{lemma:EXCLUSIVE FILTERING} are pivot filtering techniques. A distance computation is still needed for verifying each object that cannot be pruned. Consequently, validation techniques are introduced to save unnecessary verifications.

\begin{lemma}\label{lemma:PIVOT VALIDATION}
	{\bf (PIVOT VALIDATION)}. Given a pivot set $P$, a query object $q$, and a search radius $r$, if there exists, for an object $o$ in $O$, a pivot $p_i (\in P)$ satisfying $ d(o, p_i) \leq r - d(q, p_i) $, $o$ is validated to be a result object.
\end{lemma}

\begin{Proof}\rm
	Given a query object $q$, an object $o$, and a pivot $p_i$, $ d(q, o) \leq d(o, p_i) + d(q, p_i) $ because of the triangle inequality. If $ d(o, p_i) \leq r - d(q, p_i) $, then $ d(q, o) \leq r - d(q, p_i) + d(q, p_i) = r $. Thus, $o$ is guaranteed to belong to the search region, which completes the proof.
\end{Proof}

In Fig.~\ref{Mapped vector space}, object $o_2$ can be validated directly without computing the distance $ d(q, o_2)$ due to $d(o_2, o_1) = r - d(q, o_1)$ according to Lemma \ref{lemma:PIVOT VALIDATION}.

\begin{lemma}\label{lemma:RANGE-PIVOT VALIDATION}
	{\bf (RANGE-PIVOT VALIDATION)}. Given a ball partition region $R_i$ with center $R_i.p$ and radius $R_i.r$, a query object $q$, and a search radius $r$, if $ d(q, R_i.p) \leq r - R_i.r $, objects in $R_i$ are validated as result objects.
\end{lemma}

\begin{Proof}\rm
	For any object $o$ contained in $R_i$ (i.e., $d(o, R_i.p) \leq R_i.r)$, if $d(q, R_i.p) \leq r - R_i.r$, then $ d(q, o) \leq d(o, R_i.p) + d(q, R_i.p) \leq r - R_i.r + R_i.r = r $. Hence, $o$ is in the search region according to Definition \ref{defn:metricRangeQuery}, and all objects in $R_i$ are validated as result objects.
\end{Proof}

Consider the example shown in Fig.~\ref{fig:Partitioning Methods}(a), where the dashed green circle represents the search region, and the solid red circle denotes the ball region $R_i = \left\{o_6, o_7, o_8\right\}$ with center $R_i.p = o_7$ and radius $R_i.r = d(o_7, o_6)$. As $ d(q{'}, R_i.p) < r{'} - R_i.r $, $R_i$ is validated according to Lemma \ref{lemma:RANGE-PIVOT VALIDATION}.

\begin{lemma}\label{lemma:EXCLUSIVE VALIDATION}
	{\bf (EXCLUSIVE VALIDATION)}. Given a $\rho$-split function $ bps^{\rho}(c, o) $, a query object $q$, and a search radius $r$, if $ bps^{\rho+r}(c, q) = 0$ $($or $1$$)$, the query result must be contained in $R_0$ $($or $R_1$$)$.
\end{lemma}

\begin{Proof}\rm
	If $ bps^{\rho+r}(c, q) = 0$, we can get that $ d(q, c) \leq d_{med} - \rho - r $. For any object o outside $R_0$ (i.e., $d(c, o) > d_{med} - \rho), d(q, o) \geq d(o, c) - d(q, c) > d_{med} - \rho - d_{med} + \rho + r = r$, and thus, the query result must be contained in $R_0$. If $bps^{\rho+r}(c, q) = 1$, we can get that $d(q, c) > d_{med} + \rho + r$. For any object $o$ outside $ R_1$ (i.e., $d(c, o) \leq d_{med} + \rho), d(q, o) \geq d(q, c) - d(c, o) > d_{med} + \rho + r - d_{med} - \rho = r $, and hence, the query result must be contained in $R_1$.
\end{Proof}

In Fig.~\ref{fig:Partitioning Methods}(c), let the dashed green circle be the search region. The search result must be contained in the hash partition region $R_1$ due to $ bps^{\rho+r{'}}(c, q{'}) = 1 $ according to Lemma \ref{lemma:EXCLUSIVE VALIDATION}. Consequently, we only need to verify the objects in $R_1$.

{In addition to general pivot filtering principles applicable in all metric spaces, alternative filtering approaches are proposed for exact metric search, such as Ptolemaic filtering~\cite{H15, HSLB13} and Hilbert exclusion~\cite{CCVR16, CVCR19}. These alternative filtering approaches use properties in addition to the triangle inequality in order to accelerate search, while the additional properties hold for a series of metrics.}

\section{METRIC INDEXES FOR EXACT SIMILARITY SEARCH}
\label{sec:METRIC INDEXES FOR EXACT METRIC SIMILARITY SEARCH}
	In order to illustrate different designs of the metric indexes, we index the dataset shown in Fig.~\ref{fig:Partitioning Methods} using each of the metric indexes. Sections \ref{subsec:AESA Family} through~\ref{subsec:M-Index} offer an extended presentation of pivot-based indexes covered in our previous work~\cite{ref37}.
	
	\subsection{The GHT Family}\label{subsec:The GHT Family}
\begin{figure}[t]
		\center
		\subfigtopskip=0cm
		\subfigbottomskip=0cm
		\subfigcapskip=0cm
		\hspace{-4mm}
		\subfigure[GHT]{
			\label{GHT}
			\includegraphics[]{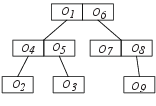}
		}\hspace{-2mm}
		\subfigure[GNAT]{
			\label{GNAT}
			\includegraphics[]{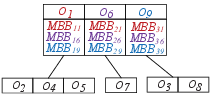}
		}\hspace{-2mm}
		\subfigure[EGNAT]{
			\label{EGNAT}
			\includegraphics[]{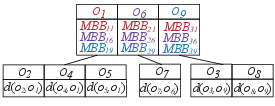}
		}\hspace{-2mm}
		\subfigure[Data Distribution]{
			\label{DataDistribution}
			\includegraphics[]{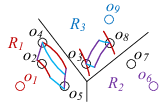}
		}
		\setlength{\abovecaptionskip}{0.2cm}
		\caption{Examples of GHT and its Variants}
		\label{fig:GHTAndItsVariants}
		\vspace{-0.5cm}
	\end{figure}

	The Generalized Hyperplane Tree (GHT){~\cite{ref115}} is a binary tree built by using the generalized partitioning recursively. In the example of GHT depicted in Fig.~\ref{GHT}, objects $o_1$ and $o_6$ in the node denote the centers of two subtrees. 
	
	GHT can be generalized to \textit{m}-ary trees, yielding Geometric Near-neighbor Access Tree (GNAT){~\cite{ref21}}. An example of GNAT is depicted in Fig.~\ref{GNAT}. When constructing GNAT, $m$ centers $ c_i$ $(1 \leq i \leq m) $ are selected each time, and objects are assigned to the nearest center.
	In addition, GNAT stores the minimum bounding box $ \textit{MBB}_{ij} = [\textit{mindist}(o, c_j), \textit{maxdist}(o, c_j)] (o \in R_i)$ of each node with respect to each centers $c_j$, as shown in Fig.~\ref{DataDistribution}, where red circles denote the MBB w.r.t. center $o_1$, the purple circles represent the MBB  w.r.t. $o_6$, while the blue circle denotes the MBB w.r.t.  $o_9$. 

	The Evolutionary Geometric Near-neighbor Access Tree (EGNAT){~\cite{ref76,ref89}} is a dynamic version of GNAT. An example is shown in Fig.~\ref{EGNAT}. It supports insertion and deletion, as well as extends GNAT to be an external memory index. In addition, to improve the pruning ability, each entry in the leaf nodes of EGNAT stores the distance from the entry to its parent entry. 
	
	\textbf{MRQ and M\textit{k}NNQ Processing.} {Query processing traverses the GHT in depth-first manner, where Lemma \ref{lemma:DOUBLE-PIVOT FILTERING} (i.e., double-pivot filtering) is used to filter unqualified nodes. In the case of GNAT, the additional MBBs stored in non-leaf nodes enable pruning using Lemma \ref{lemma:PIVOT FILTERING}. EGNAT leaf nodes store their distances to their parent entry; thus, Lemma \ref{lemma:RANGE-PIVOT FILTERING} is also employed for pruning leaf entries. M\textit{k}NNQ($q$, $k$) processing based on GHT, GNAT, and EGNAT follows the second approach introduced in Section \ref{subsec:Metric Similarity Queries}.}
	
	
	\textbf{Discussion.} The storage cost of GHT is $O(ns)$, and its construction cost is $ \Omega (n\log_{2}n) $, while GNAT has storage cost $ O(ns + mn) $ and construction cost $ \Omega (mn\log_{m} n) $, where $n$ denotes the total number of objects, $s$ represents the size of an object, and $m$ denotes the tree arity. {In the original study~\cite{ref21}, GNAT has $O(m^2 n_{node})$ MBB storage cost,  where $n_{node}$ denotes the number of non-leaf nodes. Here, we use $O(n/m)$ to estimate $n_{node}$ so that $O(mn)$ space is used to store the MBB information.}   
	Note that, GHT, GNAT, and EGNAT are unbalance trees, meaning that their worst construction cost is $O(n^2)$. If we assume that the tree structure is balanced, we get optimal construction costs of $ \Omega (n\log_{2} n) $ for GHT and $ \Omega (mn\log_{m} n) $ for GNAT and EGNAT.
	
	\subsection{BST Family}\label{subsec:BST Family}
\begin{figure}[t]
		\centering
		\subfigtopskip=0cm
		\subfigbottomskip=0cm
		\subfigcapskip=0cm
		\hspace{-6mm}
		\begin{minipage}[t]{0.45\linewidth}
			\centering
			\setlength{\abovecaptionskip}{0cm}
			\includegraphics[]{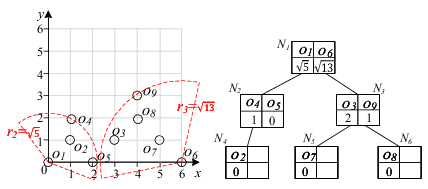}
			\caption{{BST}}
			\label{fig:BST}
		\end{minipage} \hspace{7mm}
		\begin{minipage}[t]{0.45\linewidth}
			\centering
			\setlength{\abovecaptionskip}{0cm}
			\includegraphics[]{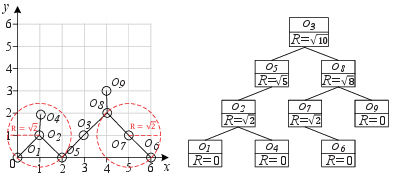}
			\caption{{SAT}}
			\label{fig:SAT}
		\end{minipage}
		\setlength{\abovecaptionskip}{0.2cm}
		\vspace{-0.5cm}
	\end{figure}

	Like GHT, {the Bisector Tree (BST)~\cite{ref69}} is a binary tree, which is constructed by inserting objects {one by one.} However, unlike GHT, BST uses a ball (i.e., a center with a radius) to represent each partition/node.
	{To improve efficiency by obtaining a relatively balanced structure, BST can be built recursively in a top-down manner using generalized hyperplane partitioning, yielding $\rm BST^{*}$.} 
	{Fig.~\ref{fig:BST} shows an example of $\rm BST^{*}$, where $N_2$ is represented by a ball with center $o_1$ and radius $r_2 = \sqrt{5}$.}  In contrast, a variant called the BU-tree{~\cite{ref72}} builds the tree in a bottom-up manner instead of top-down. In BST, a sub node may have a larger radius than its parent node. To avoid this, two variants of BST, called {MBST~\cite{ref92} and VT~\cite{ref46}}, are proposed, where VT nodes have arity 3.

	\textbf{MRQ and M\textit{k}NNQ Processing.} MRQ($q$, $r$) processing using BST proceeds as for GHT. The only difference is that BST uses Lemma \ref{lemma:RANGE-PIVOT FILTERING} (i.e., the range-pivot filtering) instead of Lemma \ref{lemma:DOUBLE-PIVOT FILTERING} to filter nodes.
	 M\textit{k}NNQ($q$, $k$) processing based on BST follows the second strategy described in Section \ref{subsec:Metric Similarity Queries}.
	
	\textbf{Discussion.} The storage costs of $\rm BST^{(*)}$ and MBST are $O(ns)$, and the corresponding construction costs are $\Omega(n\log_2n)$. The degree of balance depends on the chosen centers and the data distribution. If the centers used are chosen well, the BST and its variants can be balanced trees. 
	As the VT nodes have arity 3, its construction cost is $\Omega(n\log_3n)$. However, the BU-tree repeatedly forms a parent node from the two nodes with the minimum distance. Thus, its construction cost is $O(n^3)$.
	
	\subsection{SAT Family}\label{subsec:SAT family}
	In contrast to GNAT, the Spatial Approximation Tree (SAT){~\cite{ref84,ref85}} selects centers inspired by the Delaunay graph. An example is shown in Fig.~\ref{fig:SAT}. For each node $c$, SAT chooses a set of so-called nearest neighbors $N(c)$ as the centers of its sub-nodes. More specifically, $N(c)$ does not denote the real nearest neighbors of $c$, but consists of the objects that are closer to $c$ than to other objects in $ N(c)$, i.e., $ o \in N(c)$ iff $\forall{u} \in N(c) - \left\{o\right\}, d(o, c) < d(o, u) $. {Next, SAT uses the generalized hyperplane partitioning to construct the tree, while each node in SAT also uses the ball representation.} As an optimization, the farthest outlier data is selected as the root node instead of a randomly chosen root to maximize the hyperplane separation ~\cite{ref30}. In addition, when dividing the dataset in each iteration, the objects are sorted in ascending order of their distances to the parent entry. Another typical index called the \textit{k} nearest-neighbor graph (\textit{k}NNG) is similar as SAT. The difference is that while SAT uses $N(c)$ as sub-nodes, \textit{k}NNG uses real \textit{k} nearest neighbors of $c$ as its sub-nodes.
	
	Dynamic SAT (DSAT){~\cite{ref86,ref87,ref88,ref90,ref91}} extends SAT from being a static in-memory index to being a dynamic secondary-memory index.
	{DSAT is built by inserting objects one by one using {generalized hyperplane partitioning}, and each object is associated with an insertion timestamp.} 
	The sub-nodes are sorted in ascending order of their insertion times.	
	To better cluster each page of DSAT, DSAT with clusters (DSACLT){~\cite{ref12,ref22}} is proposed. Each node in DSACLT is stored as a single page in secondary memory. In addition, DSACLT stores \textit{k} nearest neighbors (\textit{k}NNs) of the node center in each node. Further, the distance from each \textit{k}NN to the node center is stored. Two  DSACLT versions exist, $\rm DSACLT^+$ and $\rm DSACLT^*$, where $\rm DSACLT^+$ is tree structured, while $\rm DSACLT^*$ is a list.

	
	\textbf{MRQ Processing using SAT.} MRQ($q$, $r$) processing using SAT proceeds as for GHT, using depth-first search. {In particular, Lemmas \ref{lemma:RANGE-PIVOT FILTERING} and \ref{lemma:DOUBLE-PIVOT FILTERING} are used to prune nodes.}
	Instead of applying Lemma \ref{lemma:DOUBLE-PIVOT FILTERING} only once for each sub-node, all sub-nodes of the same parent node and accessory nodes can be used for pruning. Here, the accessory nodes $\textit{AN}(o)$ of $o$ are the nodes in the path from $o$ to the root. For each sub-node $ o \in N(c) $, if $ \exists{u} \in (N(c) - \left\{o\right\}) \cup \textit{AN}(o) $ st. $d(q, o) - d(q, u) > 2 \times r $, the subtree of $o$ can be pruned.
	
	{\textbf{MRQ Processing using DSAT and DSACLT.} Unlike MRQ($q$, $r$) processing using SAT, DSAT cannot prune a whole subtree using Lemma \ref{lemma:DOUBLE-PIVOT FILTERING}. To see why, assume that a node has two sub-nodes $c_i$ and $c_j$ $(j > i)$, indicating that $c_i$ is inserted before $c_j$. Then, objects $o$ inserted before $c_j$ in the subtree of $c_i$ might belong to the subtree of $c_j$. Therefore, we still visit older nodes when a newer center exists that satisfies the condition of Lemma \ref{lemma:DOUBLE-PIVOT FILTERING}. MRQ($q$, $r$) processing using DSACLT proceeds as when using DSAT. The only difference is that, for unpruned nodes, DSACLT determines whether the stored \textit{k}NNs are in the final result by using Lemma \ref{lemma:RANGE-PIVOT FILTERING}. }
	
	
	\textbf{M\textit{k}NNQ Processing.} M\textit{k}NNQ($q$, $k$) processing using SAT follows the second approach from Section \ref{subsec:Metric Similarity Queries}. Best-first traversal is used, so that nodes are visited in ascending order of their minimum distances to the query object $q$. In the cases of DSAT and DSACLT, the nodes are visited in ascending order of their insertion times, such that these indexes are traversed in depth-first order.
	
	\textbf{Discussion.} The construction costs of SAT, DSAT, and DSACLT are $ O(n\log^2{n}/\log\log_n)$, $O(nm$ $\log_m{n})$, and $O(nm\log_{m}{n⁄k})$, respectively, where $m$ is the maximum tree-arity and $k$ denotes the number of nearest neighbors stored in each DSACLT node.  The storage costs of SAT, DSAT, and DSACLT are all $O(ns)$. Note that, the construction cost of \textit{k}NNG is $O(n^2)$ in the worst case and its storage cost is $O(ns + nk)$. Unlike other indexes (e.g., the GHT family) that only use sibling centers for pruning, the SAT family indexes can also use ancestor nodes for pruning, and thus, offer improved pruning capabilities. Due to the dynamic insertions, the pruning of DSAT is weaker than that of SAT. DSACLT stores more pre-computed distances in each node than those of SAT and DSAT, which improves its pruning capabilities over these indexes. However, the height of DSACLT is smaller than that SAT and DSAT, which reduces its pruning capabilities over these. Consequently, the combined pruning enabled by DSACLT depends on the data distribution.
	
		\begin{figure}
		\centering
		\subfigtopskip=0cm
		\subfigbottomskip=0cm
		\subfigcapskip=0cm
		\subfigure[Data distribution]{
			\label{M tree Data distribution}
			\includegraphics[]{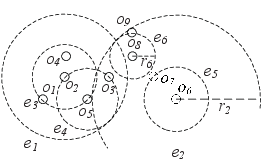}
		}
		\subfigure[The M-tree]{
			\label{The M-tree}
			\includegraphics[]{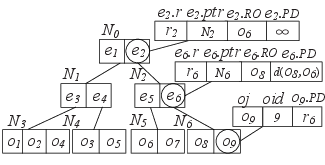}
		}
		\vspace{-0.4cm}
		\caption{Example of the M-tree}
		\label{fig:Example of M-tree}
		\vspace{-0.5cm}
	\end{figure}

	\subsection{M-tree Family}\label{subsec:M-tree Family}
	
	{The M-tree{~\cite{ref38,ref42,ref110}} is a dynamic tree that uses ball partitioning. Fig.~\ref{fig:Example of M-tree} shows an M-tree, where each non-leaf entry $e$ records a center $e.RO$, a covering radius $e.r$,  a parent distance $e.PD$ that equals the distance from $e$ to the center of its parent entry, and a pointer to the root of its sub-tree. A leaf entry stores detailed object information with the parent distance.}
	Many  M-tree variants exist (cf. Section \ref{subsec:Metric Indexes for Exact Metric Similarity Queries}) that aim to improve its construction or query efficiency. Here, we only introduce a representative variant called the PM-tree{~\cite{ref112}}. The PM-tree combines the pivot mapping and the M-tree, where the M-tree is used to cluster objects and the pivot mapping is utilized to avoid unnecessary distance computations. Hence, {unlike} the M-tree, each leaf entry of the PM-tree stores the mapped vector (i.e., the pre-computed distances to the pivots) with every object. In each non-leaf entry, the PM-tree stores an MBB that contains all the mapped vectors in its child leaf entries. 
	
		
	\textbf{MRQ and M\textit{k}NNQ  Processing using the M-tree.} {In order to answer MRQ($q$, $r$) using the M-tree, the entries are traversed depth-first, and Lemma \ref{lemma:RANGE-PIVOT FILTERING} is used for pruning. M\textit{k}NNQ($q$, $k$) using the M-tree follows the second strategy from Section \ref{subsec:Metric Similarity Queries}. The entries are traversed in best-first manner, i.e., in ascending order of their minimum distances to the query object $q$, where Lemma \ref{lemma:RANGE-PIVOT FILTERING} is employed to eliminate unqualified entries.}

\textbf{MRQ and M\textit{k}NNQ  Processing using the PM-tree.} 	MRQ($q$, $r$) and M\textit{k}NNQ($q$, $k$) using PM-tree are similar as that using M-tree, the only difference is that Lemma \ref{lemma:PIVOT FILTERING} can also be used for pruning entries due to pivot mapping technique.
	
	
	\textbf{Discussion.} The M-tree is a balanced tree, and thus, the storage cost of the M-tree is $O(ns + ns/m)$, and {its construction cost is $ O(nm\log_{m}{n})$ or $O(nm^2\log_{m}{n})$, depending on the split strategy}. The tree-arity $m$ of the M-tree depends on its disk page size to store each node. The PM-tree stores pre-computed distances w.r.t. the pivots in the tree structure. Hence, the construction cost of the PM-tree is $O(n(m..{m^2})\log_{m}{n}+nl\log_{m}{n})$, and its storage cost is $O(n(s + l) + n(s + l)/m + ls)$, where $l$ denotes the number of pivots, and it needs $O(ls)$ space to store the pivots. The PM-tree costs are relatively high compared with those of the M-tree. Note that, the M-tree and its variants (e.g., the PM-tree) store the data objects in its entries instead of in a separate file; thus, the page/node size varies for different types of data. In particular, for complex objects (e.g., the 282 dimensional vectors used in our experiments), the M-tree family indexes need a large page size.
	
	\subsection{LC Family}\label{subsec:LC Family}
	
\begin{figure} [t]
		\centering
		\subfigtopskip=0cm
		\subfigbottomskip=0cm
		\subfigcapskip=0cm
		\subfigure[Data distribution]{
			\label{LC Data distribution}
			\includegraphics[]{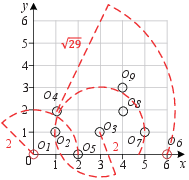}
		}
		\subfigure[LC structure]{
			\label{LC structure}
			\includegraphics[]{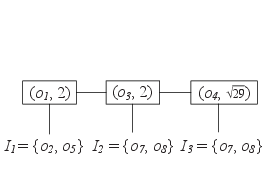}
		}
		\setlength{\abovecaptionskip}{0.2cm}
		\caption{Example of LC with Fixed Size}
		\label{fig:Example of LC with fixed size}
		\vspace{-0.5cm}
	\end{figure}

	List of Clusters (LC){~\cite{ref28,ref31}} is a list of clusters, where each cluster is represented by a center and a radius. Each cluster has a corresponding bucket, which contains objects whose distances to the center are not larger than the radius. LC has two versions, i.e., fixed radius and fixed size. Fixed radius means that the cluster radius is fixed, while fixed size indicates that the number of objects in each bucket is fixed. Fig.~\ref{fig:Example of LC with fixed size} shows an example of LC with fixed size. Two variants Dynamic LC (DLC){~\cite{ref91}} and Hierarchy of Clusters{~\cite{ref56,ref57}}  (HC) are proposed, where DLC is a dynamic version of LC, and HC utilizes a tree structure to store the list of clusters.
	
	\textbf{MRQ and M\textit{k}NNQ Processing.} MRQ($q$, $r$) using LC, DLC, or HC visits the list of clusters in sequel or visits the tree in depth-first order, {where Lemma \ref{lemma:RANGE-PIVOT FILTERING} is used for filtering buckets. If the search region is contained in the ball cluster, the search stops, and the results are returned.} M\textit{k}NNQ($q$, $k$) using LC, DLC, or HC adopts the second strategy from Section \ref{subsec:Metric Similarity Queries}, where Lemma \ref{lemma:RANGE-PIVOT FILTERING} is used for pruning.
	
	\textbf{Discussion.} The construction costs of LC and DLC are $O(n^2/m)$, where $m$ denotes the number of objects in each bucket. The construction cost of LC is very high; however, it can achieve high query efficiency due to its compact clusters. HC is built recursively in a binary way to reduce the construction cost of LC, which yields a construction cost of $O(n\log_2{n⁄m})$. The storage sizes of LC, DLC, and HC are all $O(ns)$.
	
	\subsection{D-index Family}\label{subsec:D-index Family}
	
\begin{figure} [t]
		\centering
		\includegraphics{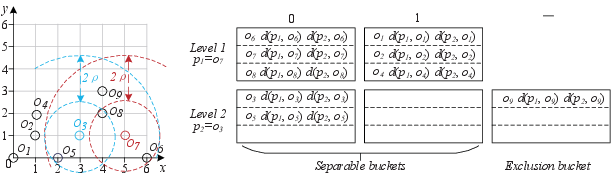}
		\setlength{\abovecaptionskip}{0.2cm}
		\caption{Example of the D-index}
		\label{fig:Example of D-index}
		\vspace{-0.5cm}
	\end{figure}

	The D-index{~\cite{ref48,ref49,ref123}} combines hash partitioning and the pivot mapping. The basic idea of the D-Index is to create a multilevel structure that uses several $ \rho$-split functions as defined in Definition \ref{defn:HASH PARTITIONING}, one for each level, to create buckets for storing objects.
	Here, the $\rho$-split functions of individual levels use the same $\rho$. In Fig.~\ref{fig:Example of D-index}, a $ \rho$-split function based on $o_7$ is used at level 1, and a $ \rho$-split function based on $o_3$ is used at level 2. {Objects in the exclusion bucket `$-$' (i.e., $o_3, o_5, o_9$) at level 1 are candidates to be divided at level 2, and the exclusion bucket of the last level forms the exclusion bucket of the D-index.}
	
\begin{figure}
		\centering
		\subfigtopskip=0cm
		\subfigbottomskip=0cm
		\subfigcapskip=0cm
		\subfigure[Data distribution]{
			\label{MB+Tree Data distribution}
			\includegraphics[]{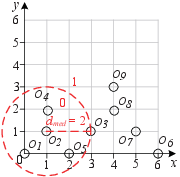}
		}
		\subfigure[block tree]{
			\label{MB+Tree block tree}
			\includegraphics[]{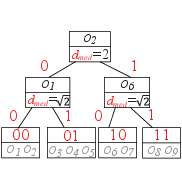}
		}
		\subfigure[$\rm B^+$-tree]{
			\label{MB+Tree B+-tree}
			\includegraphics[width=0.35\linewidth]{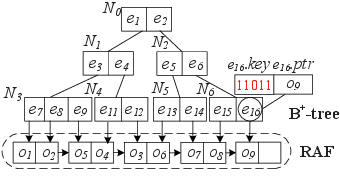}
		}
		\setlength{\abovecaptionskip}{0.2cm}
		\caption{Example of the $\rm MB^+$-tree using Hash Partitioning}
		\label{fig:Example of MB+-tree by using hash partitioning}
		\vspace{-0.5cm}
	\end{figure}

	The $\rm MB^+$-tree{~\cite{ref63}} divides the dataset into two subsets by using either hash partitioning or generalized hyperplane partitioning recursively. Fig.~\ref{fig:Example of MB+-tree by using hash partitioning} gives an example of the $\rm MB^+$-tree using hash partitioning, where $ \rho $ is set to 0. The $\rm MB^+$-tree includes two parts, i.e., block tree and $\rm B^+$-tree. The block tree stores the partition information, where each internal node records the partition center $c$ and the medium distance $d_{med}$ used for a $\rho$-split function. {
	For each object $o$, the $\rm MB^+$-tree generates a key that is formed by the partition key $pk$ and the distance key $dk$. 
	Finally, all keys are indexed by the $\rm B^+$-tree, and objects are stored in a separate random access file RAF.}

	\textbf{MRQ Processing using the D-index.} The D-index is traversed top-down. At each level, we use Lemma \ref{lemma:EXCLUSIVE FILTERING} to prune unqualified buckets.
	In addition, if the condition of Lemma \ref{lemma:EXCLUSIVE VALIDATION} holds, we can retrieve the final result in the corresponding bucket and terminate the search. Note that, when searching a bucket, Lemma \ref{lemma:PIVOT FILTERING} is used for pruning to boost efficiency.
	
	\textbf{MRQ Processing using the $ \bm{{\rm MB^+}}$-tree}. MRQ($q$, $r$) using the $\rm MB^+$-tree visits the block tree in depth-first order, and Lemma \ref{lemma:EXCLUSIVE VALIDATION} is employed to find sub-trees that needs to be visited. {When a leaf node is reached, it computes the distance key range for the query object $q$.}
	Next, all the candidates in the B$^+$-tree that belong to this leaf node and fall in this distance key range are found, and it is determined each candidate is in the final result.
	
	\textbf{M\textit{k}NNQ Processing using the D-index.} M\textit{k}NNQ($q$, $k$) using the D-index is complex. It adopts the second solution from Section \ref{subsec:Metric Similarity Queries}. However, instead of setting the search radius to infinity initially, we first set the search radius to $\rho$, and then search the \textit{k} NNs. If the upper bound distance between the current \textit{k}NNs and the query object exceeds $\rho$ (i.e., the initial search radius $\rho$ is underestimated), we need to search the D-index again to refine the result.

	\textbf{M\textit{k}NNQ Processing using the $\bm{{\rm MB^{+}}}$-tree}. M\textit{k}NNQ($q$, $k$) using the $\rm MB^{+}$-tree adopts the third strategy stated in Section \ref{subsec:Metric Similarity Queries}. It first finds \textit{k} candidate NNs according to the keys, i.e., finds \textit{k} candidates with keys nearest to the query object's key, and then calculates the current \textit{k}-th NN distance $\textit{ND}_k$ using the \textit{k} candidates. Thereafter, M\textit{k}NNQ($q$, $k$) is transformed to a metric range search MRQ($q$, $ND_k$) to find the final result.
	
	\textbf{Discussion.} The construction cost of the D-index is $O(nl)$, as it depends on the number of pivots (i.e., the number of hash functions) used. The storage cost of the D-index is $O(ns + nl + ls)$, where $O(ns)$ is the cost to store the dataset, $O(nl)$ is cost to store the pre-computed distances w.r.t. the pivots, and $O(ls)$ is the cost to store the pivots used for pivot mapping and hash partitioning. The construction cost of the $\rm MB^+$-tree is $O(n\log_2{n⁄m} + nm\log_{m}{n})$, where it needs $O(n\log_2{n⁄m})$ to build the block tree, and $O(nm\log_{m}{n})$ to build the $\rm B^+$-tree, where $m$ denotes the number of entries in both the block tree leaf node and  the $\rm B^+$-tree node. The storage cost of  the $\rm MB^+$-tree is $O((n+n/m)(s+\log_2{n⁄m}+ \log_2{n_d}))$, in which $n_d$ represents the maximum distance value of $d()$. This is because, the length of a partition key is $O(\log_2{n⁄m})$, while the length of a distance key is $O(\log_2{n_d})$. It needs $O((n+n/m)(\log_{2}{n⁄m} + \log_{2}{n_d}))$ to store  the $\rm B^+$-tree, $O(ns)$ to store  the RAF, and $O((s+\log_2{n⁄m})n/m)$ to store the block tree.
	
	\subsection{AESA Family}\label{subsec:AESA Family}
	
\begin{figure}
		\begin{minipage}[t]{0.28\linewidth}
			\centering
			\includegraphics[]{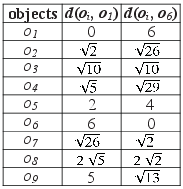}
			\caption{Example of LAESA}
			\label{fig:Example of LAESA}
			\setlength{\abovecaptionskip}{0cm}
		\end{minipage}	\hspace{-7mm}
		\begin{minipage}[t]{0.45\linewidth}
			\centering
			
			\includegraphics[width=0.9\linewidth]{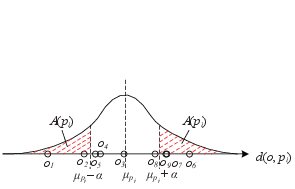}
			\caption{Illustration of $A(p_i)$}
			\label{fig:Illustration of $A(p_i)$}
			\setlength{\abovecaptionskip}{0cm}
		\end{minipage}\hspace{-3mm}
		\begin{minipage}[t]{0.28\linewidth}
			\centering
			\includegraphics[]{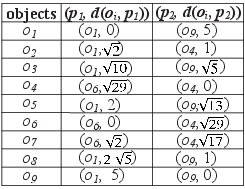}
			\caption{Example of EPT}
			\label{fig:Example of EPT}
			\setlength{\abovecaptionskip}{0cm}
		\end{minipage}
		\vspace{-0.5cm}
	\end{figure}

	AESA{~\cite{ref102}} uses a table to store the distances from every object to other objects. If $n$ is the cardinality of a dataset, the storage cost of AESA is $O(n^2)$, which is high for large datasets. As an example, if $n = 100,000$, the storage cost is 80G, which renders AESA a theoretical metric index. To further improve the query efficiency, two variants ROAESA{~\cite{ref118}} and iAESA{~\cite{ref118}} are proposed, where the pre-computed distances are visited in a particular order during the search. To reduce the storage cost of AESA, Linear AESA (LAESA){~\cite{ref79}} is developed. LAESA only stores the distances from each object to the pivots in a pivot set $P$. Fig.~\ref{fig:Example of LAESA} shows an example of LAESA when using $P = \left\{o_1, o_6\right\}$. In addition, objects in LAESA can be organized in a binary tree TLAESA{~\cite{ref80,ref113}} instead of a table.
	
	\textbf{MRQ and M\textit{k}NNQ Processing.} {MRQ($q$, $r$) processing using LAESA visits objects one by one, and prunes objects using Lemma \ref{lemma:PIVOT FILTERING}.}
 M\textit{k}NNQ($q$, $k$) processing based on LAESA follows the second strategy covered in Section \ref{subsec:Metric Similarity Queries}, {and Lemma \ref{lemma:PIVOT FILTERING} is used for pruning.}
	
	\textbf{Discussion.} The construction costs of AESA, ROAESA and iAESA are all $O(n^2)$, while the storage costs of AESA, ROAESA, and iAESA are all $O(ns + n^2)$, as they need $O(ns)$ to store the data and $O(n^2)$ to store the pre-computed distances. To reduce the cost of AESA, the construction cost of LAESA is $O(nl)$, and its storage cost is $O(ns + nl + ls)$, as LAESA needs $O(ns)$ to store the data, $O(nl)$ to store the pre-computed distances, and $O(ls)$ to store pivots. Since TLAESA is a tree structure, the construction cost of TLAESA is $O(n\log_{2}{n}+ nl)$, but its storage cost is the same as LAESA. For similarity search processing, although LAESA utilizes Lemma \ref{lemma:PIVOT FILTERING} to avoid certain unnecessary distance computations, it still needs to scan the full table to find the result set, incurring additional scanning cost.
	
	\subsection{EPT}\label{subsec:EPT}
	Unlike LAESA that utilizes the same pivots for each object, Extreme Pivot Table (EPT){~\cite{ref103}} selects different pivots for different objects in order to achieve better search performance. EPT consists of a set of pivot groups. Each group $G$ contains $g$ pivots $ p_i (1 \leq i \leq g) $, according to which the entire dataset $O$ is partitioned into $g$ parts $A(p_i)$, such that $ A(p_i) \cap A(p_j) = \emptyset$ $(i \ne j)$ and $ \cup_{p_i \in G} A(p_i) = O $. An object $o$ belongs to $ A(p_i)$ iff $|d(o, p_i) - \mu_{p_i}| \geq \alpha $, where $ \mu_{p_i} $ is the expected value of $d(o, p_i)$. For instance, in Fig.~\ref{fig:Illustration of $A(p_i)$}, $A(p_i) = \left\{o_1, o_2, o_7, o_9, o_6\right\}$. {However, it is difficult to obtain $ \alpha $, so EPT tries to maximize $ \alpha $.} {In order to further improve the efficiency of EPT, EPT* is equipped with a new pivot selection algorithm (PSA)~\cite{ref37}.}

	\textbf{MRQ and M\textit{k}NNQ Processing.} Like LAESA, EPT, and $\rm EPT^*$ use tables to store pre-computed distances. The only difference is that EPT and $\rm EPT^*$ utilize different pivots for different objects, while LAESA uses the same pivots for every object. Consequently, MRQ and M\textit{k}NNQ processing on EPT or $\rm EPT^*$ are the same as those on LAESA.
	
	\textbf{Discussion.} The construction cost of EPT is $O(nlg)$, and its storage cost is $O(nl + ns + lgs)$, as EPT needs $O(ns)$ to store the data, $O(nl)$ to store the pre-computed distances, and $O(lgs)$ to store the pivots. EPT has $l$ groups while each group contains $g$ pivots, and thus, it needs $O(lgs)$ to store all the pivots. For each object, we will select the best pivot among each pivot group, and hence, it needs $O(nl)$ to store the pre-computed distances between objects and the best pivots. To get high quality pivots, the construction cost of $\rm EPT^*$ is $O(nll_cn_s)$, where $l_c$ is the number of candidate pivots and $n_s$ denotes the cardinality of the sample set. The storage cost of $\rm EPT^*$ is $O(ns + nl + l_cs)$, as $\rm EPT^*$ needs $O(ns)$ to store the data, $O(nl)$ to store the pre-computed distances, and $O(l_cs)$ to store all the pivots. $\rm EPT^*$ achieves better similarity search performance than EPT contributed by the higher quality pivots selected by PSA. 
	Nonetheless, the pivot selection is costly, and thus, the construction cost of $\rm EPT^*$ is very high.
	
	
		\begin{figure}[b]
			\vspace{-3mm}
		\begin{minipage}[t]{0.25\linewidth}
			\centering
			\setlength{\abovecaptionskip}{0cm}
			\includegraphics[scale=1]{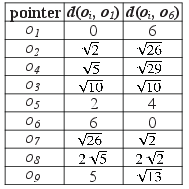}
			\caption{CPT}
			\label{fig:Example of CPT}
		\end{minipage}
		\begin{minipage}[t]{0.3\linewidth}
			\centering
			\setlength{\abovecaptionskip}{0cm}
			\includegraphics[scale=1.1]{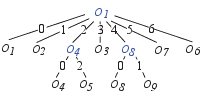}
			\caption{BKT}
			\label{fig:BKT}
		\end{minipage}
		\begin{minipage}[t]{0.3\linewidth}
			\centering
			\setlength{\abovecaptionskip}{0cm}
			\includegraphics[scale=1.1]{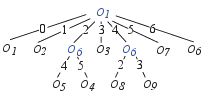}
			\caption{FQT}
			\label{fig:FQT}
		\end{minipage}
		\vspace{-0.5cm}
	\end{figure}

	\subsection{CPT}\label{subsec:CPT}
	LAESA and EPT store the distance table and the data in main memory, and similarity query processing needs to scan the whole table. However, when the size of the dataset exceeds the capacity of the main memory, it is necessary to store the dataset on disk, and it becomes attractive to cluster the data to improve I/O efficiency.
	The CPT{~\cite{ref82}} uses an M-tree to cluster and store the objects on disk. The CPT consists of two parts, i.e., the distance table (depicted in Fig.~\ref{fig:Example of CPT}) and the M-tree (depicted in Fig.~\ref{fig:Example of M-tree}). 
	Note that, the distance table includes the pointers to the leaf entries in the M-tree, in order to enable loading of the corresponding objects for verification.

	\textbf{MRQ and M\textit{k}NNQ Processing.} MRQ and M\textit{k}NNQ processing using CPT is similar as LAESA. The difference is when an object is not pruned by Lemma \ref{lemma:PIVOT FILTERING}, the object must be read from disk.
	
	\textbf{Discussion.} The construction cost of CPT is $O(n(m..{m^2})\log_{m}n + nl) $, as it needs $O(nl)$ to construct the table and $O(n(m..{m^2})\log_{m}{n})$ to build the M-tree. The storage cost of CPT is $O(ns + ns/m + nl + ls)$, as it needs $O(ns + ns/m)$ to store the M-tree, $O(nl)$ to store the pre-computed distances, and $O(ls)$ to store the pivots. Using CPT, we can avoid loading the whole dataset into main memory during query processing. However, CPT incurs the I/O cost to load objects from disk. In addition, the distance table is stored in main memory, meaning that the applicability of CPT is still limited to datasets for which the distance table fits in main memory.
	
	\subsection{BKT}\label{subsec:BKT}
	BKT{~\cite{ref23}} is a tree structure designed for discrete distance functions. It chooses a pivot as the root, and maintains the objects having the distance $i$ to the pivot in its $i^{th}$ sub-tree.
	Fig.~\ref{fig:BKT} gives an example BKT, constructed based on the objects from Fig.~\ref{fig:Partitioning Methods} and the discrete distance function $L_{\infty}$-norm. However, for other metric index examples, Euclidean distance ($L_2$-norm) is used.
	Note that, the continuous distance range can be partitioned into discrete ranges used for indexing. For example, if the continuous distance function range is [0, 30], we can divide it into three disjoint ranges [0, 10), [10, 20), [20, 30] in order to simulate the discrete distance function. Hence, BKT can be adapted to support both discrete and continuous distance functions.

	\textbf{MRQ and M\textit{k}NNQ Processing.} {In order to compute MRQ($q$, $r$), the nodes in BKT are traversed in depth-first fashion, and  Lemma \ref{lemma:PIVOT FILTERING} is used.
	In order to compute M\textit{k}NNQ($q$, $k$), the nodes in BKT are traversed in best-first manner, i.e., in ascending order of their minimum distances to the query object $q$, and Lemma \ref{lemma:PIVOT FILTERING} is again used to prune unqualified nodes. Here,  M\textit{k}NNQ($q$, $k$) follows the second strategy from Section \ref{subsec:Metric Similarity Queries}.}
	
	\textbf{Discussion.} BKT is an unbalanced tree. The construction cost of BKT is $O(nl)$, where $l$ denotes the height of BKT. To compare different pivot-based metric methods, we use the same number of pivots, and thus, the height of BKT is set to $l$ in this paper, and {the construction cost is $O(nl)$ instead of $O(nlogn)$}. However, we cannot use the same set of pivots for BKT. Like in the case of other pivot-based methods, BKT randomly selects the pivots for its sub-trees. If BKT uses the same pivots as other pivot-based metric indexes, it produces FQT as discussed below. The storage cost of BKT is $O(ns + ln_d)$, where $n_d$ denotes the number of discrete values in the domain of distance function $d()$. {Here, we store the associated distances (e.g., 0, 1, 2, 3, 4, 5, 6 in Fig.~\ref{fig:BKT}) for each sub-tree, which results in an additional storage cost of $O(ln_d)$.} To avoid empty sub-trees for large distance domains, each sub-tree covers a range of distance values, which are stored together. 
	
	\subsection{FQ Family}\label{subsec:FQ Family}
	Unlike BKT, FQT{~\cite{ref10}} utilizes the same pivot at the same level. Fig.~\ref{fig:FQT} shows an example of FQT, where $o_1$ is used for the first level, and $o_6$ is used for the second level. FQT is an unbalanced tree. Hence, FHQT{~\cite{ref11}} is proposed, where objects are stored in the leaves, and all the leaves are at the same level. In addition, FHQT can also be stored as FQA{~\cite{ref32}} {using a table}. Note that, FQA and LAESA are the same. Although FQT is designed for a discrete distance function, it can also be extended to support a continuous distance function, as in the case of BKT.
	
	\textbf{MRQ and M\textit{k}NNQ Processing.} MRQ and M\textit{k}NNQ processing using FQT are the same as when using BKT.
	
	\textbf{Discussion.} Similar to BKT, the construction cost of FQT is $O(nl)$, and the storage cost of FQT is $O(ns + nl)$, where $l$ is the height of FQT. This is because, in order to utilize the same set $P$ of pivots as other pivot-based metric indexes, the height of FQT is set to the number of pivots, and $p_i \in P$ is set as the pivot for the $i^{th}$ level. With well-chosen pivots, FQT is expected to perform better than BKT.

	\subsection{VPT Family}\label{subsec:VPT Family}
\begin{figure}[t]
		\centering
		\subfigtopskip=0cm
		\subfigbottomskip=0cm
		\subfigcapskip=0cm		
		\begin{minipage}[b]{0.55\linewidth}
			\subfigure[VPT]{
				\label{VPT}
				\includegraphics[width=0.45\linewidth]{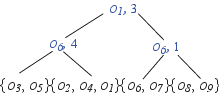}
			}
			\subfigure[MVPT]{
				\label{MVPT}
				\includegraphics[width=0.45\linewidth]{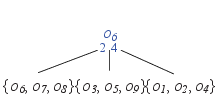}
			}
			\setlength{\abovecaptionskip}{0cm}
			\caption{Example of the VPT Family}
			\label{Examlpe of VPT Family}
		\end{minipage} \hspace{-5mm}
		\begin{minipage}[b]{0.45\linewidth}		
			\includegraphics[width=\linewidth]{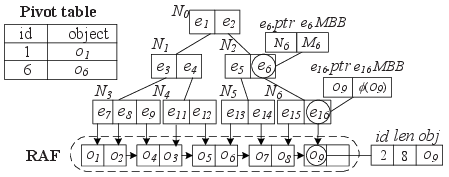}
			\setlength{\abovecaptionskip}{0cm}
			\caption{Example of the OmniR-tree}
			\label{fig:Example of OmniR-tree}
		\end{minipage}	
		\vspace{-0.5cm}
	\end{figure}

	VPT{~\cite{ref114,ref115,ref122}} is a binary tree. It chooses a pivot $p$ as the root, and selects a medium value $d_{med}$ so that the objects $o$ with $ d(o, p) \leq d_{med} $ are put in the left sub-tree, while the remaining objects are put in the right sub-tree.
	Fig.~\ref{VPT} depicts an example of VPT, where $L_{\infty}$-norm is used. Note that, the pivots for the nodes at the same level can be different. In order to be able to compare the efficiency of different indexes using the same set of pivots, nodes of VPT at the same level share the same pivot. To support insertions and deletions of VPT, a dynamic version DVPT{~\cite{ref58}} is designed.
	VPT can be generalized to \textit{m}-ary trees, yielding MVPT{~\cite{ref18,ref19}}. Each time, MVPT selects $m-1$ medium values $d_1, d_2, \cdots, d_{m-1}$. 
	Fig.~\ref{MVPT} gives an example of MVPT, where $m$ is 3. In addition, MVPT can also use multiple pivots (instead of using only one pivot) to partition each node.
	
	\textbf{MRQ and M\textit{k}NNQ Processing.} MRQ and M\textit{k}NNQ processing using VPT are similar as the processing using BKT.
	
	\textbf{Discussion.} Unlike BKT and FQT, MVPT is a balanced tree. The construction costs of VPT and DVPT are $O(n\log_{2}{n})$, while the construction cost of MVPT is $O(n\log_{m}{n})$. As $m$ grows, the pruning ability first increases and then drops. This occurs because, with larger $m$ values, more compact sub-trees are obtained at every tree level. Nevertheless, larger $m$ values result in lower MVPT tree levels, indicating that fewer pivots are available for pruning. The storage costs of VPT, DVPT, and MVPT are all $O(ns)$. MVPT only stores medium values to partition the sub-trees, which incurs lower storage cost than BKT and FQT.
	
	\subsection{Omni-family}\label{subsec:Omni-family}
	
	The Omni-family{~\cite{ref20,ref66}} utilizes an existing {secondary-memory} index, e.g., a sequential file, a $\rm B^+$-tree, or an R-tree, to index the vectors after the pivot mapping. A sequential file stores the pre-computed distances of objects in the order of their identifiers; a $\rm B^+$-tree is used to index the pre-computed distances for each pivot; or an R-tree is used to index the pre-computed distances w.r.t. all the pivots. 
	Fig.~\ref{fig:Example of OmniR-tree} depicts an example of the OmniR-tree, including the pivot table that stores the pivots, the R-tree that indexes the pre-computed distances, and the random access file (RAF) that stores the objects. Note that, unlike the M-tree family, the Omni-family uses a separate RAF to store the objects. This is done to avoid the impact of the object size.
	
	\textbf{MRQ and M\textit{k}NNQ Processing.} To answer MRQ($q$, $r$), entries in the R-tree are traversed in depth-first fashion, {and Lemma \ref{lemma:PIVOT FILTERING} is used for pruning.} 
	 To answer M\textit{k}NNQ($q$, $k$), {the second strategy from Section \ref{subsec:Metric Similarity Queries} is used, and} entries in the R-tree are traversed in best-first manner, i.e., in ascending order of their minimum distances to the query object $q$. Lemma \ref{lemma:PIVOT FILTERING} is utilized for pruning. 

	\textbf{Discussion.} The Omni-family includes the Omni-sequential-file, the $\rm OmniB^+$-tree, and the OmniR-tree. The Omni-sequential-file can be regarded as LAESA stored on disk, which incurs substantial I/O during search as the data is unclustered. The $\rm OmniB^+$-tree needs one $\rm B^+$-tree for each pivot, resulting in redundant storage and I/O during search. The OmniR-tree utilizes MBBs to cluster the data, and uses pivot filtering to achieve high query efficiency. The construction cost of the OmniR-tree is $O(nml\log_{m}{n})$, as it needs $O(nl)$ to do the pivot mapping and $O(nml\log_{m}{n})$ to build the R-tree. The storage cost of the OmniR-tree is $O(ns + nl + nl/m + ls)$, as it needs $O(ns)$ to store the data in the RAF, $O(nl + nl/m)$ to store the pre-computed distances in the R-tree, and $O(ls)$ to store the pivots in the pivot table.

	\subsection{M-index}\label{subsec:M-Index}
	
		\begin{figure}[t]
		\centering
		\subfigtopskip=0cm
		\subfigbottomskip=0cm
		\subfigcapskip=0cm
		\subfigure[Data partitioning]{
			\label{M-Indexx Hyperplane partitioning}
			\includegraphics[scale=0.9]{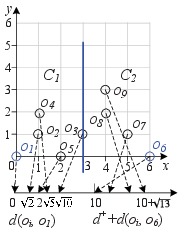}
		}
		\subfigure[M-$\rm index^*$ structure]{
			\label{M-Indexx structure}
			\includegraphics[scale=0.88]{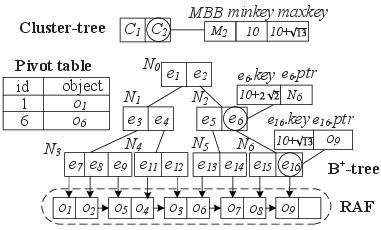}
		}
		\subfigure[MBB]{
			\label{M-Indexx MBB}
			\includegraphics[]{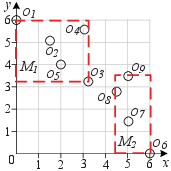}
		}
		\setlength{\abovecaptionskip}{0.1cm}
		\caption{{Example of the M-$\rm index^*$}}
		\label{fig:Example of the M-index*}
		\vspace{-0.5cm}
	\end{figure}

	The M-index{~\cite{ref93}} combines generalized hyperplane partitioning and pivot mapping. Given a set $P$ of pivots, each object $o$ is assigned to the partition of its nearest pivot $p_i (\in P)$, and is mapped to the real number $ key(o) = d(p_i, o) + (i - 1) \times n_d $, where $n_d$ is the maximum distance in a certain metric space. In Fig.~\ref{M-Indexx Hyperplane partitioning}, if $P = \left\{o_1, o_6\right\}$, we obtain clusters $C_1$ and $C_2$. After the partitioning, a cluster tree (as depicted in Fig.~\ref{M-Indexx structure}) is used to maintain the information of the clusters, a $\rm B^+$-tree is employed to index the mapped real numbers, and an RAF is used to store the objects with the pre-computed distances to all the pivots. If more pivots are used, the cluster-tree can be extended to a dynamic tree.
	{To accelerate search using the M-index, the minimum and maximum mapped numbers of MBBs are added to the cluster tree in the M-index, yielding the $\rm M^*$-index{~\cite{ref37}}.}
	
	\textbf{MRQ and M\textit{k}NNQ Processing using the M-index.} To answer MRQ($q$, $r$), the entries in the cluster tree are traversed in depth-first fashion, {where Lemma \ref{lemma:DOUBLE-PIVOT FILTERING} is used to prune intermediate entries and Lemma \ref{lemma:PIVOT FILTERING} is used to prune objects.}
	To answer M\textit{k}NNQ($q$, $k$), by taking the first strategy from Section \ref{subsec:Metric Similarity Queries}, a range query with a small search radius is performed first. Then, the search radius is increased gradually until $k$ closest objects are found.
	
	\textbf{MRQ and M\textit{k}NNQ Processing using the M-$\bm{{\rm index^*}}$.} {The M-$\rm index^*$ includes MBB information for each cluster in the M-index. Thus, the pivot filtering of Lemma \ref{lemma:PIVOT FILTERING} can be applied when traversing the cluster-tree to filter unqualified clusters. In addition, Lemma \ref{lemma:PIVOT VALIDATION} can be integrated to validate the objects.} To answer M\textit{k}NNQ($q$, $k$), the M-$\rm index^*$ can use the second strategy instead of the first strategy based on the MBBs. More specifically, the cluster-tree is traversed in best-first manner, i.e., clusters are visited in ascending order of their distances to the query object $q$.
	
	\textbf{Discussion.} The construction cost of the M-index is $\Omega(nl\log_{l}{n/m}) +O(mn\log_{m}{n})$. Here, we need the optimal cost $\Omega(nl\log_{l}{n⁄m})$ to construct the dynamic unbalanced cluster tree and cost $O(mn\log_{m}{n})$ to construct the $\rm B^+$-tree, where $m$ denotes the number of entries in each cluster or each $\rm B^+$-tree node. The storage cost of the M-index is $O(ns + nl + n + n/m + ls)$, as it needs $O(ns + nl)$ cost to store the RAF, $O(n + n/m)$ cost to store the $\rm B^+$-tree, $O(n/m)$ cost to store the cluster tree, and $O(ls)$ cost to store the pivot table. When integrating the MBB into the cluster tree, the storage of the M-$\rm index^*$ is increased to $O(ns + nl + n + n/m +nl/m + ls)$. However, the efficiency of MRQ and M\textit{k}NNQ on the M-$\rm index^*$ is improved. Since the M-$\rm index^{(*)}$ can use both Lemmas \ref{lemma:PIVOT FILTERING} and~\ref{lemma:DOUBLE-PIVOT FILTERING} for pruning, it can achieve high performance in terms of distance computations. Nonetheless, it needs to visit $\rm B^+$-tree multiple times for all unpruned clusters, making the I/O cost relatively high.

	\subsection{SPB-tree}\label{subsec:SPB-Tree}
\begin{figure}[t]
		\centering
		\subfigtopskip=0cm
		\subfigbottomskip=0cm
		\subfigcapskip=0cm
		\subfigure[SPB-tree structure]{
			\label{SPB-tree structure}
			\includegraphics[]{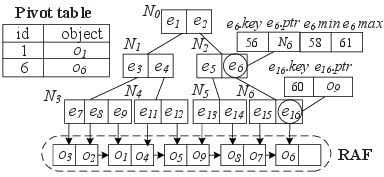}
		}
		\subfigure[Hilbert mapping]{
			\label{SPB-tree Hilbert mapping}
			\includegraphics[]{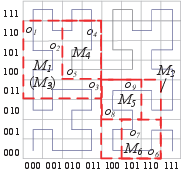}
		}
		\setlength{\abovecaptionskip}{0.1cm}
		\caption{Example of the SPB-tree}
		\label{fig:Example of SPB-tree}
		\vspace{-0.5cm}
	\end{figure}

	The SPB-tree{~\cite{ref33,ref34}} utilizes the two-stage mapping, i.e., the pivot mapping and the space-filling curve (SFC) mapping, to map objects into SFC values (i.e., integers) while (to some extent) maintaining spatial proximity. Then, a $\rm B^+$-tree is used to store the SFC values. Fig.~\ref{fig:Example of SPB-tree} depicts an example of SPB-tree, where Fig.~\ref{SPB-tree Hilbert mapping} illustrates the Hilbert mapping after the pivot mapping. {Each non-leaf entry $e$ in the $\rm B^+$-tree stores the SFC values of the two corners of its MBB.}
	
	\textbf{MRQ and M\textit{k}NNQ Processing.} To answer MRQ($q$, $r$), the entries in the $\rm B^+$-tree are traversed in depth-first fashion, {and Lemmas \ref{lemma:PIVOT FILTERING} and \ref{lemma:PIVOT VALIDATION} are used to prune and validate entries/objects.}
To answer M\textit{k}NNQ($q$, $k$), {by following the second strategy from Section \ref{subsec:Metric Similarity Queries},} the entries in the $\rm B^+$-tree are traversed in best-first manner, i.e., in ascending order of their minimum distances to the query object $q$, and Lemma \ref{lemma:PIVOT FILTERING} is used for pruning.
	
	\textbf{Discussion.} The construction cost of the SPB-tree is $O(n(l^2..l^3) + n(m+l)\log_{m}{n})$. In particular, the SPB-tree needs $O(nl^2)$ or  $O(nl^3)$ to do the pivot mapping and the space filling curve mapping ({$O(l)$ and $O(l^2)$ time are used for the Z-order curve and Hilbert curve respectively to transform each object}), and takes $O(n(m+l)\log_{m}{n})$ in building the $\rm B^+$-tree with MBB. The storage cost of the SPB-tree is $O(ns + n + n/m + ls)$, as it needs $O(ns)$ space to store the real data in RAF, $O(n + n/m)$ to store the mapped values in the $\rm B^+$-tree, and $O(ls)$ to store the pivot table. We employ the SFC mapping to reduce the storage cost and  maintain spatial proximity, resulting in improved I/O and index storage costs. However, for continuous distance functions, the continuous distances are approximated by discrete ones to perform  the SFC mapping, which decreases the pruning power. In addition, during similarity search, we need to do the Hilbert transformation to get the pre-computed distances; hence, additional CPU cost is incurred.

	\section{EXPERIMENTAL COMPARISION AMONG METRIC INDEXES}
\label{sec:EXPERIMENTAL COMPARISION AMONG METRIC INDEXES}
	We detail the experimental setting and then cover experiments in main-memory and {secondary-memory} settings in turn. {In addition, we conduct a significance evaluation of the metric indexes.}
	
	\subsection{Experimental Settings}\label{subsec:Experimental Settings}
	As discussed in Section \ref{subsec:Metric Similarity Queries}, analytical studies that do not take the effect of pruning into account offer limited insight into the real-world search performance of metric indexes. Thus, we opt to compare experimentally 19 representative metric indexes, including BST (belonging to the BST family), GNAT and EGNAT (belonging to the GHT family), SAT and DSACLT (belonging to the SAT family), M-tree and PM-tree (belonging to the M-tree family), LC (belonging to the LC family), $\rm MB^+$-tree and D-index (belonging to the D-index family), LAESA (belonging to the AESA family), $\rm EPT^*$, CPT, BKT, FQT (belonging to the FQ family), MVPT (belonging to the VPT family), OmniR-tree (belonging to the Omni-family), SPB-tree, and M-$\rm index^*$. We implemented all indexes and associated similarity search algorithms in C++, and experiments were conducted on an Intel Core i7-7700 3.6GHz PC with 16GB memory. {All code is publicly available\footnote{Code is available at  https://github.com/ZJU-DAILY/Metric\_Index.}.}
	
	We employ three real datasets, namely, \textit{LA, Words,} and \textit{Color}. \textit{LA}\footnote{LA is available at http://www.dbs.informatik.uni-muenchen.de/$\sim$seidl.} consists of geographical locations in Los Angeles, and the $L_2$-norm is utilized to measure similarity. \textit{Words}\footnote{Words is available at http://icon.shef.ac.uk/Moby/.} contains proper nouns, acronyms, and compound words taken from the Moby project, and the edit distance is used to compute the distances between words. \textit{Color}\footnote{Color is available at http://cophir.isti.cnr.it/.}  consists of standard MPEG-7 image features extracted from Flickr, and the similarity between two features is measured by the $L_1$-norm. {In addition, we generate a dataset \emph{Synthetic}, where the values are integers and the $L_{\infty}$-norm is employed. More specifically, in \textit{Synthetic}, 5 of 20 dimension values are generated randomly, while the remaining values are linear combinations of the previous ones.} Table \ref{tab:Datasets used in the Experiments} summarizes statistics of the datasets.
	\begin{table} [H]
		\centering
		\small
		\setlength{\abovecaptionskip}{0cm}
		\caption{Datasets used in the Experiments}
		\label{tab:Datasets used in the Experiments}
		\begin{tabular}{lllll}
			\toprule
			\textbf{Datasets} & \textbf{Cardinality} & \textbf{Dimensionality} & \textbf{Intrinsic Dimensionality} & \textbf{Measurement}\\
			\midrule
			\textit{LA} & 1,073,727  & 2  & {5.4} & $L_2$-norm\\
			\textit{Words} & 611,756 & 1$\sim$34  & {1.2} & Edit distance\\
			\textit{Color} & 1,000,000 & 282  & {6.5}  & $L_1$-norm\\
			\textit{Synthetic} & 1,000,000 & 20  & {6.6}  & $L_\infty$-norm\\
			\bottomrule
		\end{tabular}
		\vspace{-0.3cm}
	\end{table}
	
	\begin{table} [H]
		\centering
		\small
		\setlength{\abovecaptionskip}{0cm}
		\caption{Parameter Setting}
		\label{tab:Parameter Setting}
		\begin{tabular}{lll}
			\toprule
			\textbf{Parameter} & \textbf{Values} & \textbf{Default}\\
			\midrule
			\textit{Number of pivots $l$} & 3, 5, 10, 15, 20 & 5\\
			\textit{Search radius $r$ (selectivity of range queries)} & 2$\%$, 4$\%$, 8$\%$, 16$\%$, 32$\%$ & 8$\%$\\
			\textit{Number of $k$} & 5, 10, 20, 50, 100 & 20\\
			\bottomrule
		\end{tabular}
	\end{table}

	We investigate the similarity query performance using the indexes when varying the parameters listed in Table \ref{tab:Parameter Setting}.
	
	\begin{itemize}\setlength{\itemsep}{-\itemsep}		
	\item{} {The number $l$ of pivots is an integer in [1, +$\infty$), and we find that a small $l$ yields the best performance for all four datasets. Hence, we chose five representative values (specifically 3, 5, 10, 15, 20) for $l$.}
	\item{}	{Following a previous study~\cite{ref66}, we use the selectivity of metric range queries to set the search radius $r$ that controls the search region. In particular, the value of radius $r$ denotes the percentage of objects in the dataset that are result objects of a metric range query. As different datasets have different distance ranges, it is difficult to use the same value across different datasets. In addition, the data distribution might be skewed in real-life data, which makes it possible for even a large radius to only cover few objects (and a small radius to cover many objects). Thus, we use selectivity values (i.e., 2\%, 4\%, 8\%, 16\%, and 32\%) instead of fixed values (e.g., 1 km, 2 km, etc.) to better evaluate the performance of metric range queries.}
	\item{}	{The value $k$ in the M$k$NN query is an integer in [1, +$\infty$). In real-life applications, a small number of result objects is usually retrieved by an M$k$NN query. For example, in recommendation systems, we cannot provide too many recommendations for users. Thus, we vary $k$ from 5 to 100, which aligns with previous studies \cite{ref32,ref66}.}
\end{itemize}
	\noindent
	In each experiment, one parameter is varied, and the others are fixed at their default values. As explained in Section \ref{subsec:Metric Similarity Queries}, the main performance metrics include the number of page accesses (\textit{PA}), the number of distance computations (\textit{compdists}), and the running time. Each reported measurement is an average over 100 random queries. To facilitate a fair comparison, we use the same set of random queries for all indexes 
	
	When comparing the performance of pivot-based methods, their use of different pivot selection strategies renders the comparison challenging. For example, the OmniR-tree~\cite{ref66} utilizes the HF algorithm to select outliers as pivots, while the SPB-tree uses the HFI algorithm to select pivots that maximize the similarity between the original metric space and the vector space obtained by using the pivots. Since the similarity query performance depends highly on the pivots used~\cite{ref5,ref24}, we utilize the same pivot selection strategy for the pivot-based indexes. Specifically, we use the HFI algorithm~\cite{ref33,ref37}. This does, however, not apply to the $\rm EPT^*$, BKT, and GNAT families. As discussed in Sections \ref{subsec:The GHT Family}, \ref{subsec:EPT}, and \ref{subsec:BKT}, GNAT uses centers in the same level as pivots, $\rm EPT^*$ utilizes different pivots for different objects, and BKT needs to select pivots randomly in its sub-trees.
	
	\begin{figure} [b]
		\centering
		\subfigtopskip=0cm
		\subfigbottomskip=0cm
		\subfigcapskip=0cm
		\includegraphics[width=0.6\linewidth]{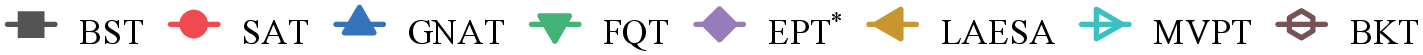}
		\\
		\subfigure[{LA}]{
			\includegraphics[width=0.3\linewidth]{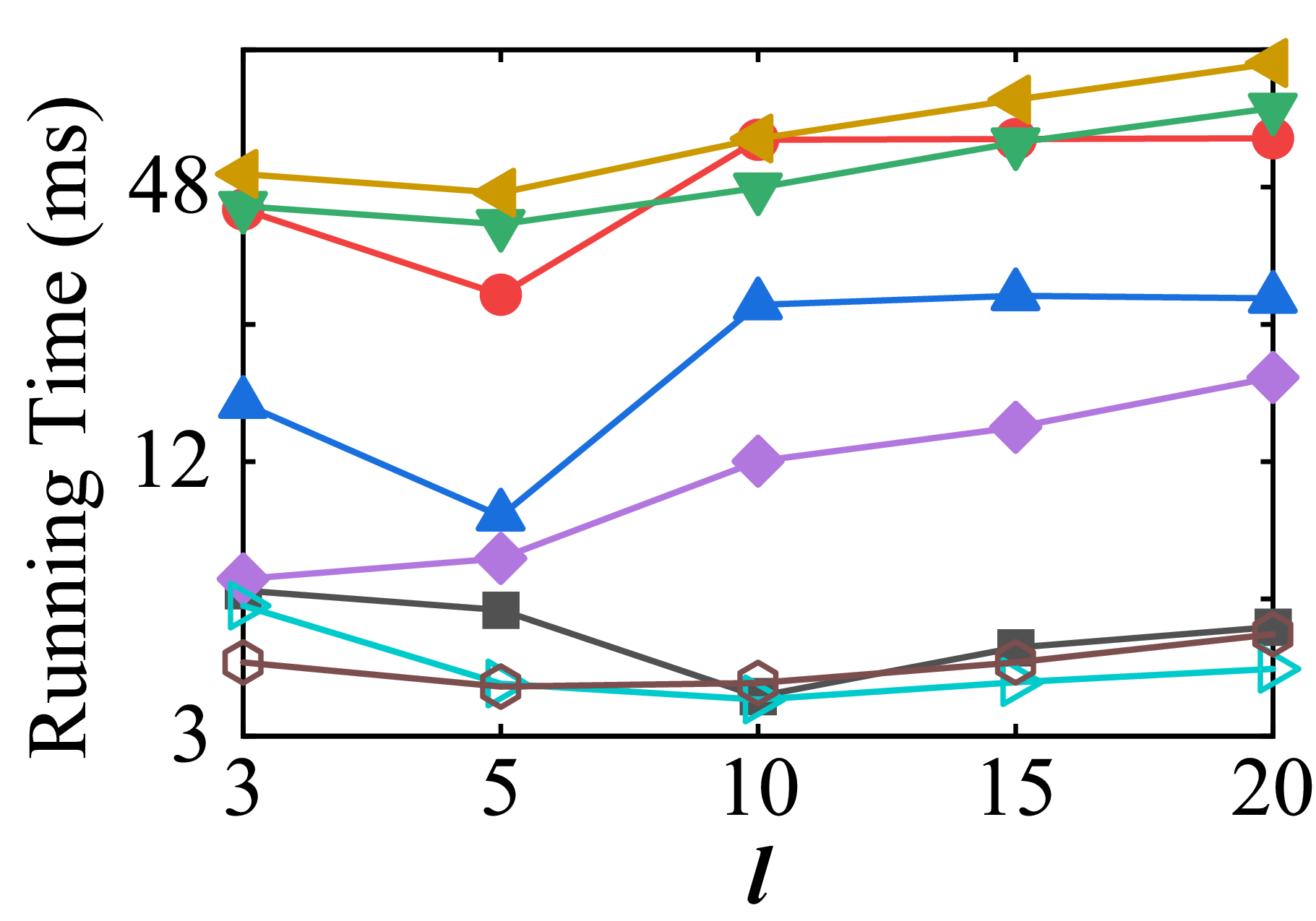}
		}
		\subfigure[LA]{
			\includegraphics[width=0.3\linewidth]{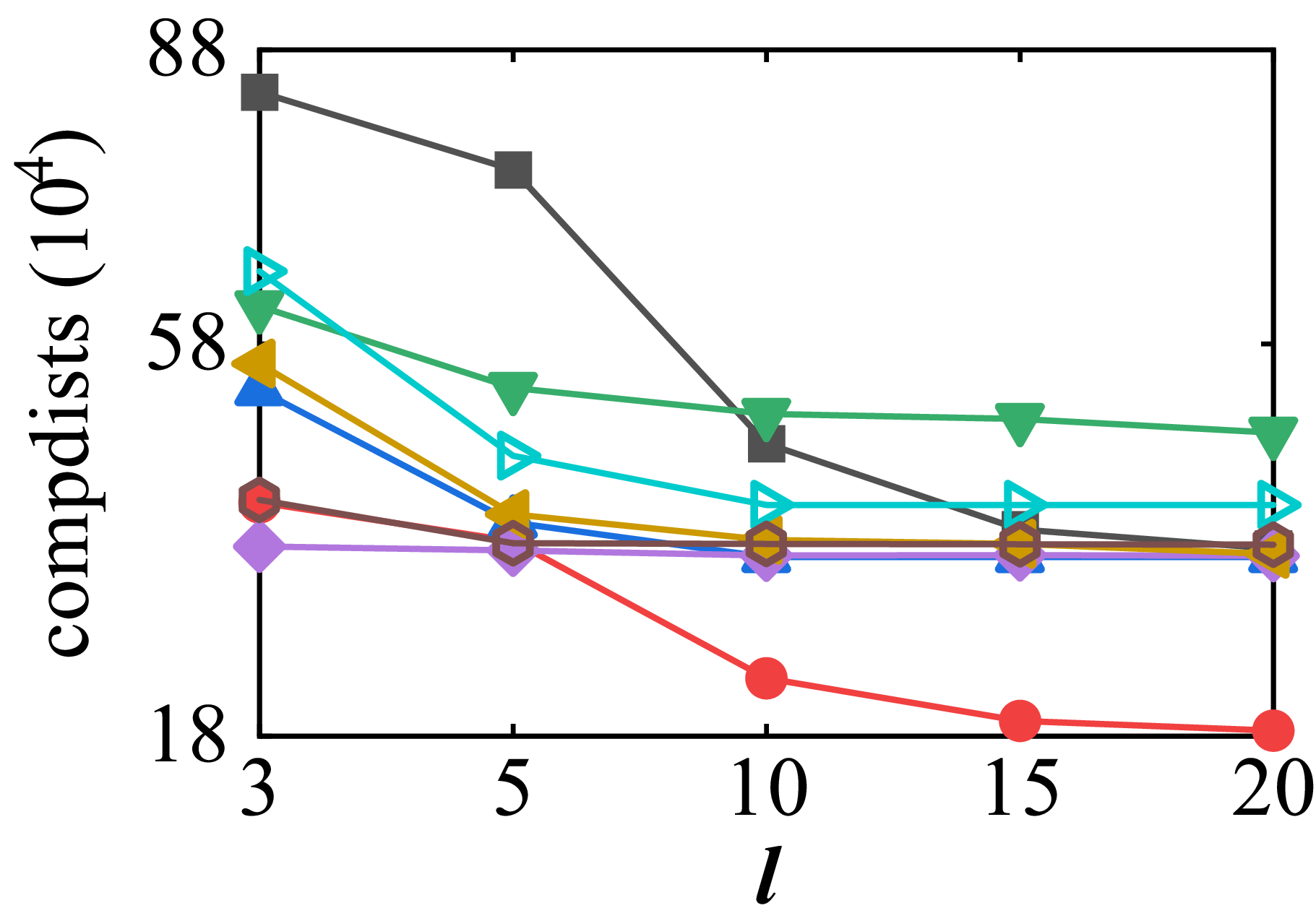}
		}
		\\
		\subfigure[Words]{
			\includegraphics[width=0.3\linewidth]{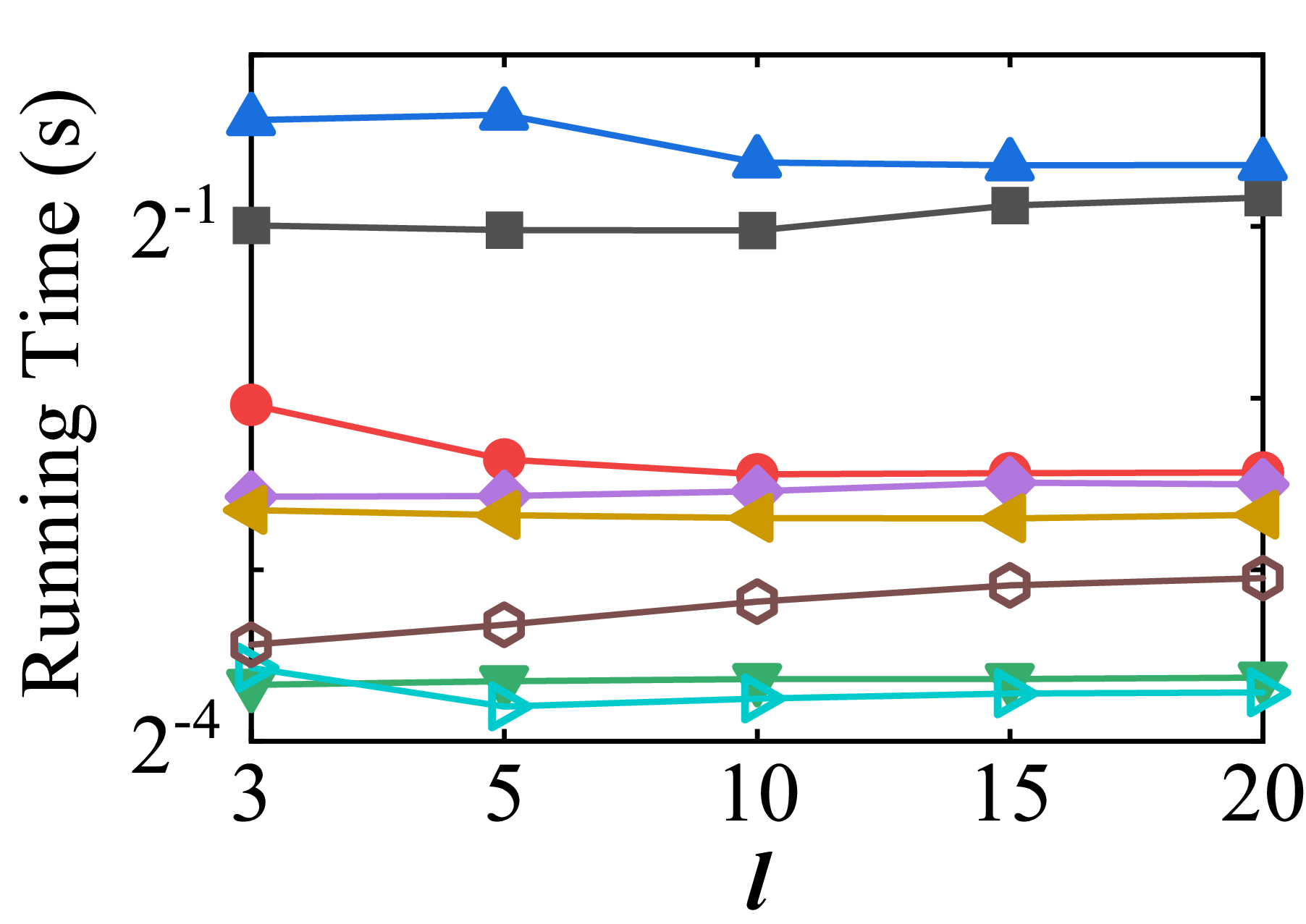}
		}
		\subfigure[Words]{
			\includegraphics[width=0.3\linewidth]{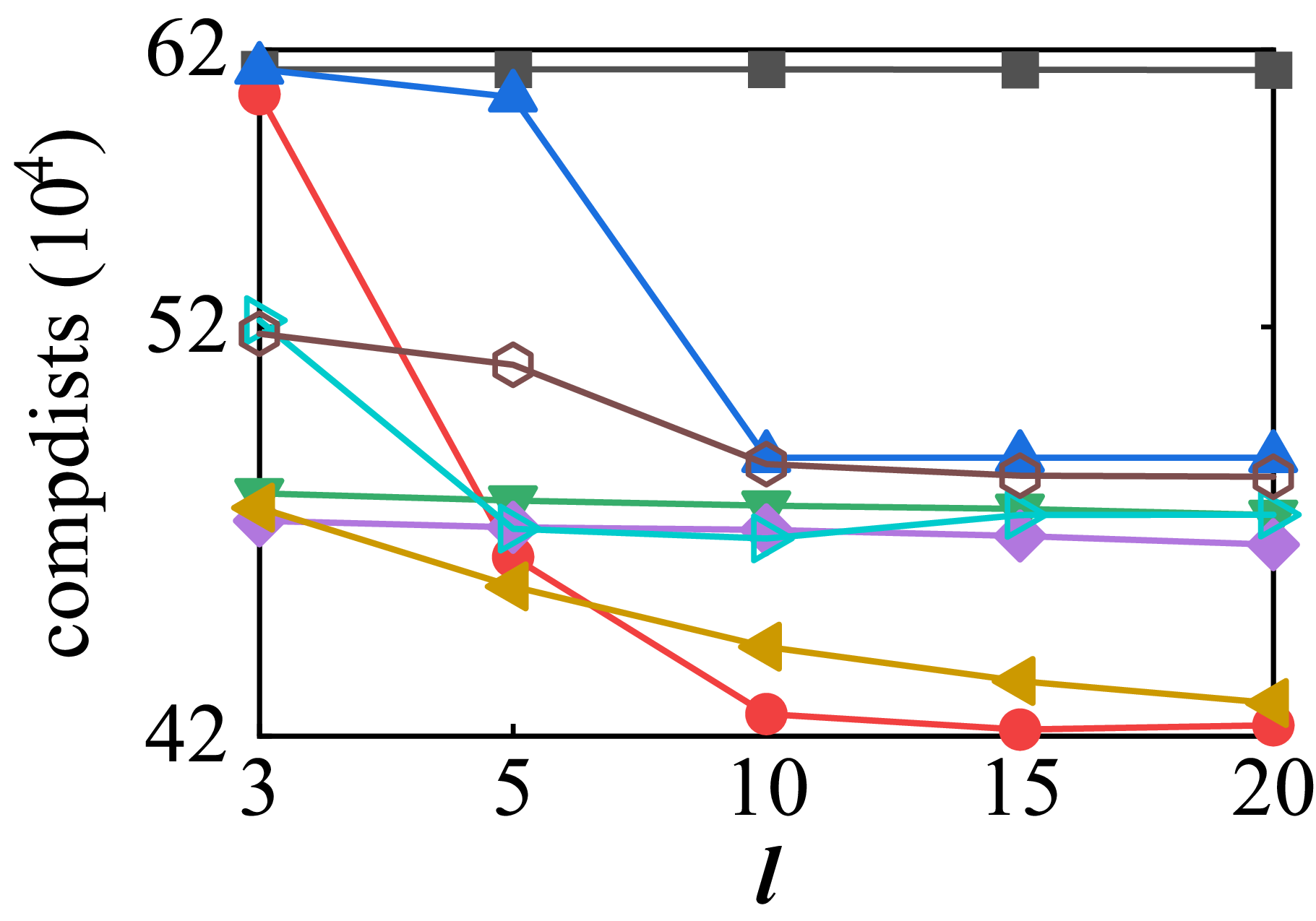}
		}
		\\
		\subfigure[Color]{
			\includegraphics[width=0.3\linewidth]{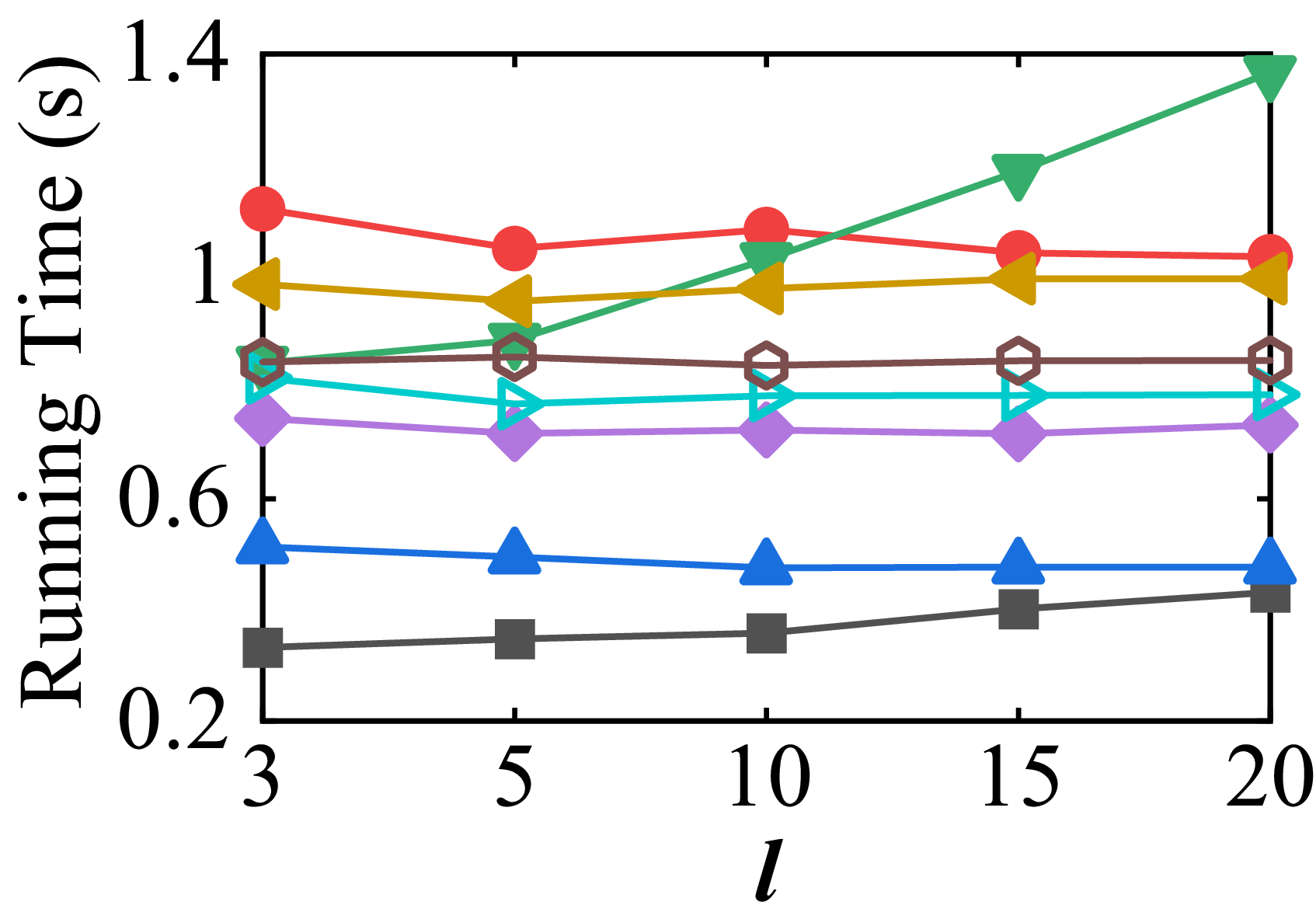}
		}
		\subfigure[Color]{
			\includegraphics[width=0.3\linewidth]{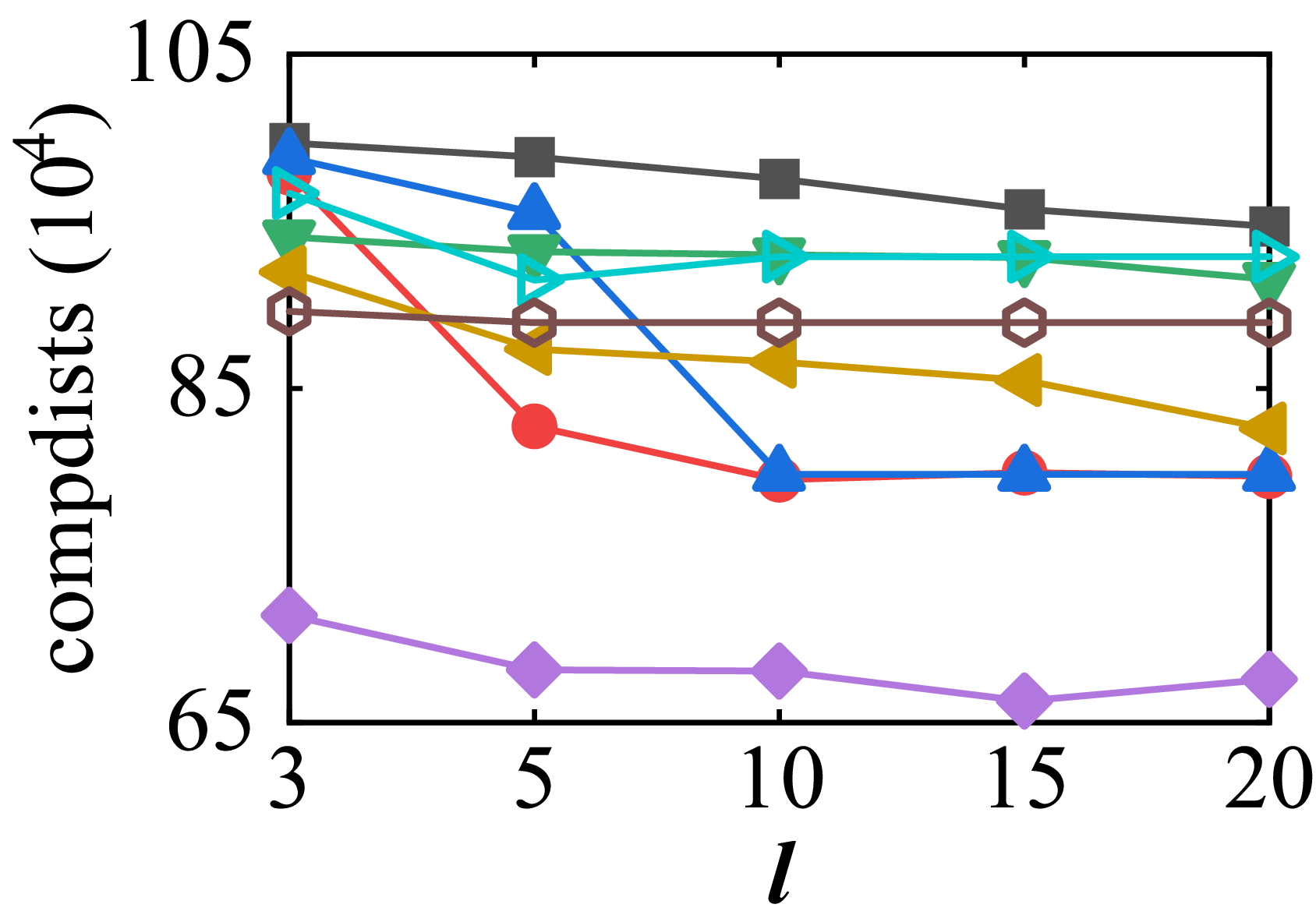}
		}
		\\
		\subfigure[Synthetic]{
			\includegraphics[width=0.3\linewidth]{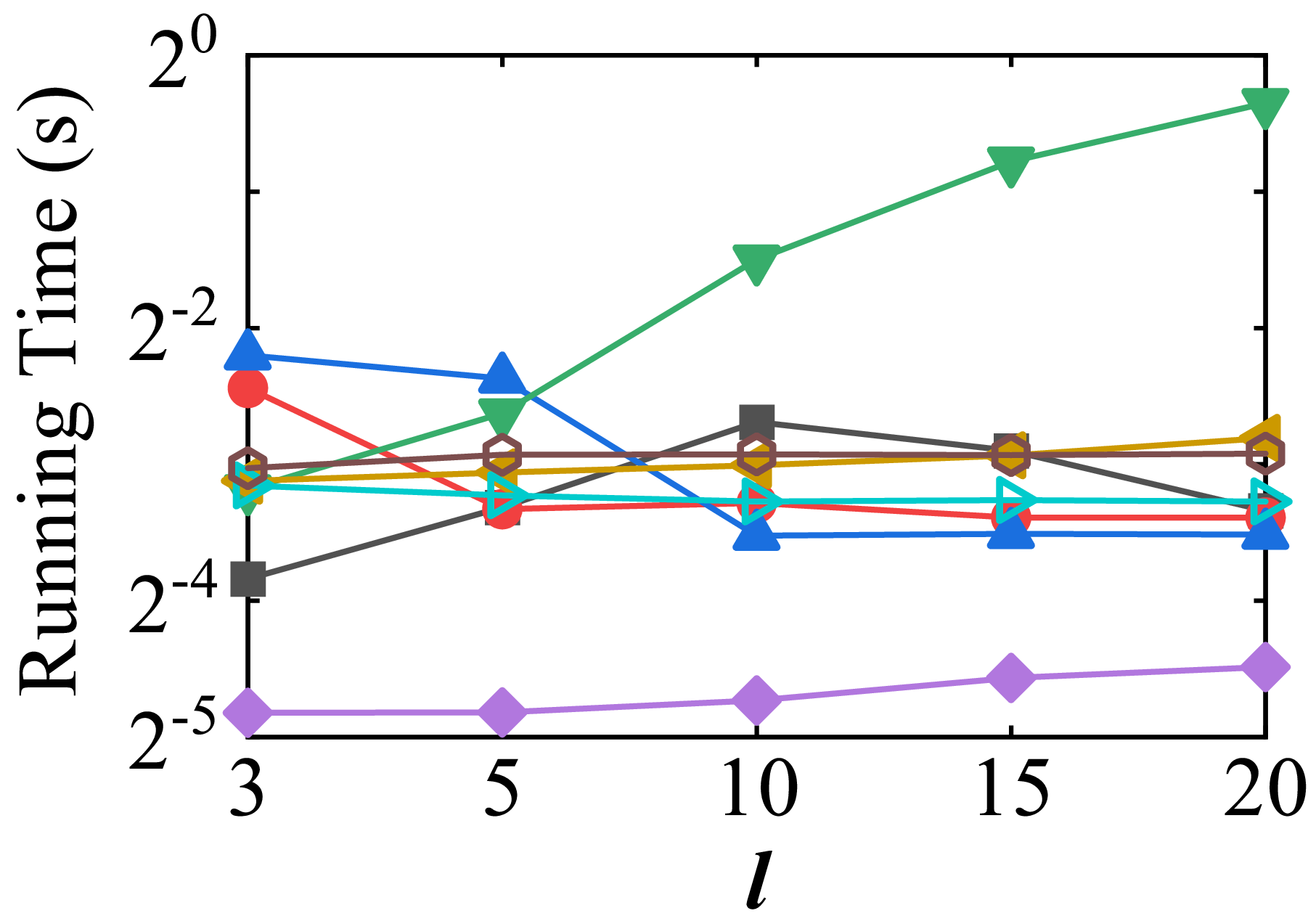}
		}
		\subfigure[Synthetic]{
			\includegraphics[width=0.3\linewidth]{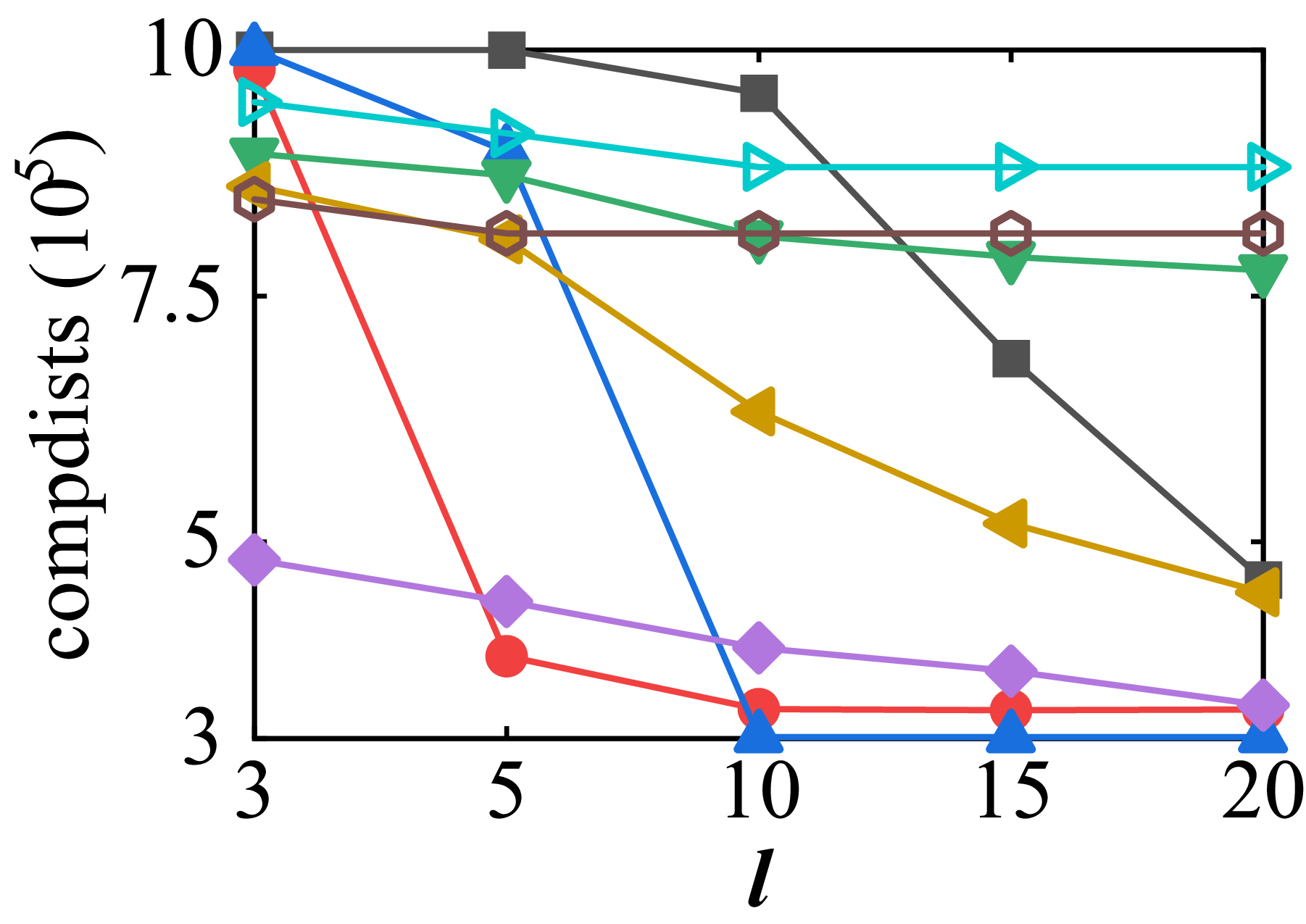}
		}
		\setlength{\abovecaptionskip}{0.1cm}
		\caption{Main-Memory based Metric Index Comparison Using MRQ Queries}
		\label{fig:Main-Memory based Metric Index Comparison Using MRQ queries}
		\vspace{-0.5cm}
	\end{figure}
	
	\subsection{Comparison among Main-memory based Metric Indexes}
	\label{subsec:Comparison among Main-memory based Metric Indexes}
	We compare the performance of the main-memory based metric indexes using MRQ and M\textit{k}NN queries, which includes compact-partitioning based indexes (i.e., BST and SAT), pivot-based indexes (i.e., LAESA, $\rm EPT^*$, BKT, FQT, and MVPT), and the hybrid index GNAT. When comparing the pivot-based indexes and compact-partitioning based indexes, the index pruning ability depends on the number of pivots and the number of centers, respectively. Hence, we set the number of pivots used in the pivot-based indexes equaling to the height of compact-partitioning based methods. Then, the number of pivots and the number of centers used for pruning each object in pivot-based and compact-partitioning methods, respectively, are the same. GNAT and MVPT are multi-arity trees. Here, we set arity to 5, which yields the best results across a range of values tested.

	\begin{figure}
	\centering
	\subfigtopskip=0cm
	\subfigbottomskip=0cm
	\subfigcapskip=0cm
	\includegraphics[width=0.6\linewidth]{Main_memory.eps}
	\\
	\subfigure[{LA}]{
		\includegraphics[width=0.3\linewidth]{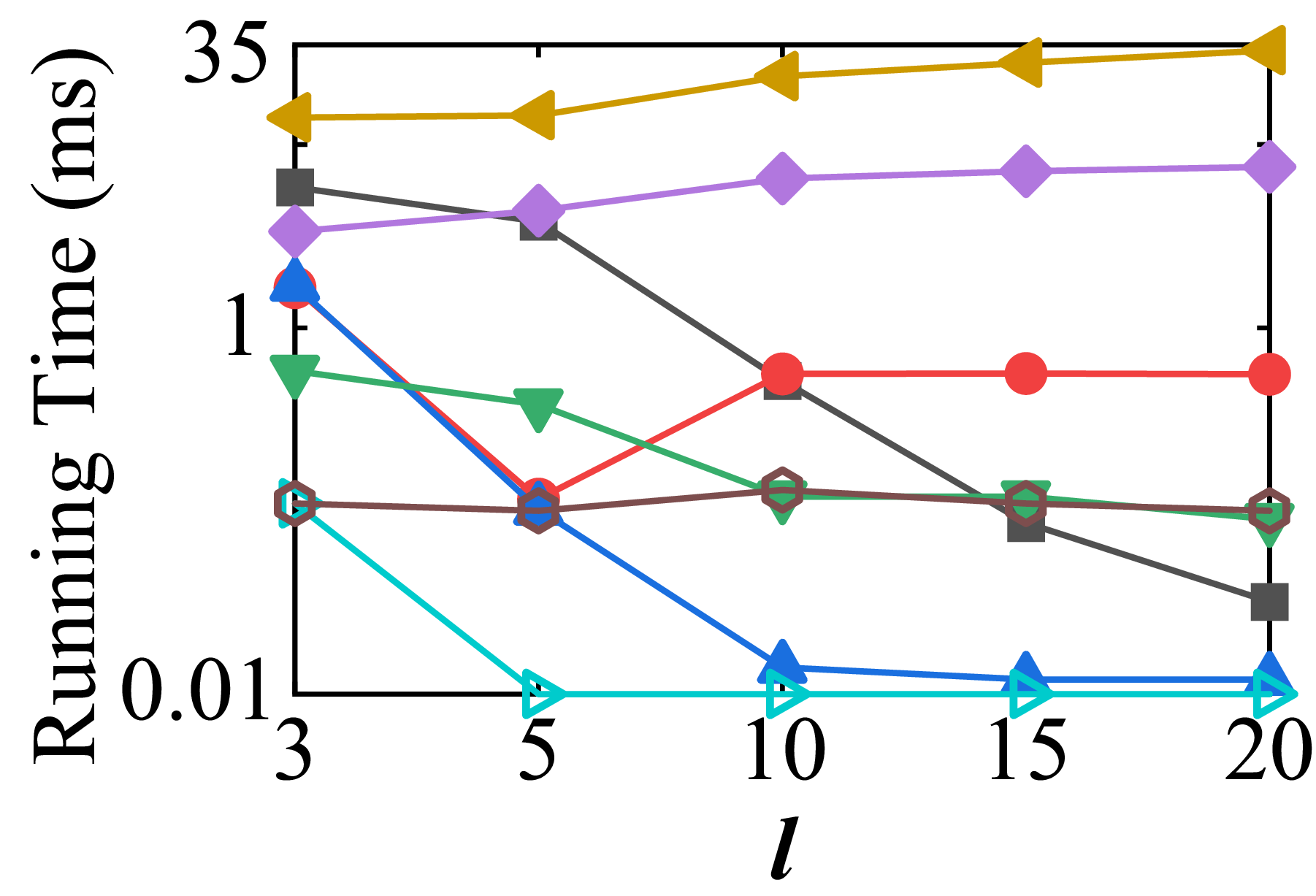}
	}
	\subfigure[LA]{
		\includegraphics[width=0.3\linewidth]{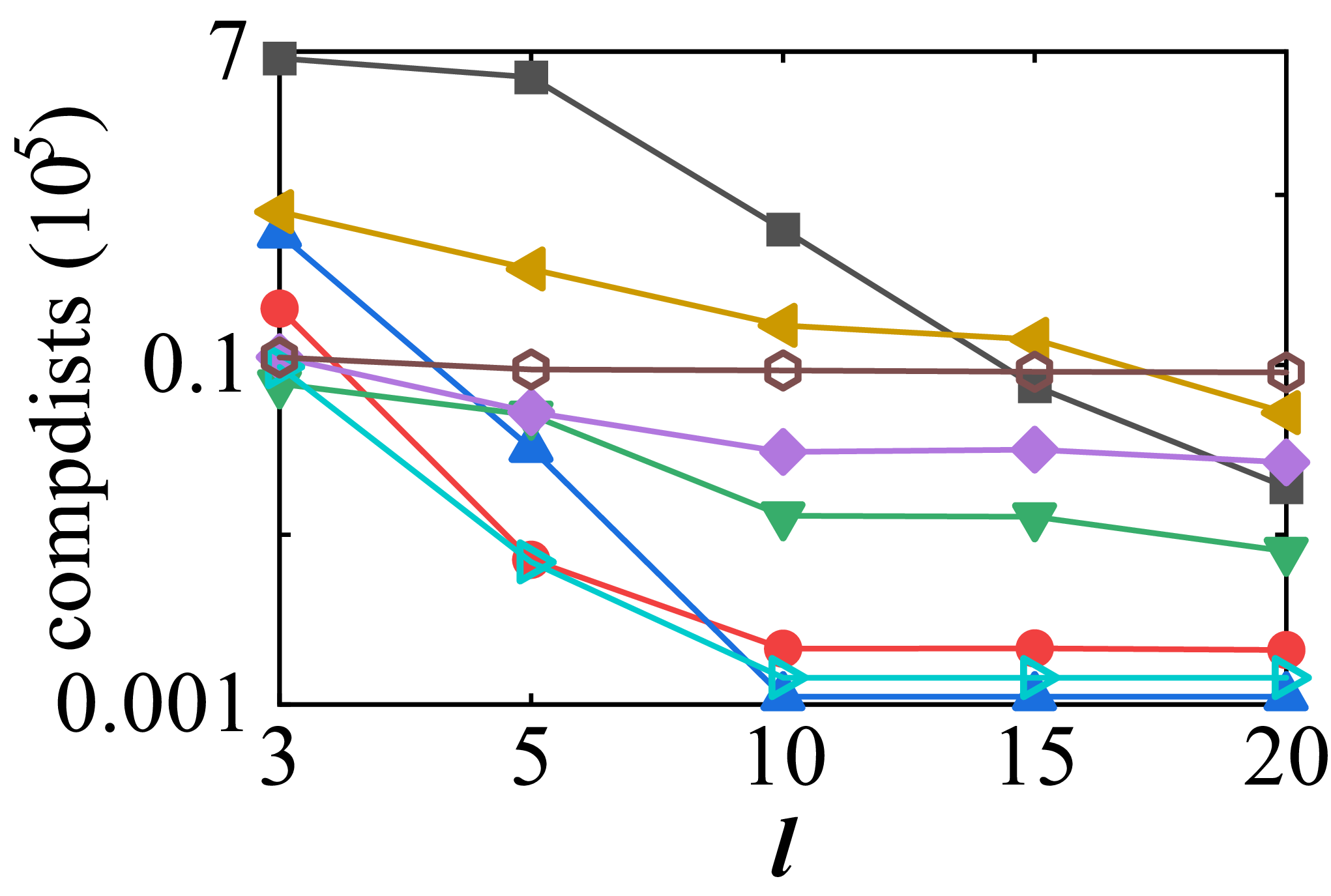}
	}
	\\
	\subfigure[Words]{
		\includegraphics[width=0.3\linewidth]{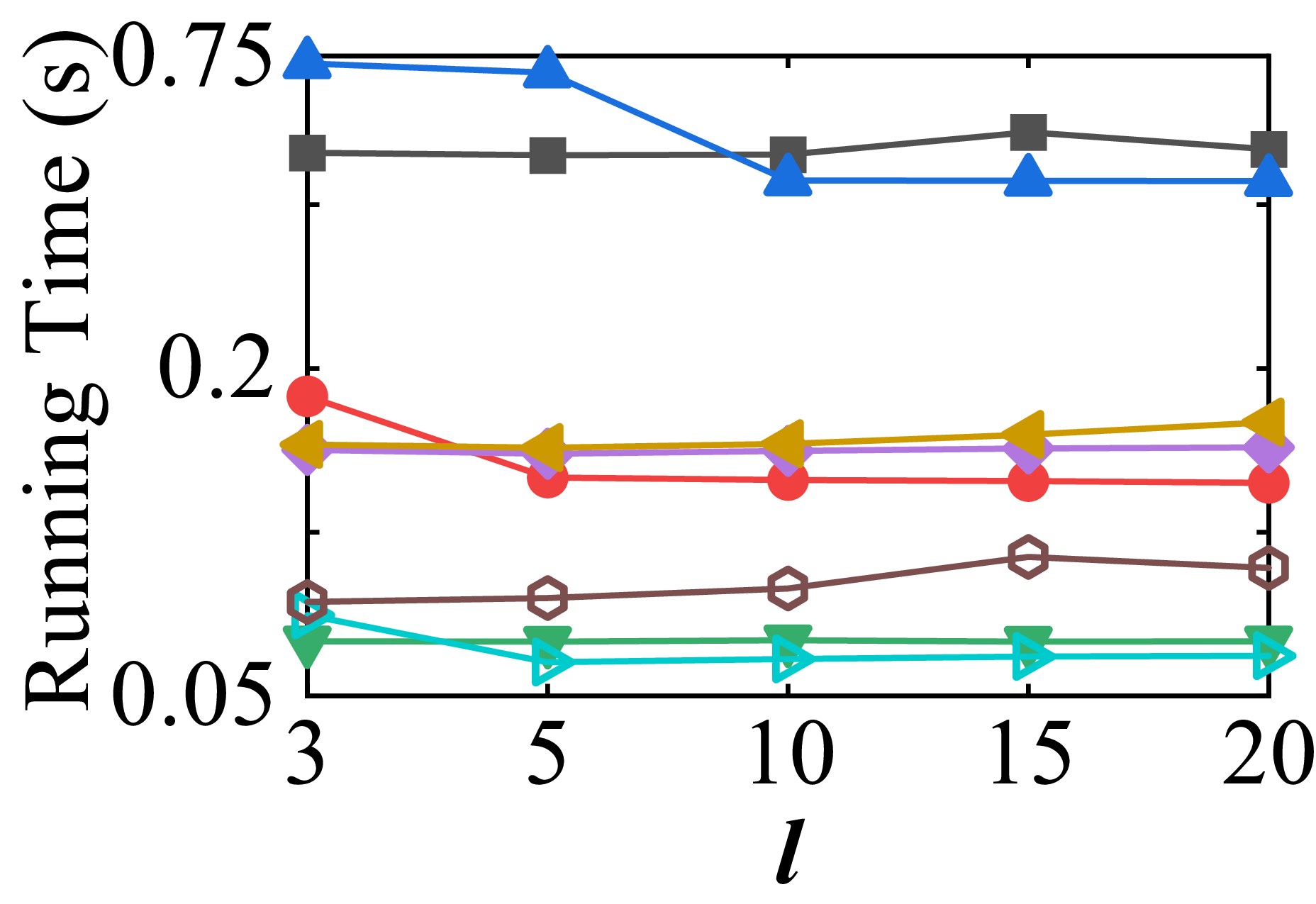}
	}
	\subfigure[Words]{
		\includegraphics[width=0.3\linewidth]{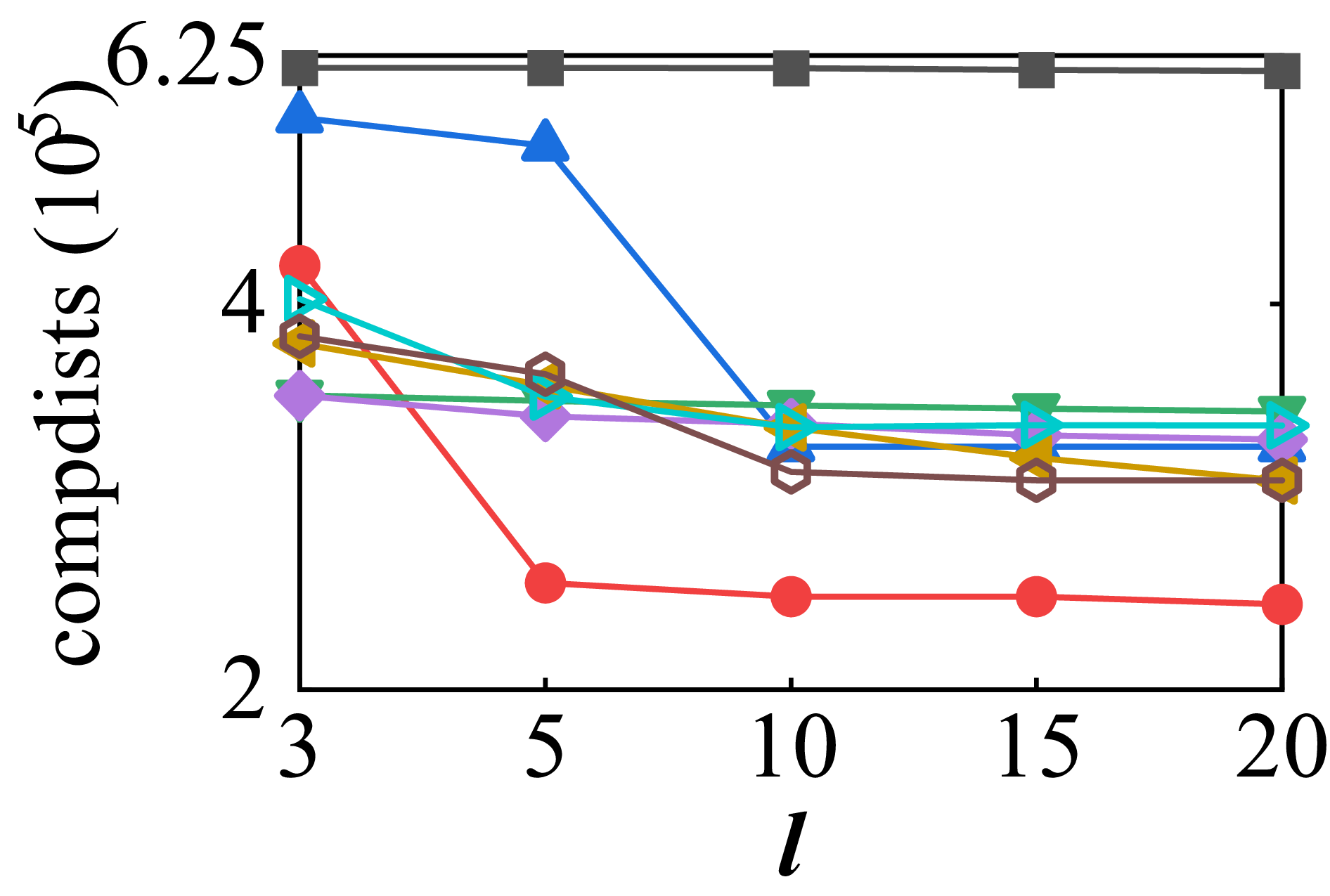}
	}
	\\
	\subfigure[Color]{
		\includegraphics[width=0.3\linewidth]{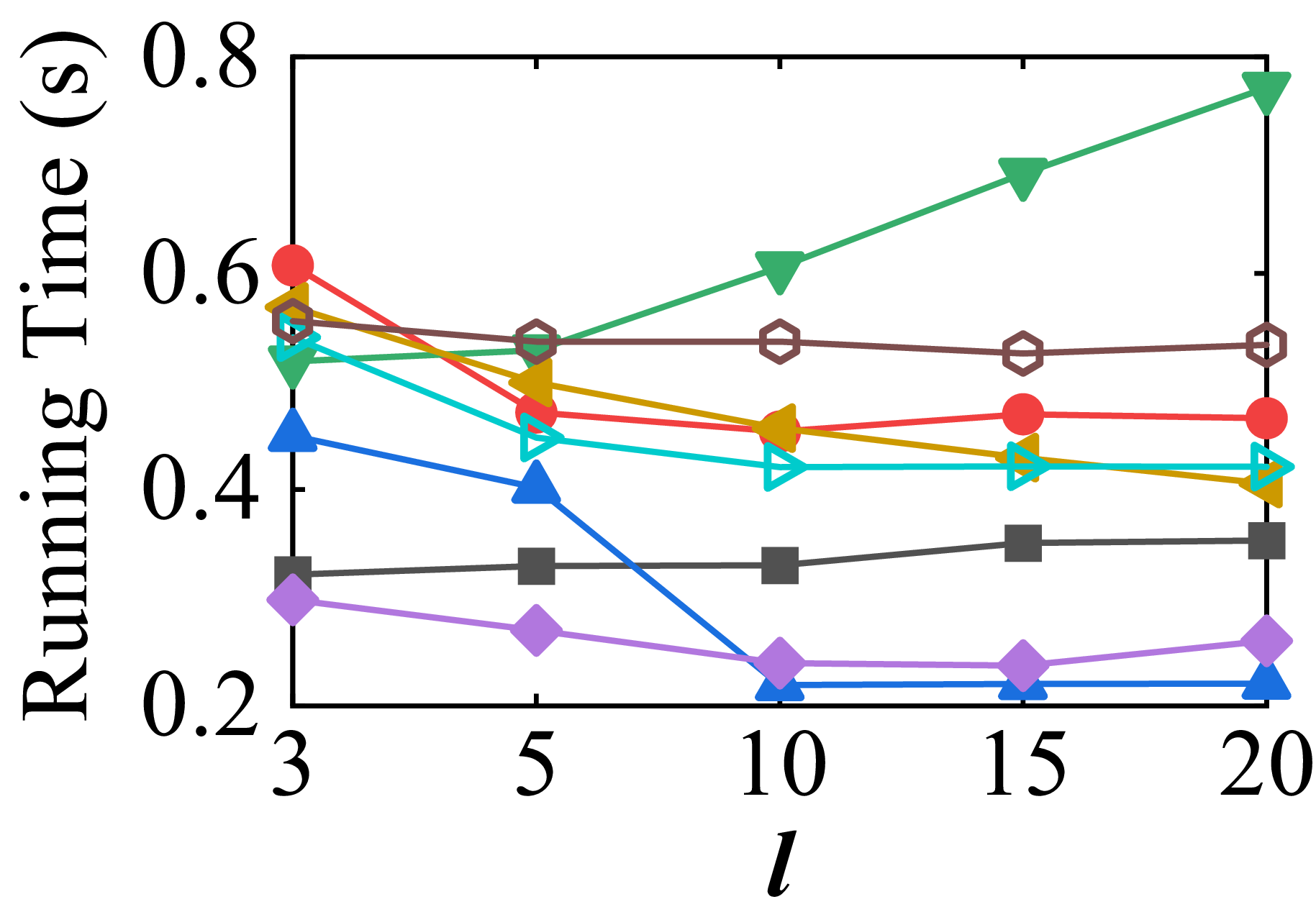}
	}
	\subfigure[Color]{
		\includegraphics[width=0.3\linewidth]{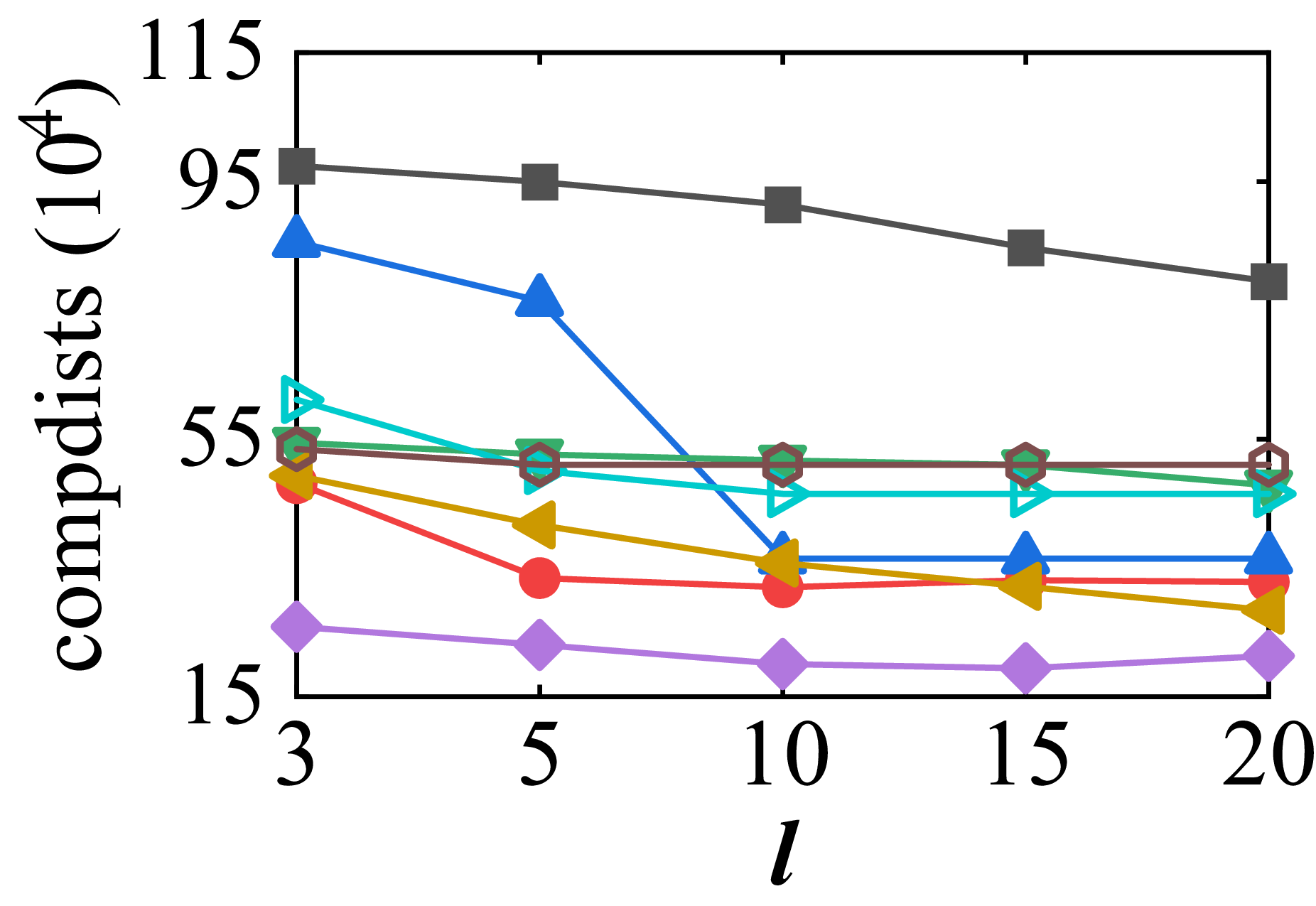}
	}
	\\
	\subfigure[Synthetic]{
		\includegraphics[width=0.3\linewidth]{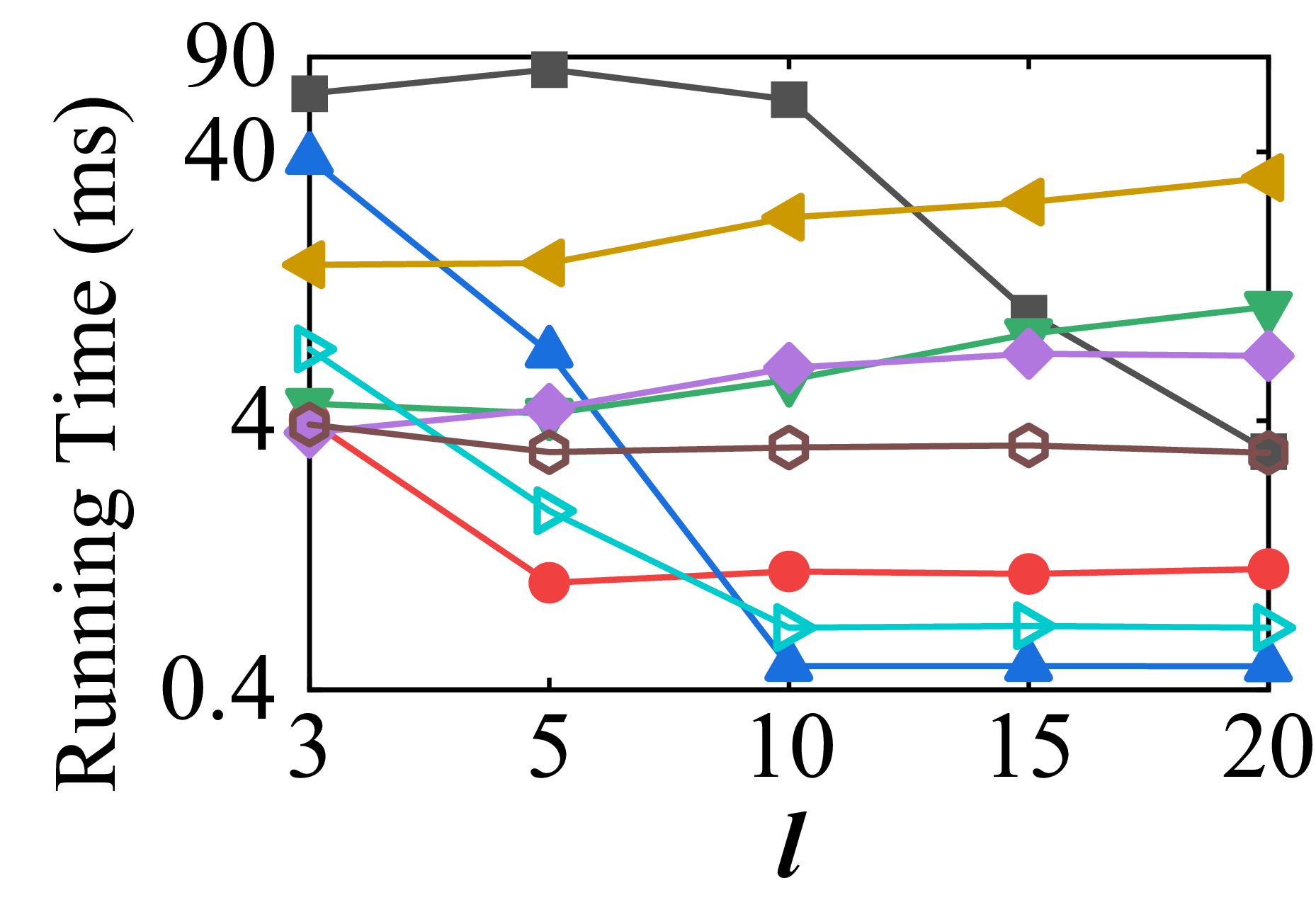}
	}
	\subfigure[Synthetic]{
		\includegraphics[width=0.3\linewidth]{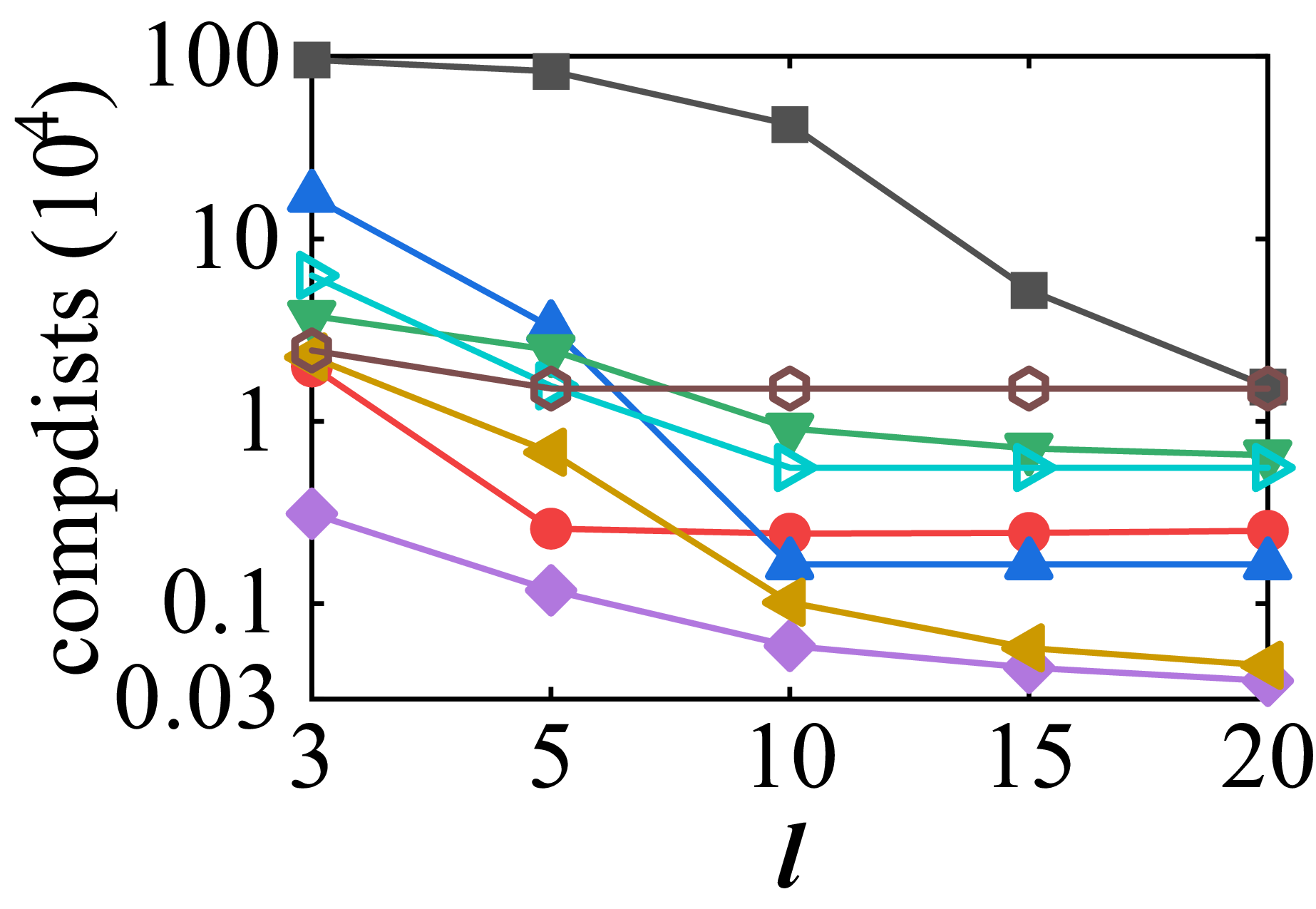}
	}
	\setlength{\abovecaptionskip}{0cm}
	\caption{Main-Memory based Metric Index Comparison Using M\textit{k}NN Queries}
	\label{fig:Main-Memory based Metric Index Comparison Using MkNN queries}
	\vspace{-0.5cm}
\end{figure}

	Fig.~\ref{fig:Main-Memory based Metric Index Comparison Using MRQ queries} and Fig.~\ref{fig:Main-Memory based Metric Index Comparison Using MkNN queries} show the MRQ and M\textit{k}NN performance (\textit{compdists} and running time) on four datasets, respectively. 
	The first observation is that, the query cost, \textit{compdists} as well as running time, first drops and then remains stable or increases when the number of pivots or the tree height increase. On the one hand, the more pivots/the higher the tree, the stronger pruning ability. On the other hand, using more pivots increases the CPU cost of filtering the search space. Hence, the best number of pivots for an index depends on the distance distribution of the dataset and the structure of the index. Table \ref{tab:Main-memory-based Metric Indexes Ranking} summarizes the performance metric (including \textit{compdists} and running time) ranking of all main-memory based indexes on the four datasets using MRQ and M\textit{k}NN queries.

	\textbf{Compdists Performance Analysis.} As summarized in Table \ref{tab:Main-memory-based Metric Indexes Ranking}, SAT, $\rm EPT^*$, and GNAT perform best in terms of distance computations, while BST and FQT perform the worst. The number of distance computations depends on the selected pivots and the pivot based filtering and validation techniques used. However, to achieve a fair comparison among the index structures, we use the same set pivots for all the metric indexes except $\rm EPT^*$ and BST as explained earlier. Thus, the reasons behind the distance computation performance observation are that: 1) $\rm EPT^*$ selects different pivots for each object in order to improve its pruning, while the other indexes use the same set of pivots for all objects in the dataset; 2) GNAT is a hybrid index that uses two different pivot-based filtering techniques, which improves its pruning ability; 3) SAT utilizes its ancestors for pruning while the others use only the centers at the same level for pruning, which improves the pruning ability of SAT; 4) BST uses random pivots, which yields relatively weak filtering; and 5) FQT is a unbalanced tree, indicating that relatively few pivots are used for pruning. As stated in previous work, the pivot-based metric indexes can achieve better performance in terms of distance computations when compared with compact-partitioning based methods~\cite{ref32}. However, if we use the same set of pivots and the same pruning and validation techniques, the number of distance computations across these two categories are similar. A key reason why the distance computation performance of the pivot-based metric indexes better is that they (e.g., LAESA, $\rm EPT^*$) can support unlimited number of pivots, while the pivot number for compact-partitioning methods is limited by the dataset size (i.e., if little data is left, no further partitions/no additional centers are needed to build the index).
	
		\begin{table}
		\centering
		\setlength{\abovecaptionskip}{0cm}
		\caption{Main-memory-based Metric Indexes Ranking}
		\label{tab:Main-memory-based Metric Indexes Ranking}
		\begin{tabular}{ccccc}
			\toprule
			\multirow{2}{*}{\textbf{Main-memory based Metric Index}} & \multicolumn{2}{c}{\textbf{\textit{\quad Compdists} ranking}} & \multicolumn{2}{c}{\textbf{Running time ranking}} \\ \cline{2-5}
			~ & \textbf{\quad \quad MRQ} & \textbf{M\textit{k}NN} & \textbf{\quad MRQ} & \textbf{M\textit{k}NN} \\
			\midrule
			\textbf{BST}     & \quad \quad 8 & 8 &\quad 3 & 7\\
			\textbf{SAT}     &\quad \quad 1 & 1 & \quad 6 & 4\\
			\textbf{LAESA}   & \quad \quad 4 & 4 & \quad 7 & 8\\
			\textbf{$\bm{{\rm EPT^*}}$} & \quad \quad 2 & 2& \quad 2 & 4\\
			\textbf{BKT}     & \quad \quad 5 & 6 & \quad  4 & 3\\
			\textbf{FQT}     & \quad \quad  6 & 7 & \quad 7 & 6\\
			\textbf{MVPT}    & \quad \quad 6 & 5 & \quad  1 & 1\\
			\textbf{GNAT}    & \quad \quad 3 & 2 & \quad  4 & 2\\
			\bottomrule
		\end{tabular}
		\vspace{-0.3cm}
	\end{table}

	\textbf{CPU Performance Analysis.} As summarized in Table \ref{tab:Main-memory-based Metric Indexes Ranking}, MVPT, GNAT, and $\rm EPT^*$  achieve the best performance in terms of running time, while LAESA and FQT perform the worst. Although LAESA needs few distance computations, the CPU performance is bad. This is because it needs to scan the entire table to find the final result without opportunities for bulk pruning. Next, MVPT and GNAT achieve better CPU performance as they are balanced trees while the others (SAT, BKT, and FQT) are unbalanced. Although BST employs a balanced tree structure, its CPU cost is high because its tree is a binary tree and is weak in terms of pruning. Hence, the CPU performance not only depends on the number of distance computations needed during the search, but also on the index structure.
	
	\subsection{Comparison among Disk-based Metric Indexes}
	\label{subsec:Comparison among Disk-based Metric Indexes}
	We proceed to compare the MRQ and M\textit{k}NN query performance of all {secondary-memory based} metric indexes. These indexes include compact-partitioning based methods (i.e., DSACLT, LC, $\rm MB^+$-tree, and M-tree), pivot-based methods (i.e., OmniR-tree and SPB-tree), and hybrid methods (i.e., D-index, EGNAT, PM-tree, CPT, and M-$\rm index^*$). For the {secondary-memory based} indexes (the $\rm B^+$-tree, R-tree, and M-tree, etc.), the tree height depends on the page/node size. We fix the page size to 4KB. The index height can then be calculated based on the index structure and the data distribution. However, CPT, the M-tree, the PM-tree, LC, DSACLT, and EGNAT all store the data objects directly in their index structures. Thus, they need a large page size for high-dimensional data, and they are configured to use 40KB pages when applied to \emph{Color} and \emph{Synthetic} datasets as default. However, a 40KB page can be regarded as 10 4KB pages for one tree node or cluster. In addition, the number of pivots is fixed to 5, and all the pivot-based methods and hybrid methods (except EGNAT) use the same set of pivots for fair comparison. This is because, EGNAT uses the centers in each tree-level as the pivots for this tree-level due to its design.

	\begin{figure}
		\centering
		\subfigtopskip=0cm
		\subfigbottomskip=0cm
		\subfigcapskip=0cm
		\includegraphics[width=\linewidth]{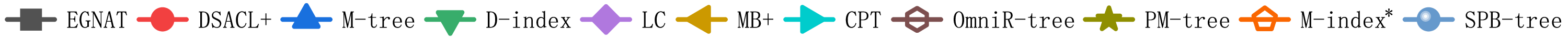}
		\\
		\subfigure[LA]{
			\includegraphics[width=0.3\linewidth]{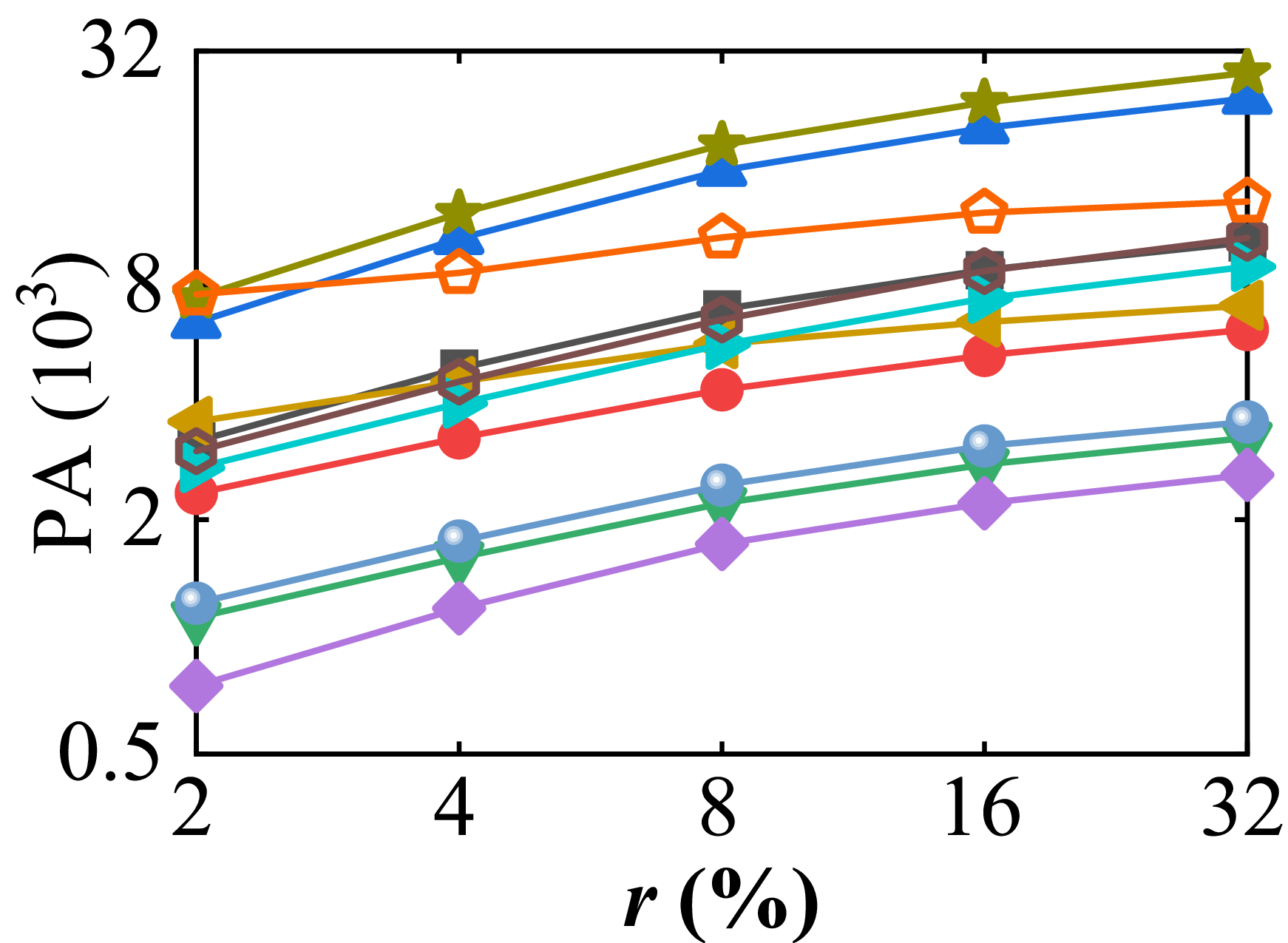}
		}
		\subfigure[LA]{
			\includegraphics[width=0.3\linewidth]{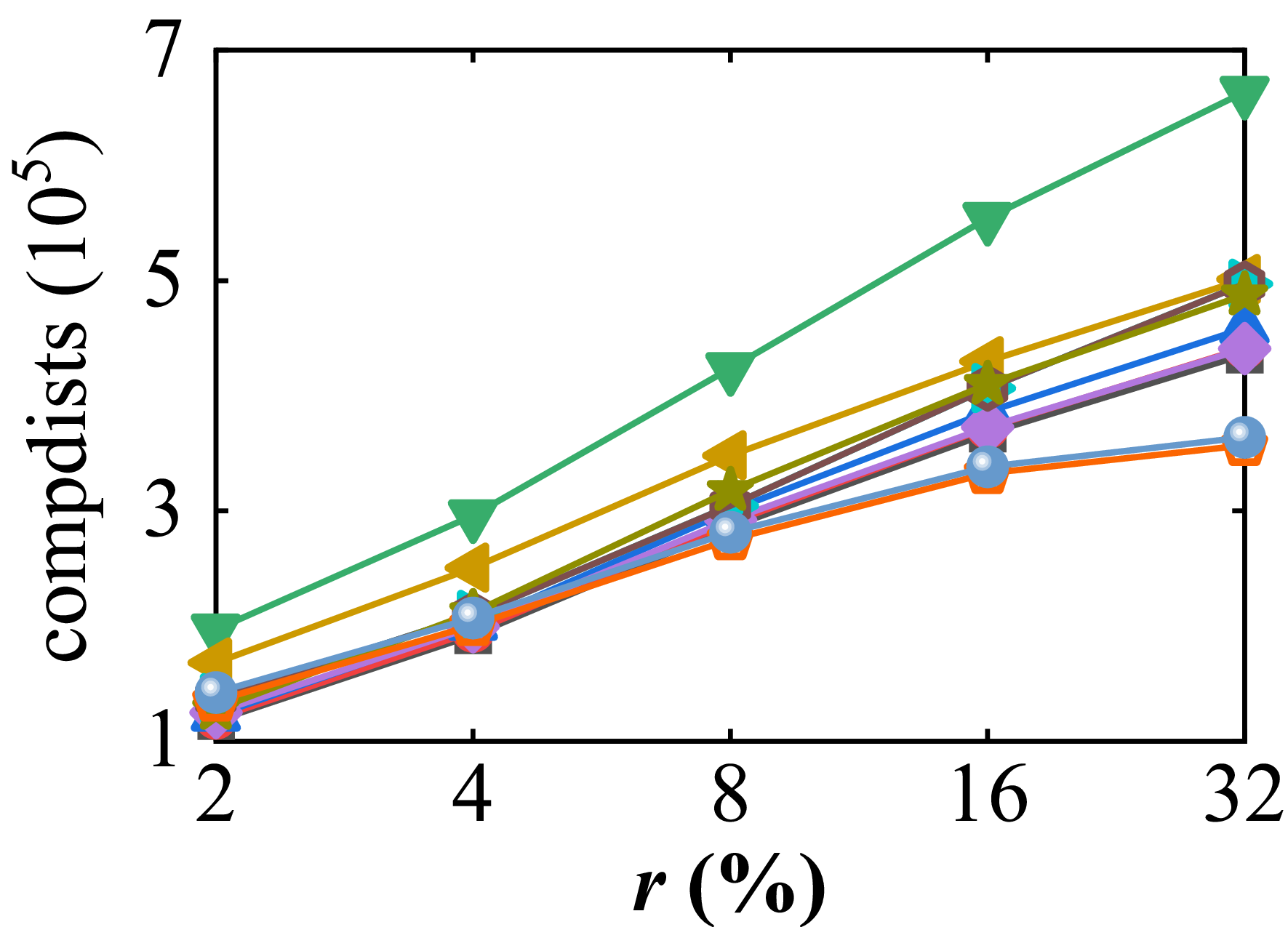}
		}
		\subfigure[LA]{
			\includegraphics[width=0.3\linewidth]{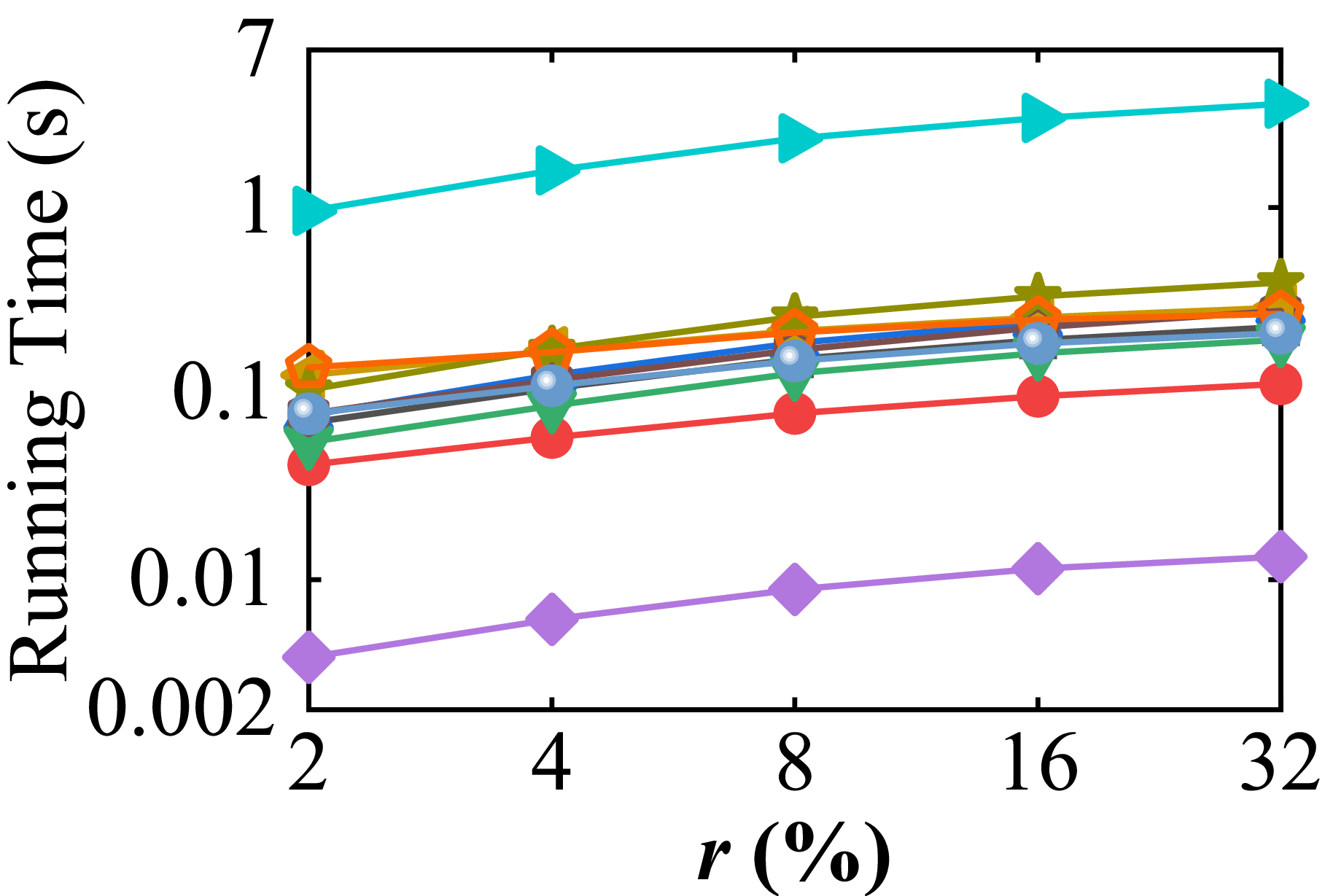}
		}
		\\
		\subfigure[Words]{
			\includegraphics[width=0.3\linewidth]{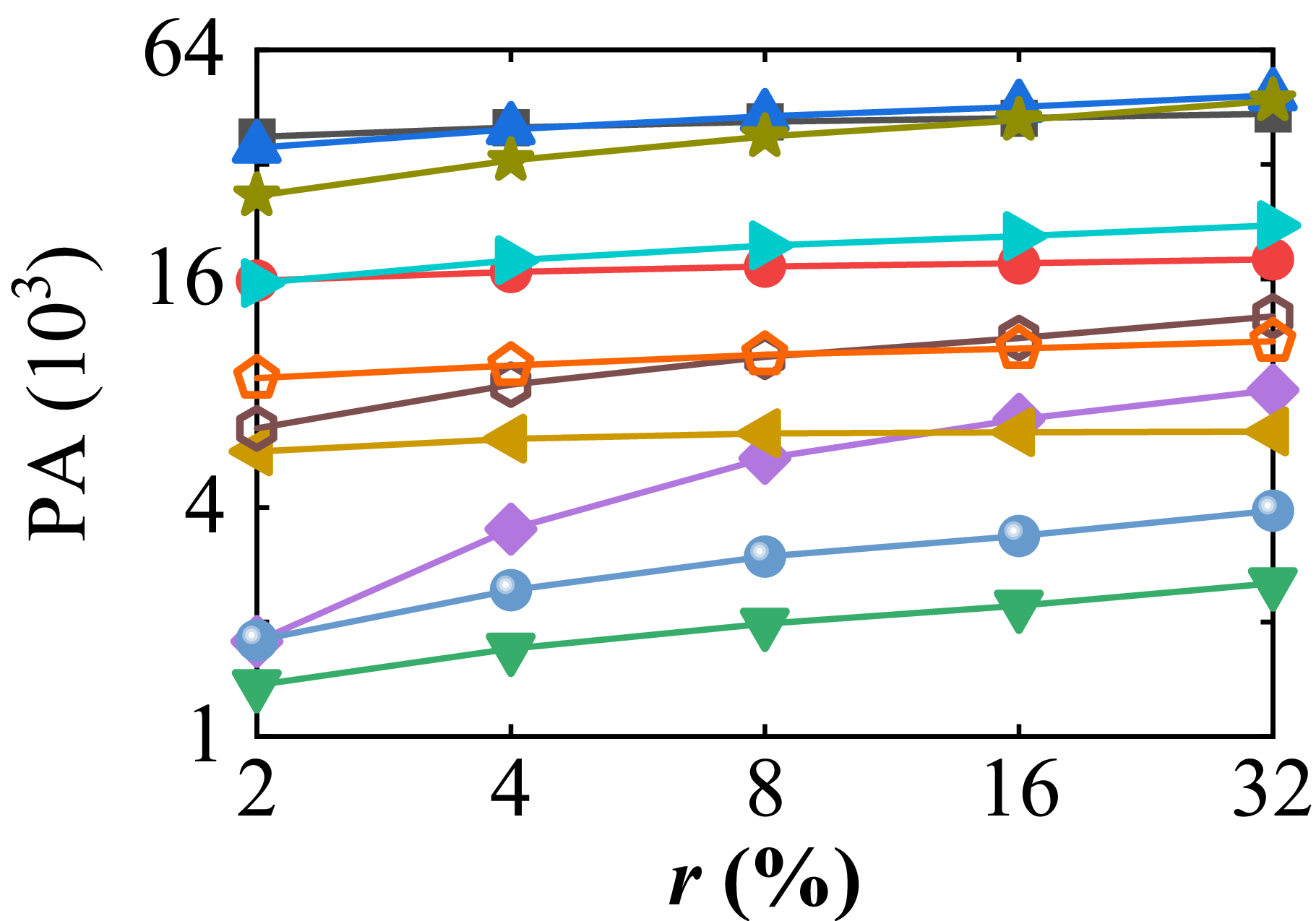}
		}
		\subfigure[Words]{
			\includegraphics[width=0.3\linewidth]{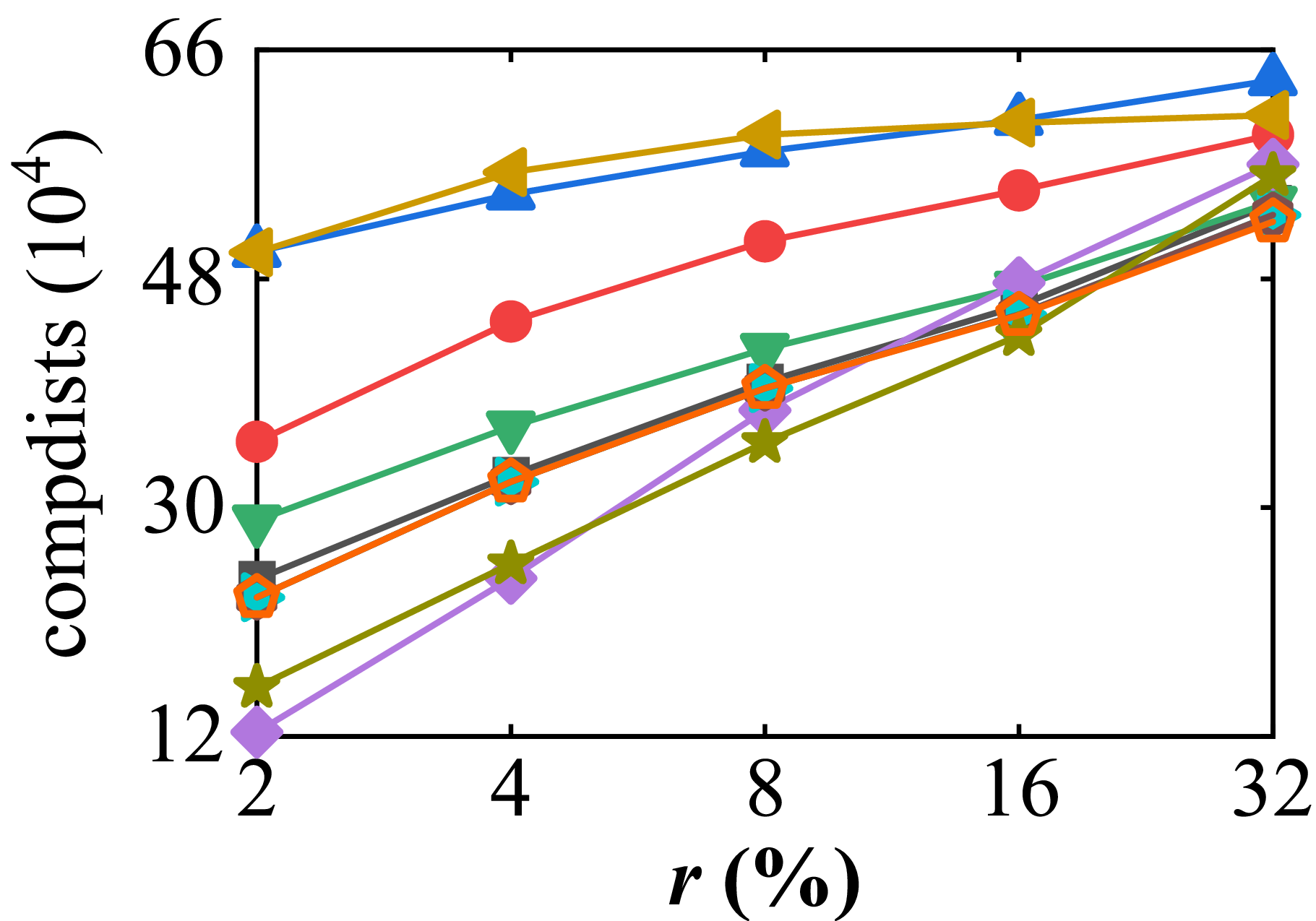}
		}
		\subfigure[Words]{
			\includegraphics[width=0.3\linewidth]{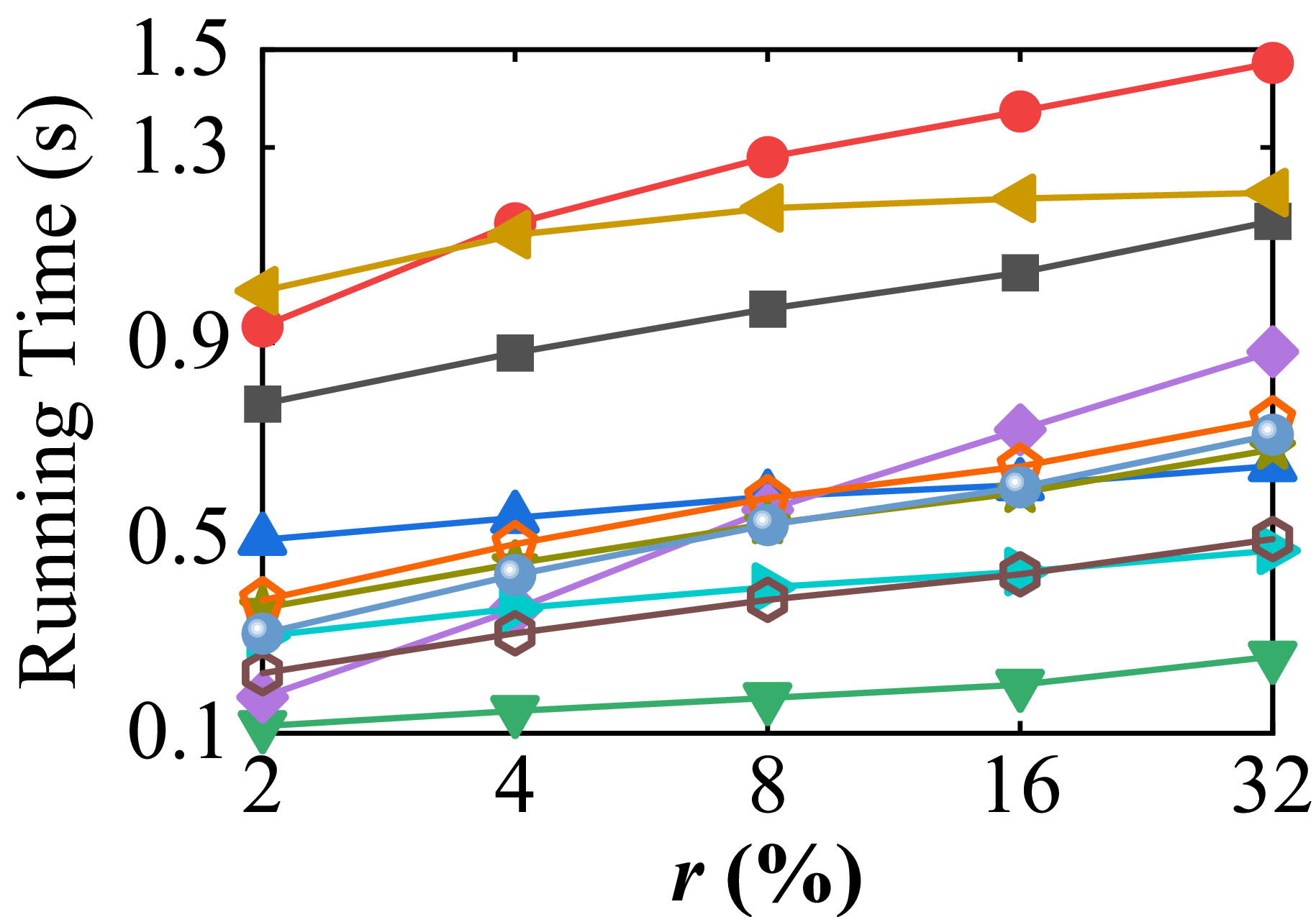}
		}
		\\
		\subfigure[Color]{
			\includegraphics[width=0.3\linewidth]{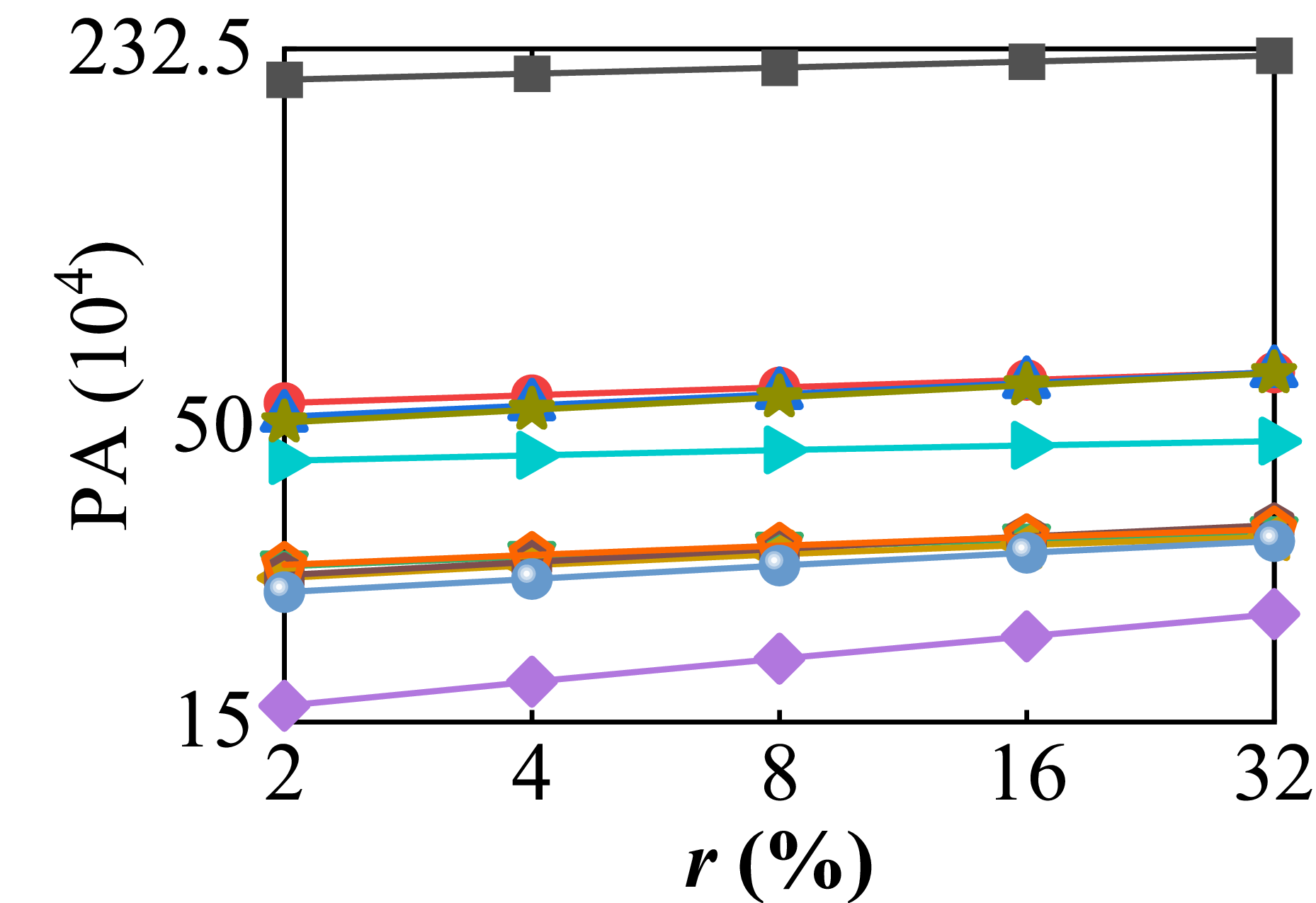}
		}
		\subfigure[Color]{
			\includegraphics[width=0.3\linewidth]{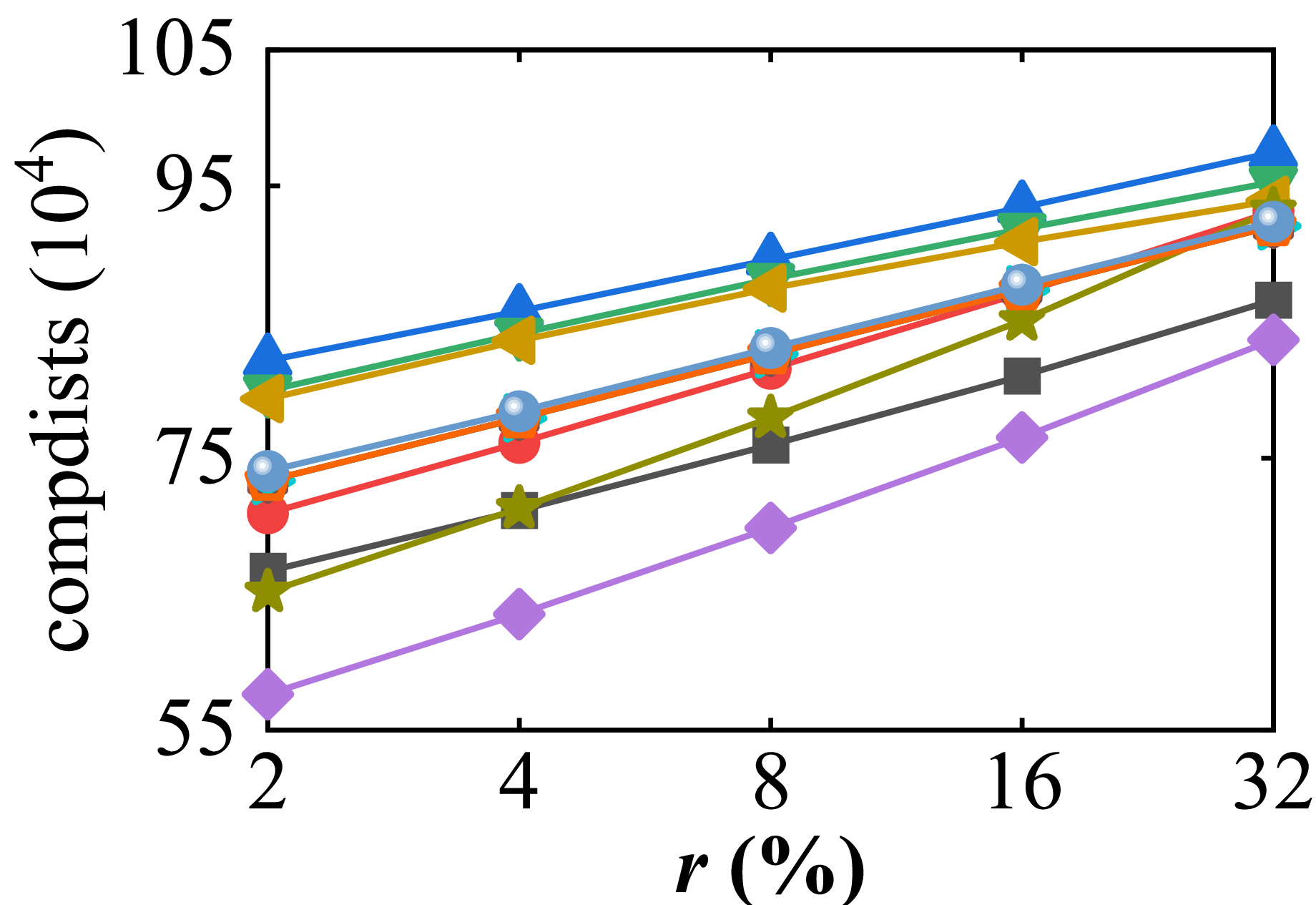}
		}
		\subfigure[Color]{
			\includegraphics[width=0.3\linewidth]{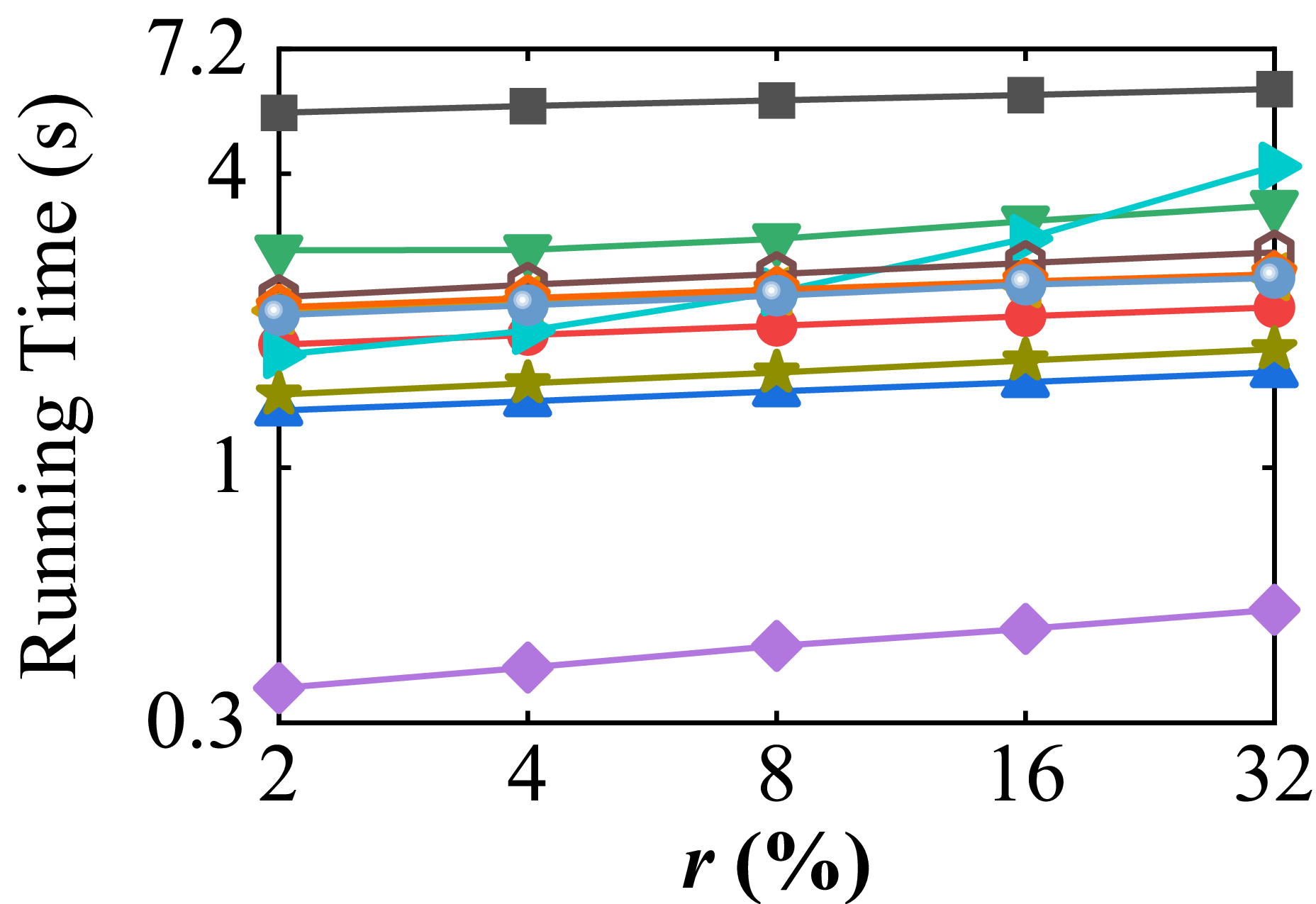}
		}
		\\
		\subfigure[Synthetic]{
			\includegraphics[width=0.3\linewidth]{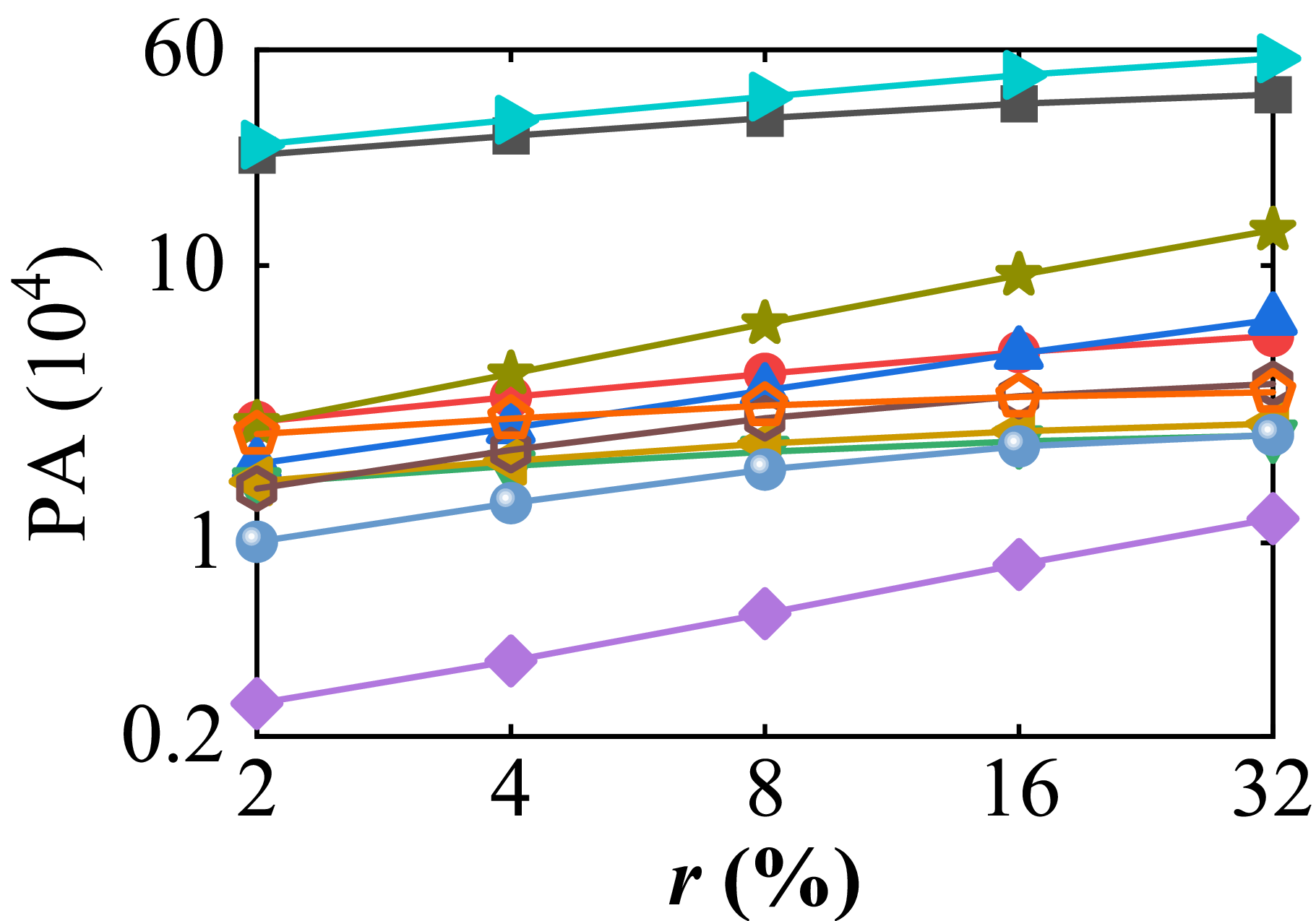}
		}
		\subfigure[Synthetic]{
			\includegraphics[width=0.3\linewidth]{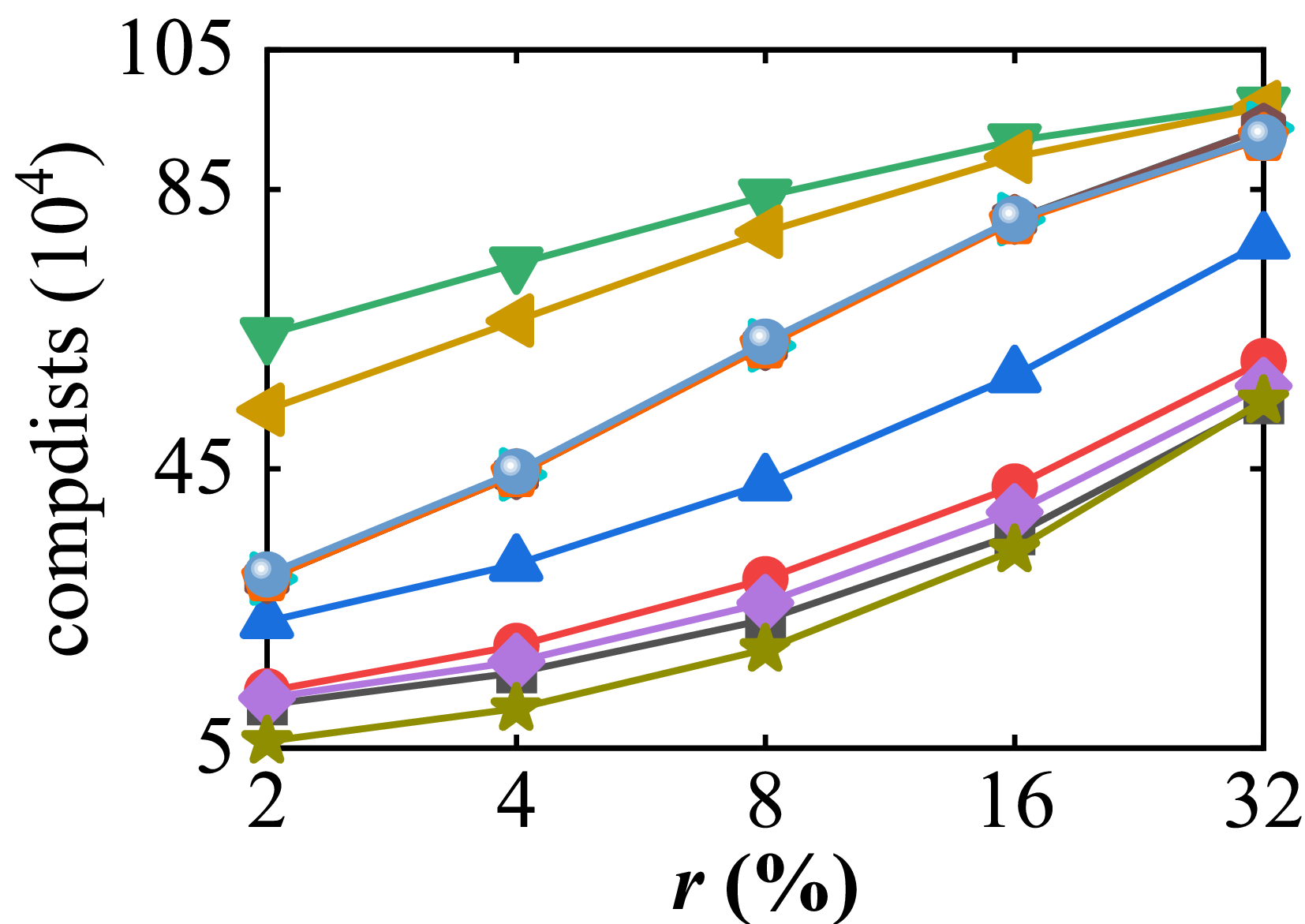}
		}
		\subfigure[Synthetic]{
			\includegraphics[width=0.3\linewidth]{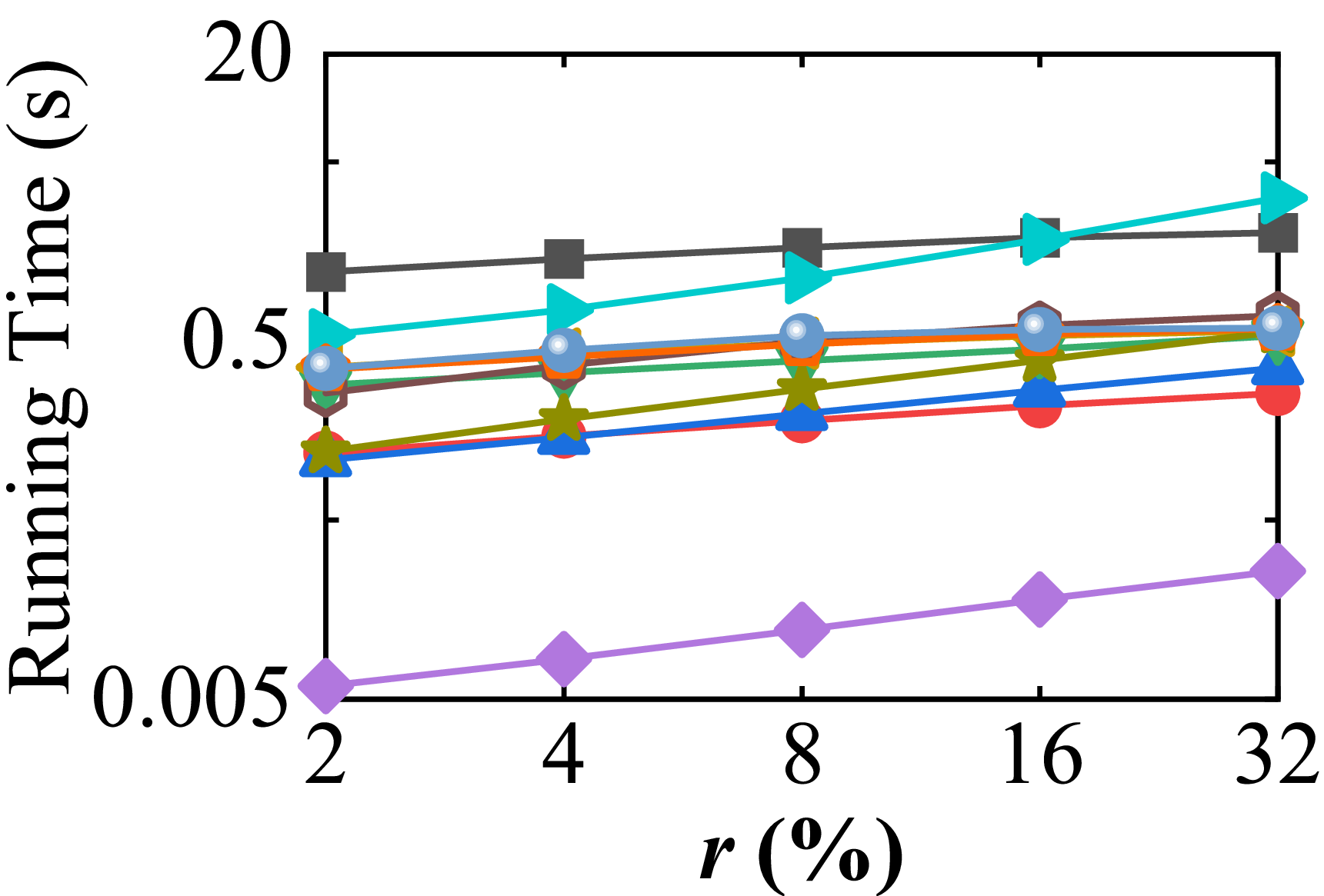}
		}
		\setlength{\abovecaptionskip}{0.1cm}
		\caption{{Secondary-memory based} Metric Index Comparison Using MRQ Queries}
		\label{fig:Disk-based Metric Index Comparison Using MRQ queries}
		\vspace{-0.3cm}
	\end{figure}

\begin{figure}
		\centering
		\subfigtopskip=0cm
		\subfigbottomskip=0cm
		\subfigcapskip=0cm
		\includegraphics[width=\linewidth]{External_memory.eps}
		\\
		\subfigure[LA]{
			\includegraphics[width=0.3\linewidth]{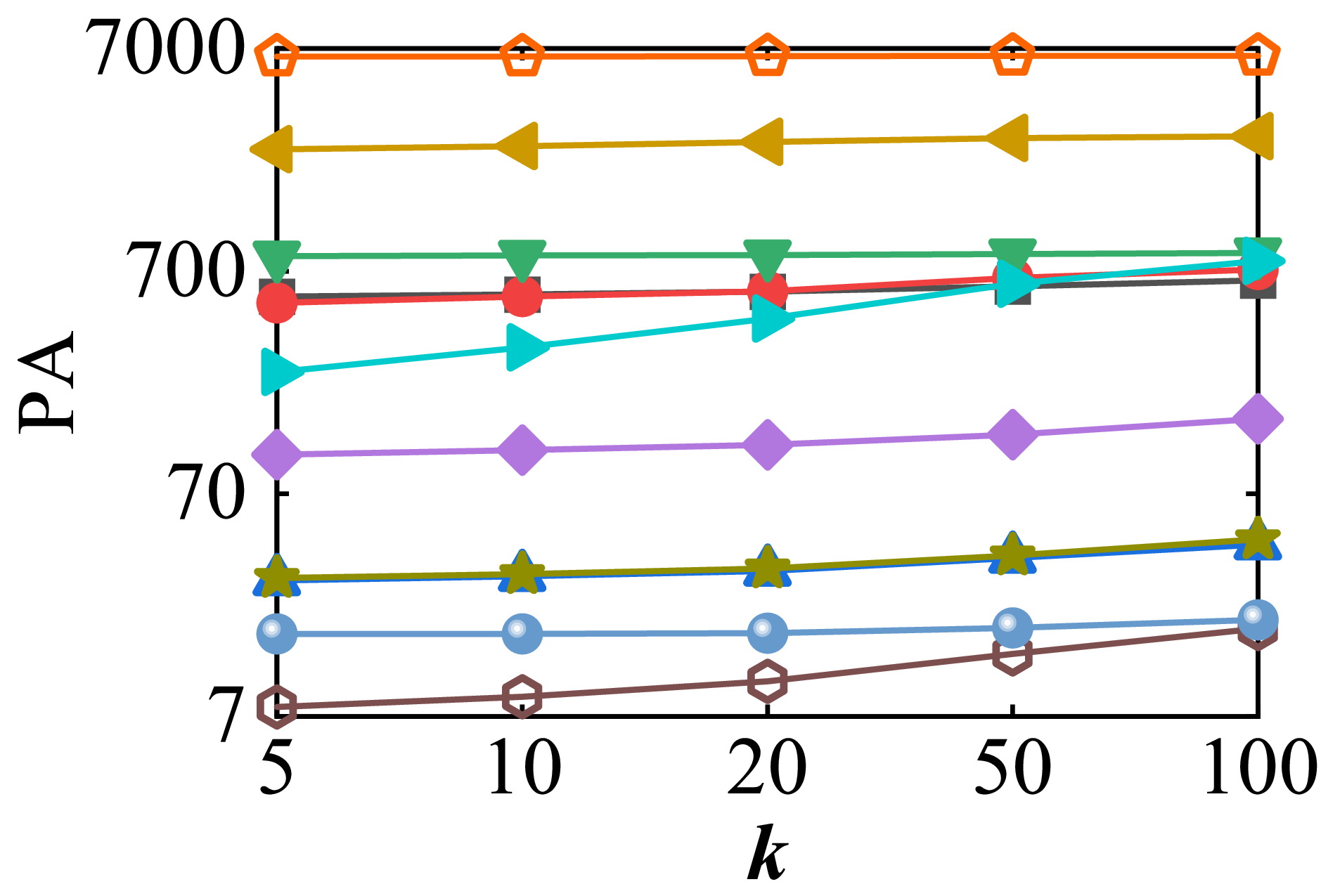}
		}
		\subfigure[LA]{
			\includegraphics[width=0.3\linewidth]{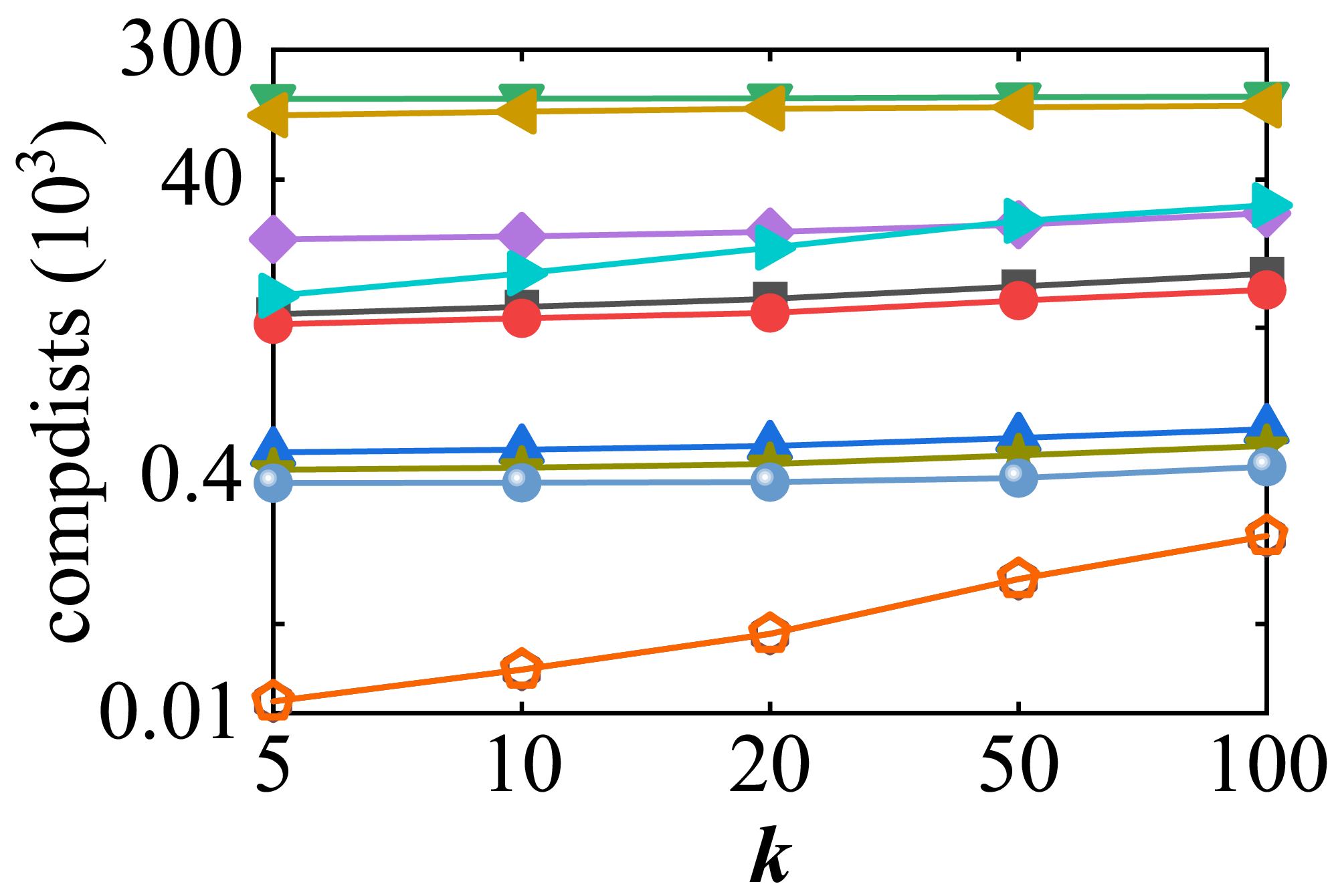}
		}
		\subfigure[LA]{
			\includegraphics[width=0.3\linewidth]{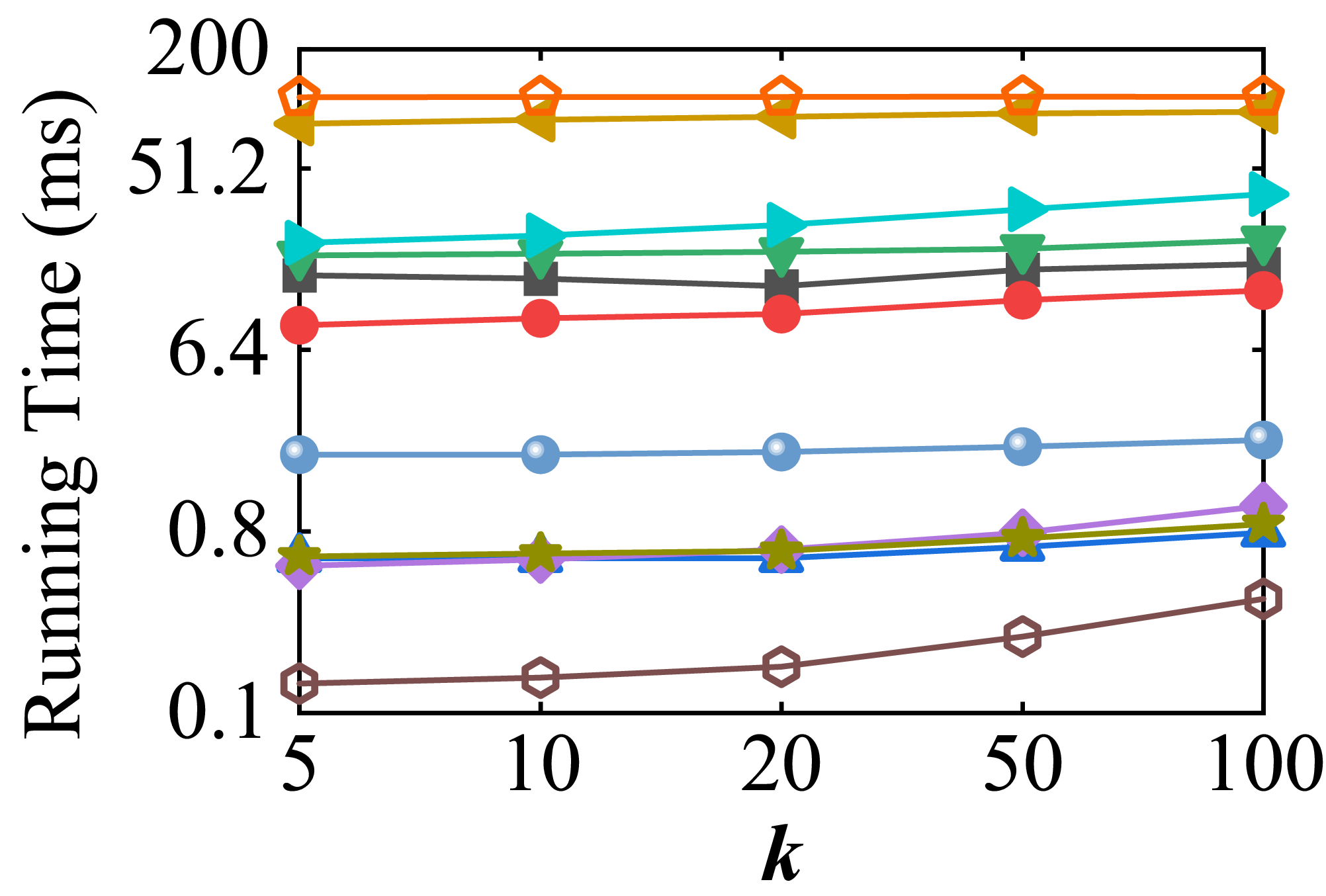}
		}
		\\
		\subfigure[Words]{
			\includegraphics[width=0.3\linewidth]{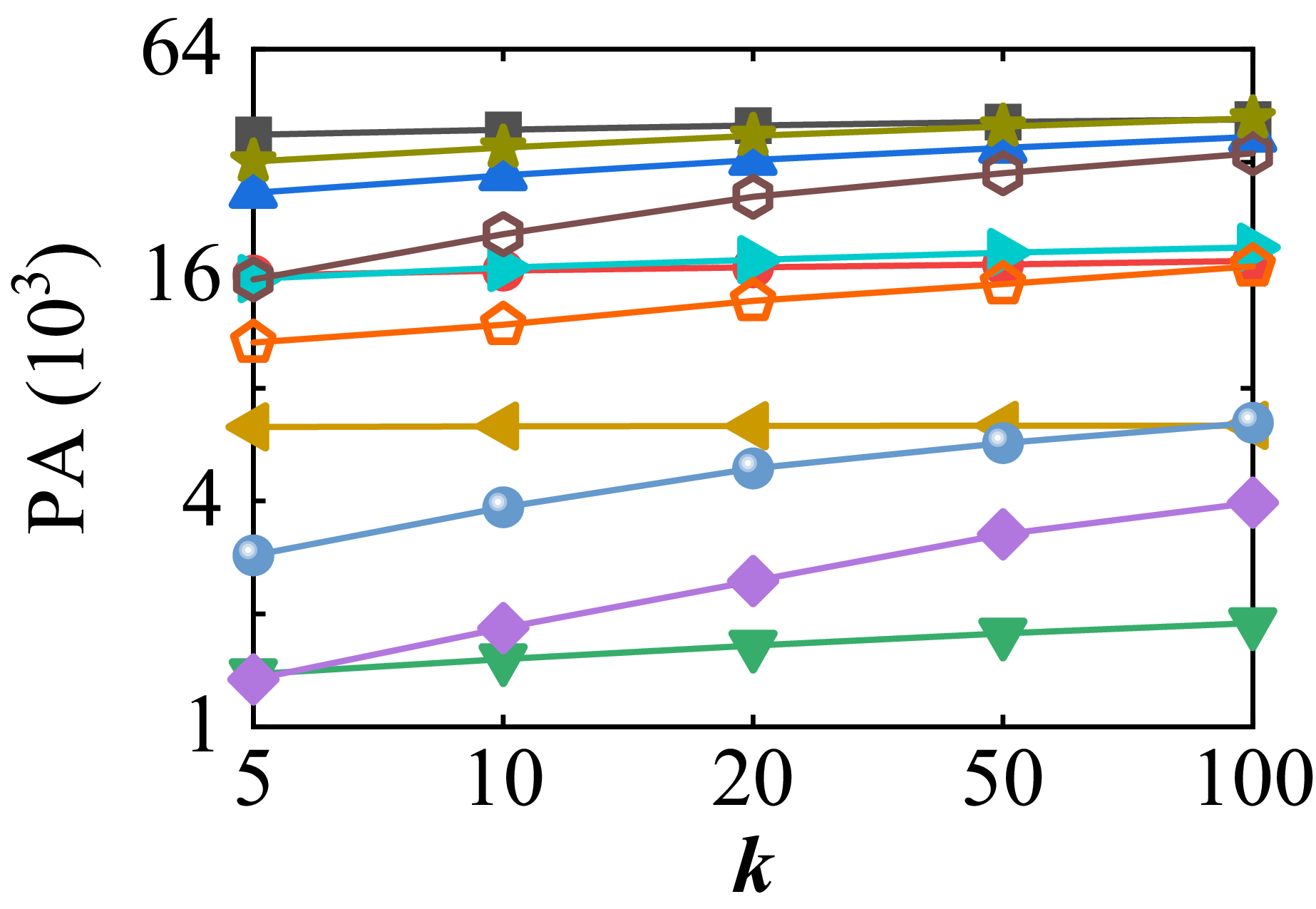}
		}
		\subfigure[Words]{
			\includegraphics[width=0.3\linewidth]{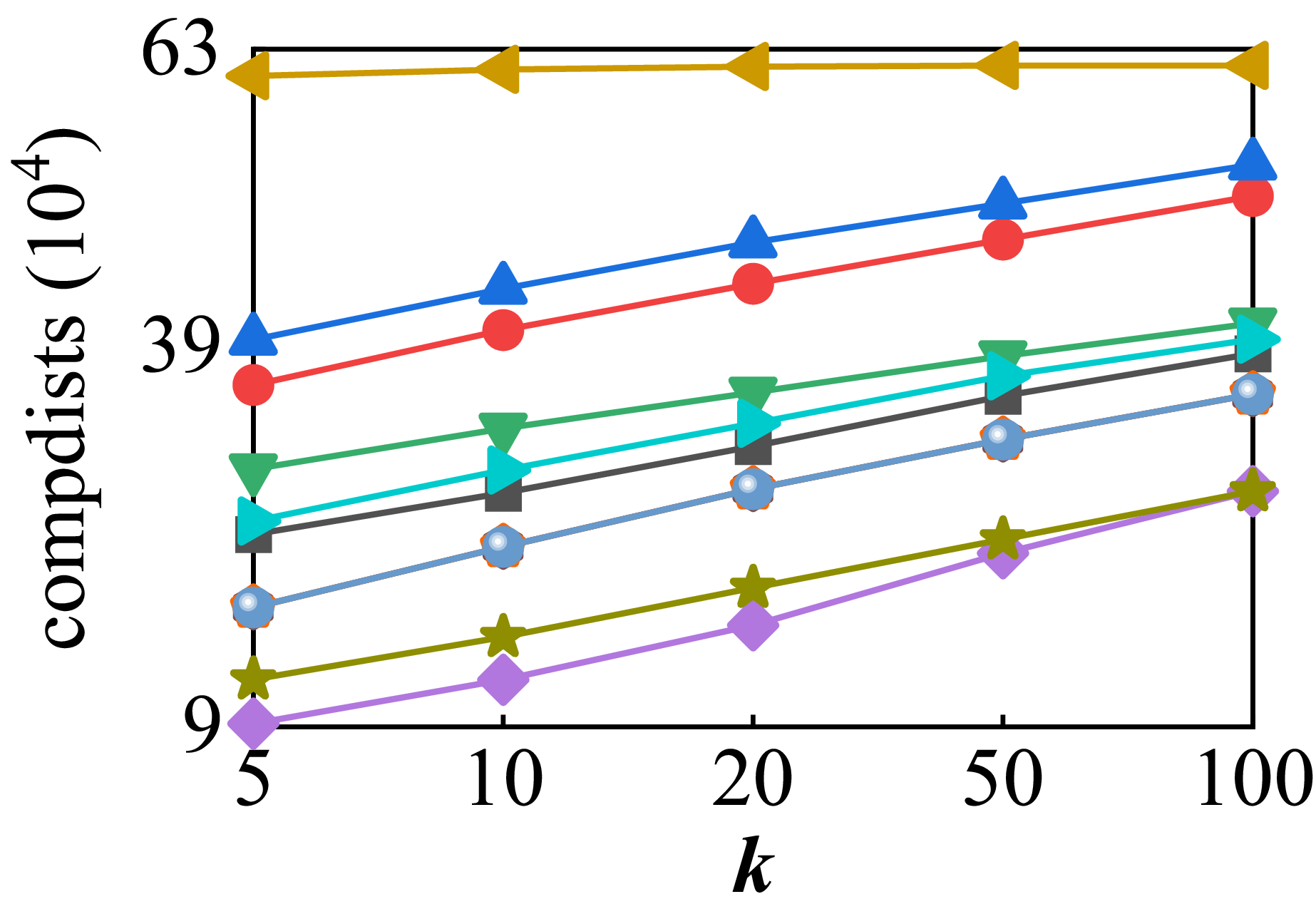}
		}
		\subfigure[Words]{
			\includegraphics[width=0.3\linewidth]{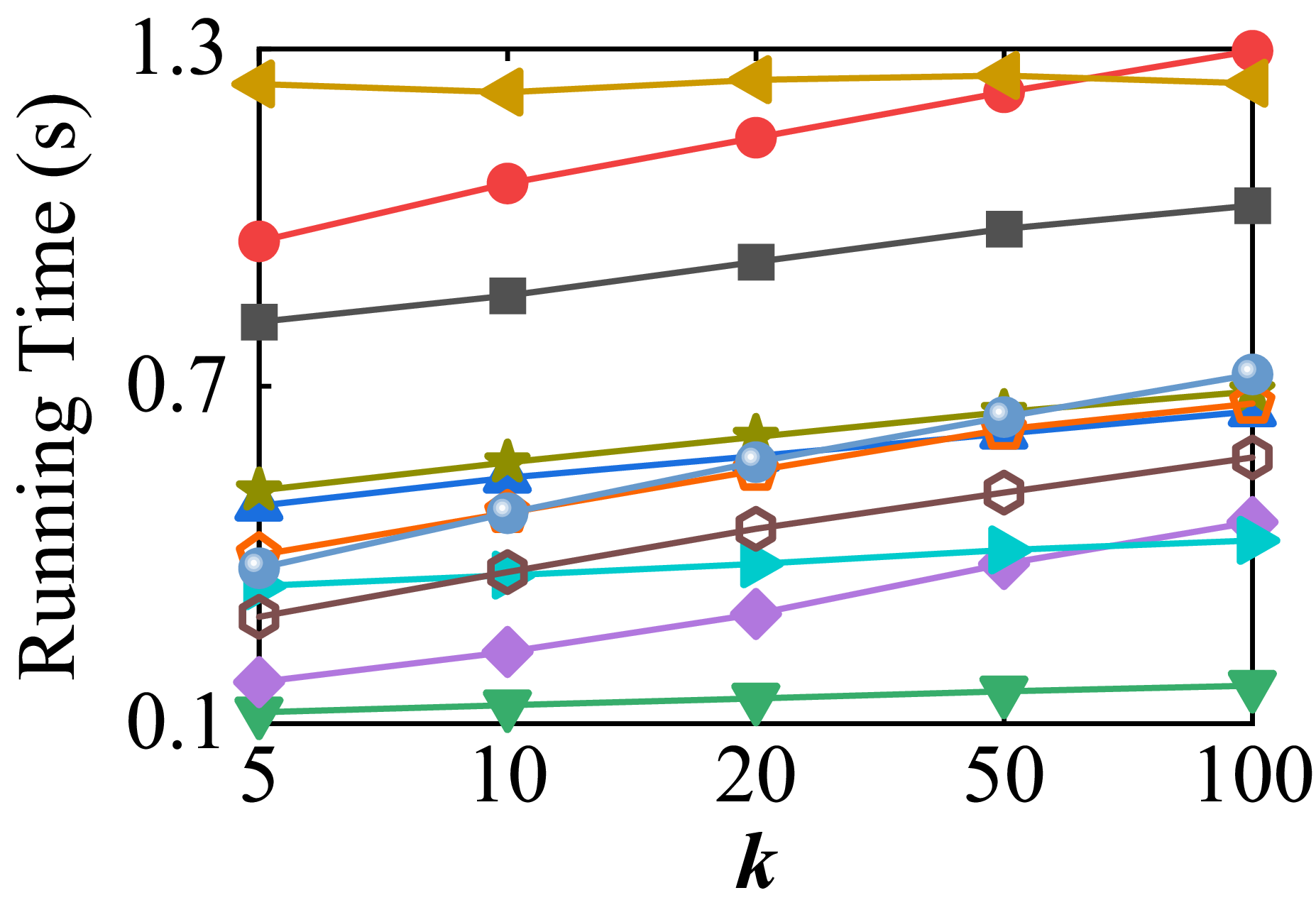}
		}
		\\
		\subfigure[Color]{
			\includegraphics[width=0.3\linewidth]{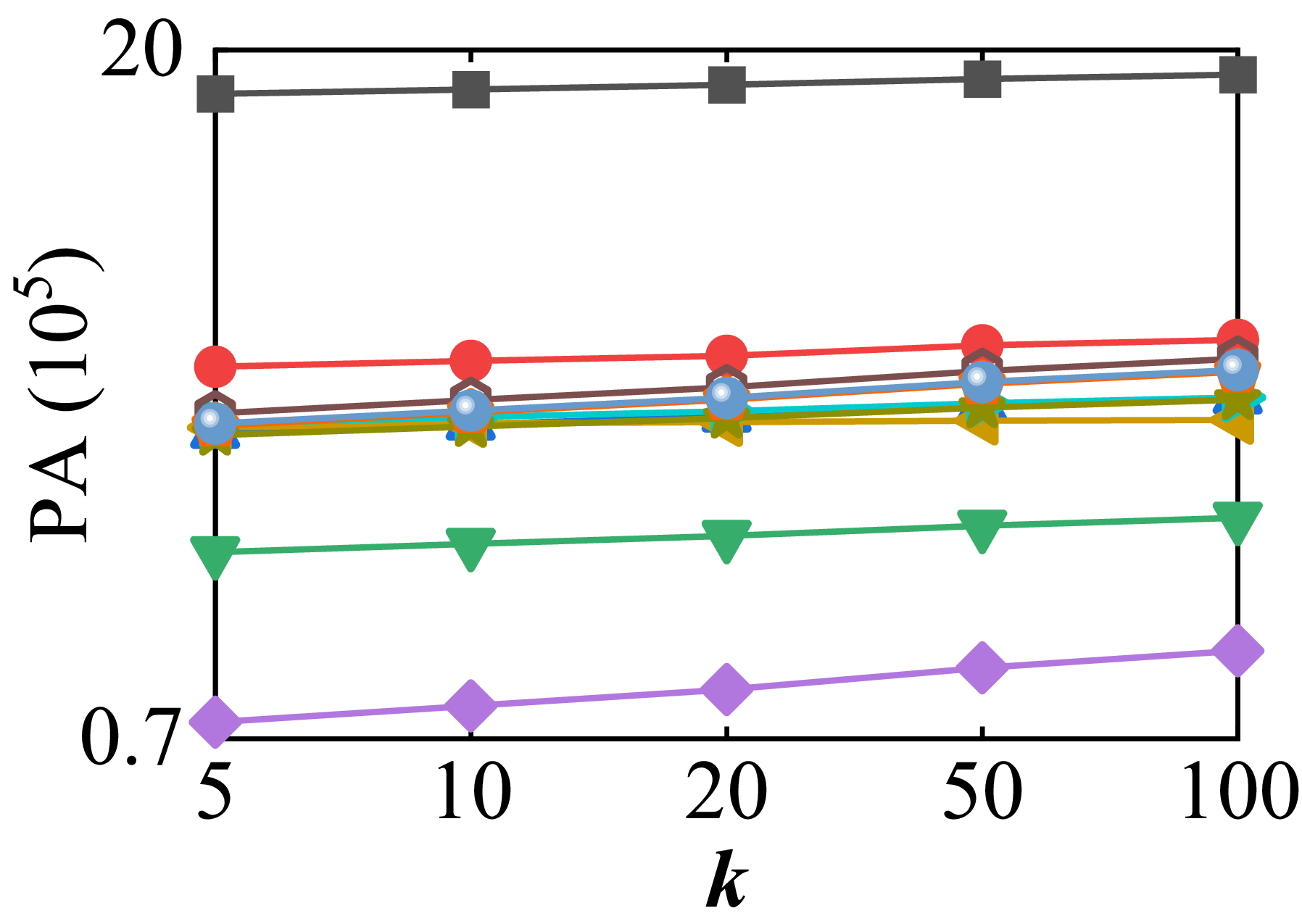}
		}
		\subfigure[Color]{
			\includegraphics[width=0.3\linewidth]{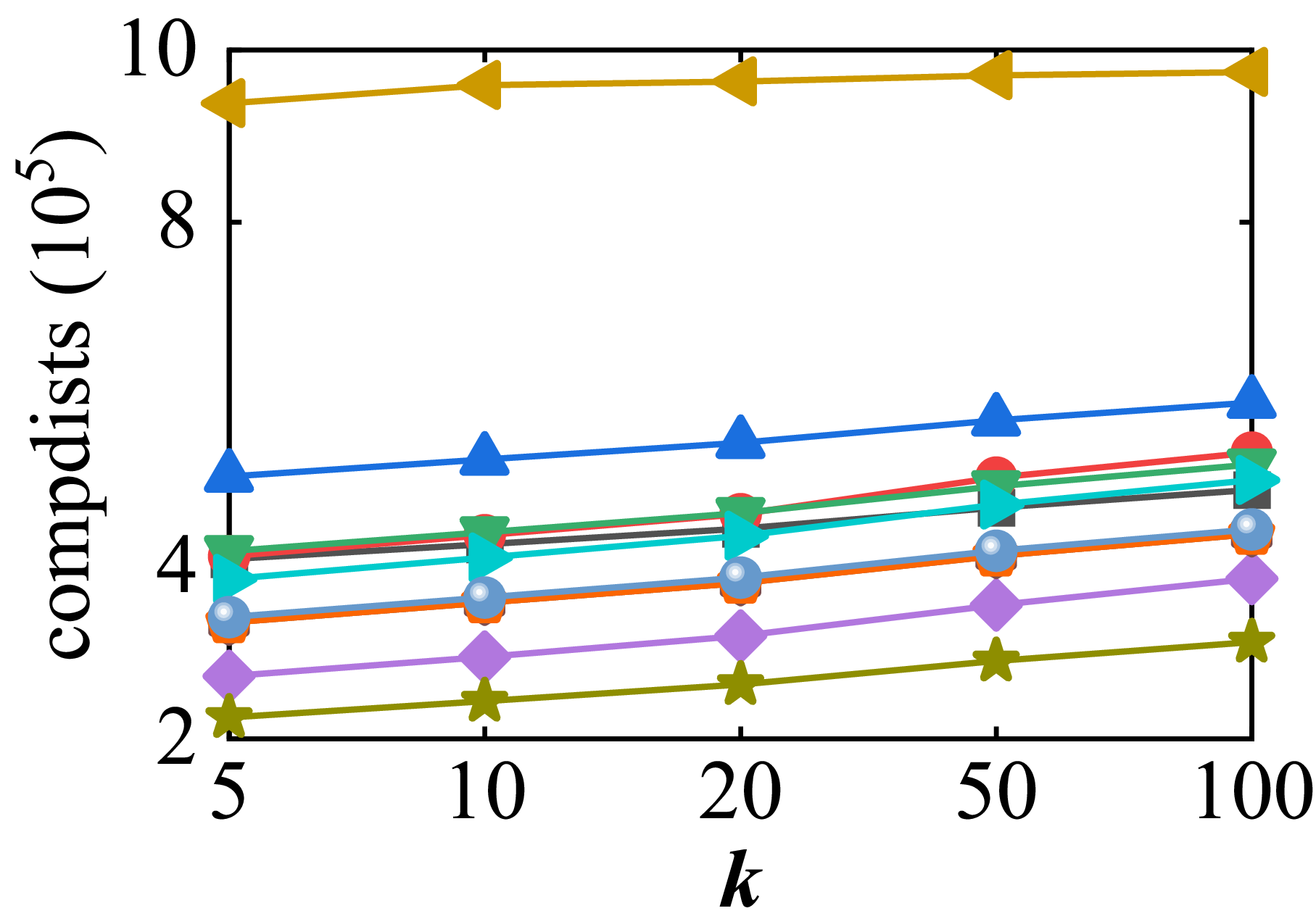}
		}
		\subfigure[Color]{
			\includegraphics[width=0.3\linewidth]{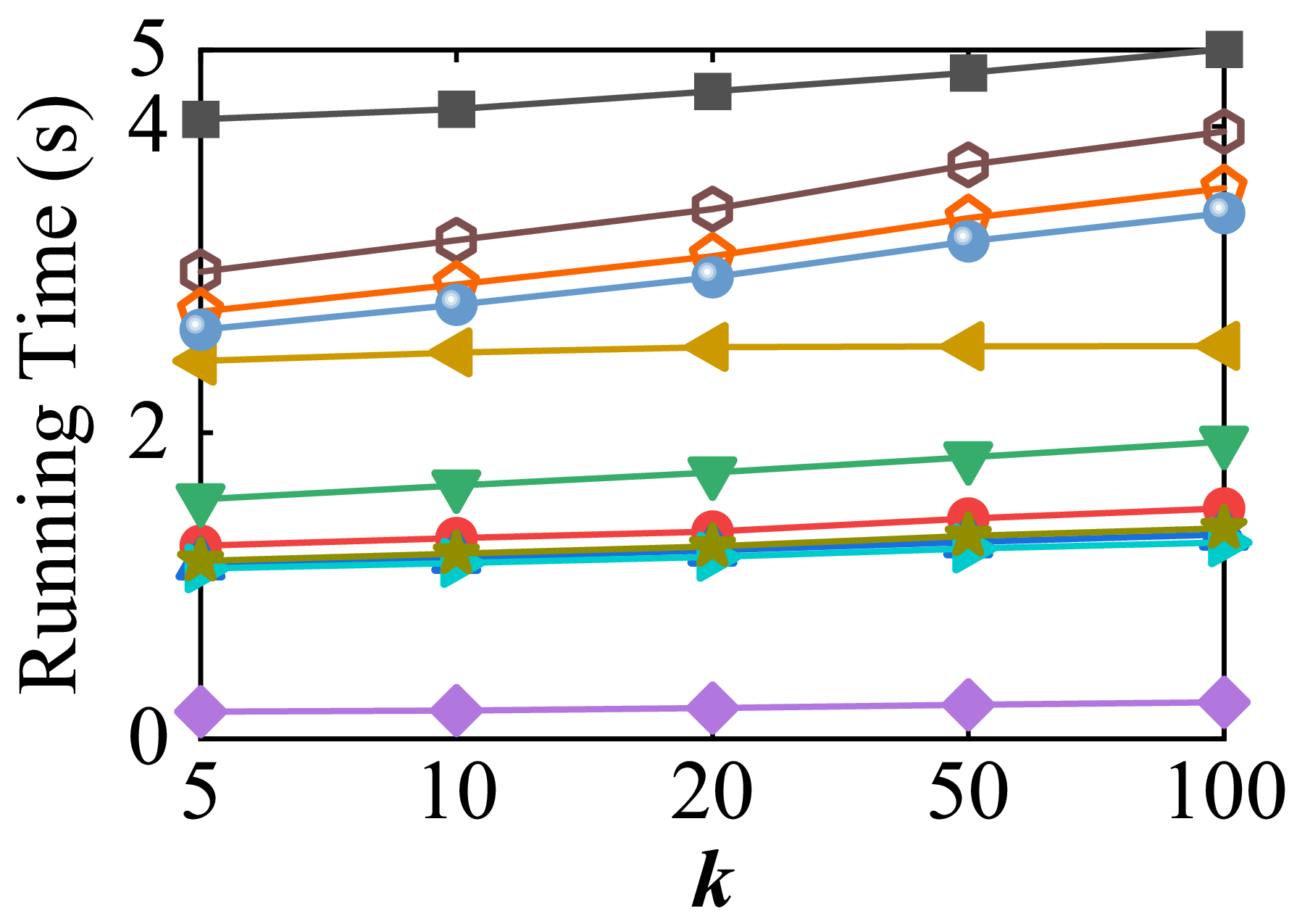}
		}
		\\
		\subfigure[Synthetic]{
			\includegraphics[width=0.3\linewidth]{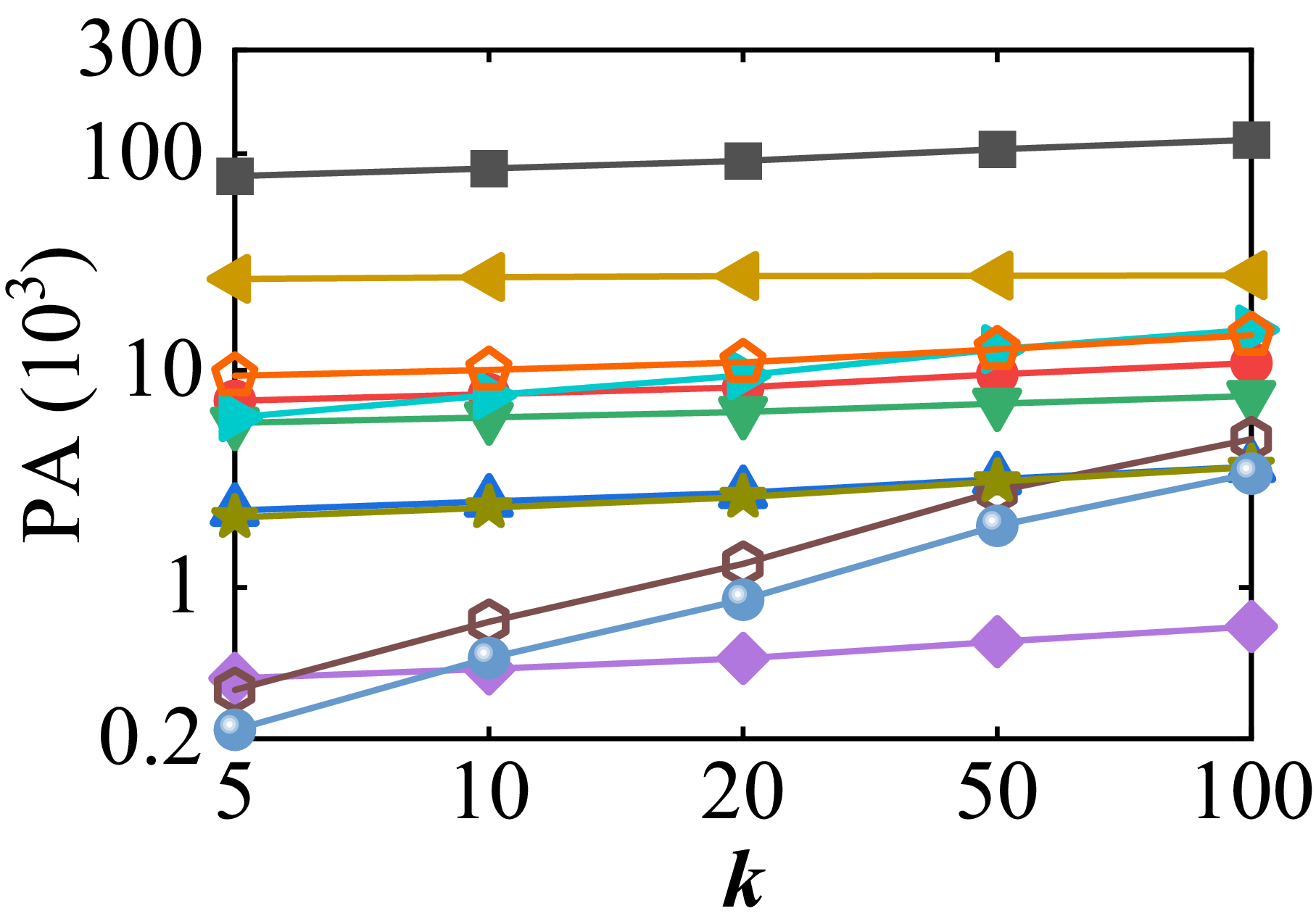}
		}
		\subfigure[Synthetic]{
			\includegraphics[width=0.3\linewidth]{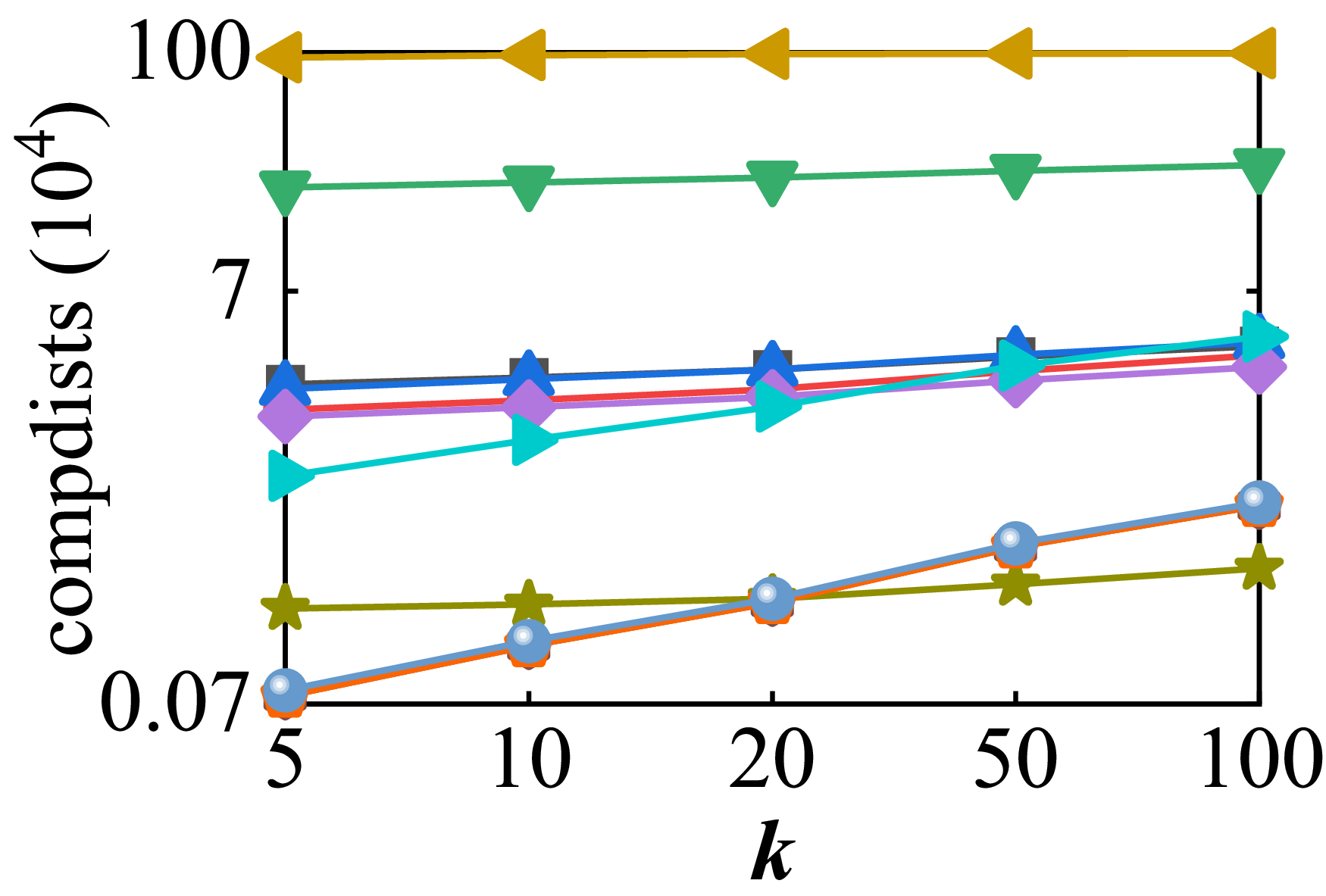}
		}
		\subfigure[Synthetic]{
			\includegraphics[width=0.3\linewidth]{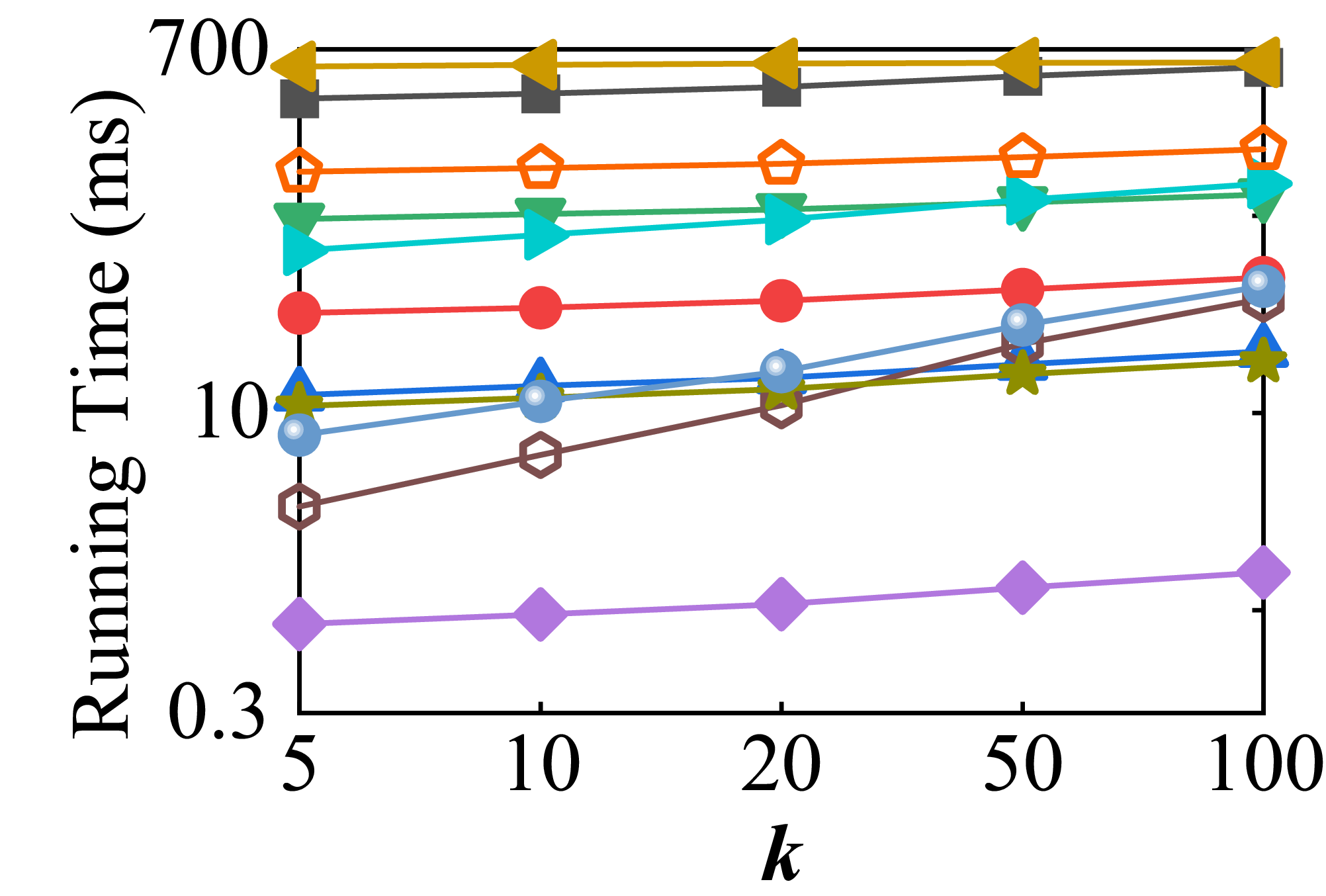}
		}
		\setlength{\abovecaptionskip}{0.1cm}
		\caption{{Secondary-memory based} Metric Index Comparison Using M\textit{k}NN Queries}
		\label{fig:Disk-based Metric Index Comparison Using MkNN queries}
		\vspace{-0.3cm}
	\end{figure}

	Fig.~\ref{fig:Disk-based Metric Index Comparison Using MRQ queries} and Fig.~\ref{fig:Disk-based Metric Index Comparison Using MkNN queries} show the MRQ and M\textit{k}NN query performance of the {secondary-memory based} metric indexes when applied to the four datasets, respectively. As expected, the query cost (including \textit{PA}, \textit{compdists}, and running time) increases with the growth of $k$ and $r$ due to the resulting larger search spaces. Table \ref{tab:Disk-based metric Indexes Ranking} displays the performance metric (including \textit{PA}, \textit{compdists}, and running time) ranking of the {secondary-memory based} metric indexes over four datasets using MRQ and M\textit{k}NN queries. In the following, we provide observations on the ranking with respect to three different performance metrics.
	
	\begin{table}
		\centering
		\setlength{\abovecaptionskip}{0cm}
		\caption{{Secondary-memory based} metric Indexes Ranking}
		\label{tab:Disk-based metric Indexes Ranking}
		\begin{tabular}{ccccccc}
			\hline
			\multirow{2}{*}{\tabincell{c}{\textbf{{Secondary-memory}} \\ \textbf{{based Metric Index}}}} & \multicolumn{2}{c}{\textbf{PA ranking}} & \multicolumn{2}{c}{\textbf{Compdists ranking}} & \multicolumn{2}{c}{\textbf{Running time ranking}} \\\cline{2-7}
			~ & \textbf{MRQ} & \textbf{M\textit{k}NN} & \textbf{MRQ} & \textbf{M\textit{k}NN} & \textbf{MRQ} & \textbf{M\textit{k}NN} \\
			\midrule
			\textbf{LC}          & 1 & 1 & 1 & 5 & 1 & 1 \\
			\textbf{DSACLT}      & 7 & 9 & 5 & 8 & 2 & 8\\
			\textbf{M-Tree}      &9 & 5 & 9 & 9 & 4 & 2\\
			\textbf{$\bm{{\rm MB^+}}$-tree} &  4 & 7 & 10 & 11 & 10 & 11\\
			\textbf{OmniR-tree}  &  5 & 4 & 8 & 2 & 7 & 3\\
			\textbf{SPB-tree}    & 2 & 2 & 6 & 4 & 5 & 7\\
			\textbf{D-index}     & 3 & 3 & 11 & 10&2& 6\\
			\textbf{EGNAT}       & 11 & 11 & 2 & 7 &11 & 10\\
			\textbf{PM-tree}     & 9 & 5 & 2 & 3 & 5 & 4\\
			\textbf{CPT}         & 8 & 7 &7 & 6 & 9 & 5\\
			\textbf{M-$\bm{{\rm index^*}}$} &  6 & 10 & 4 & 1 & 8 & 9\\
			\bottomrule
		\end{tabular}
		\vspace{-0.3cm}
	\end{table}
	
	\textbf{I/O Performance Analysis.} As summarized in Table \ref{tab:Disk-based metric Indexes Ranking}, LC, the SPB-tree, and the D-index achieve the best performance in terms of \textit{PA}, followed by the OmniR-tree, the $\rm MB^+$-tree, the M-tree, the PM-tree, and CPT, while EGNAT, DSACLT, and the M-$\rm index^*$ perform the worst. The I/O cost depends on whether the data is well clustered and on \textit{compdists} needed during search. The compact-partitioning based methods can achieve better I/O performance compared with the pivot-based methods and the hybrid methods, because they are able to cluster the data well and because the pivot-based and hybrid methods need to store additionally large number of pre-computed distances. As LC is partitioning-based method that can cluster the data well, they achieve good I/O performance. Although the M-tree, the $\rm MB^+$-tree, and DSACLT cluster the data in compact partitions, their \textit{compdists} during search are high (i.e., corresponding to low pruning ability). The SPB-tree, a pivot-based method, uses a space-filling curve to cluster the data and reduce the storage needed for pre-computed distances. Thus, it achieves good I/O performance.
	
	\textbf{Compdists Performance Analysis.} As summarized in Table \ref{tab:Disk-based metric Indexes Ranking}, the M-$\rm index^*$, the OmniR-tree, the PM-tree, and the  SPB-tree achieve the best performance in terms of \emph{compidsts} for M\textit{k}NN queries, while LC, EGNAT, the PM-tree, and the M-$\rm index^*$ perform the best for MRQ queries. Thus, the {secondary-memory based} index performance varies slightly across the two types of queries. This is because the search radius for M\textit{k}NN queries is very small, while the search radius for MRQ is large in the experiments. More specifically, LC is constructed and searched in order of data occurrences, while the other indexes can search in best-first order to find NNs quickly and reduce the search space quickly. Hence, LC is better for MRQ with large search radius when compared to M\textit{k}NN queries with small search radius. Overall, the pivot-based and hybrid methods achieve better \textit{compdists} performance than the compact-partitioning methods. This is because we employ high-quality pivots in pivot-based and hybrid methods, while the pruning ability of centers in compact-partitioning methods is weak. Also note that the construction cost of LC, a compact-partitioning method, is $O(n^2)$, enabling it to cluster the data well to trade higher construction cost for better query performance. However, the construction cost of LC is excessive when the cardinality of the dataset is very large. As a result, LC is not a practical choice. Next, although the D-index is a hybrid method that uses pre-computed distances to prune the search space, the hashing partitioning makes it hard to control the quality of clusters, which depends on the split parameter $ \rho $ and the search radius.
	
	\textbf{Running Time Performance Analysis.} The running time depends on \textit{PA}, \textit{compdists}, and the CPU cost of pruning, which is used to evaluate the overall performance. As summarized in Table \ref{tab:Disk-based metric Indexes Ranking}, LC and the M-tree perform the best in terms of running time, followed by the OmniR-tree, the PM-tree, the D-index, the SPB-tree, and DSACLT, while EGNAT, the $\rm MB^+$-tree, the M-$\rm index^*$, and CPT perform the worst. Here, LC has good running time performance, because it performs well in terms of both \textit{PA} and \textit{compdists}. The M-$\rm index^*$ and EGNAT, which have good \textit{compdists} performance, exhibit weak running time  performance due to high I/O cost during search.  Note that, although the PM-tree achieve better \textit{compdists} performance and similar \textit{PA} performance compared with the M-tree, the M-tree performs slightly better in terms of running time performance due to the relatively high pruning CPU cost required by the PM-tree. 
Although the SPB-tree shows a very good performance on \textit{PA}, it incurs additional CPU cost (i.e., space filling curve transformation) during the search, resulting in a relatively high CPU cost.
	
	\subsection{Effect of Intrinsic Dimensionality}
{In order to vary the intrinsic dimensionality (ID), we change the dimensionality of \textit{Synthetic} dataset (denoted as \textit{num}) among 3, 5, 7, and 9. Specifically, \textit{num} of the 20-dimension values in \textit{Synthetic} is generated randomly, while the remaining values are linear combinations of the previous ones. The corresponding IDs are 3.7, 6.8, 10.5, and 14.1, respectively.}
	
\begin{figure}
		\centering
		\subfigtopskip=0cm
		\subfigbottomskip=0cm
		\subfigcapskip=0cm
		\includegraphics[width=0.6\linewidth]{Main_memory.eps}
		\\
		\subfigure[Compdists]{
			\includegraphics[width=0.3\linewidth]{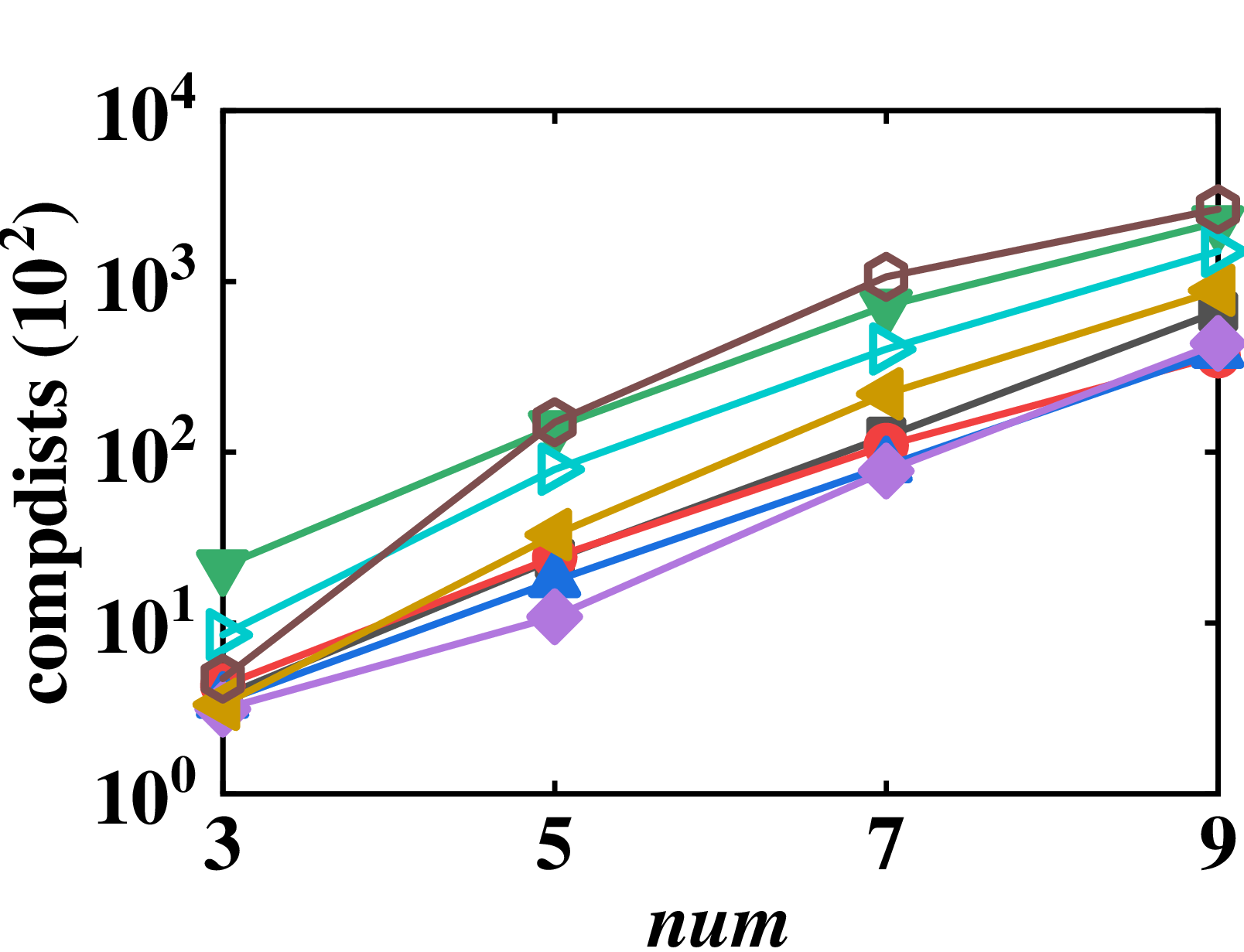}
		}
		\subfigure[Running Time]{
			\includegraphics[width=0.3\linewidth]{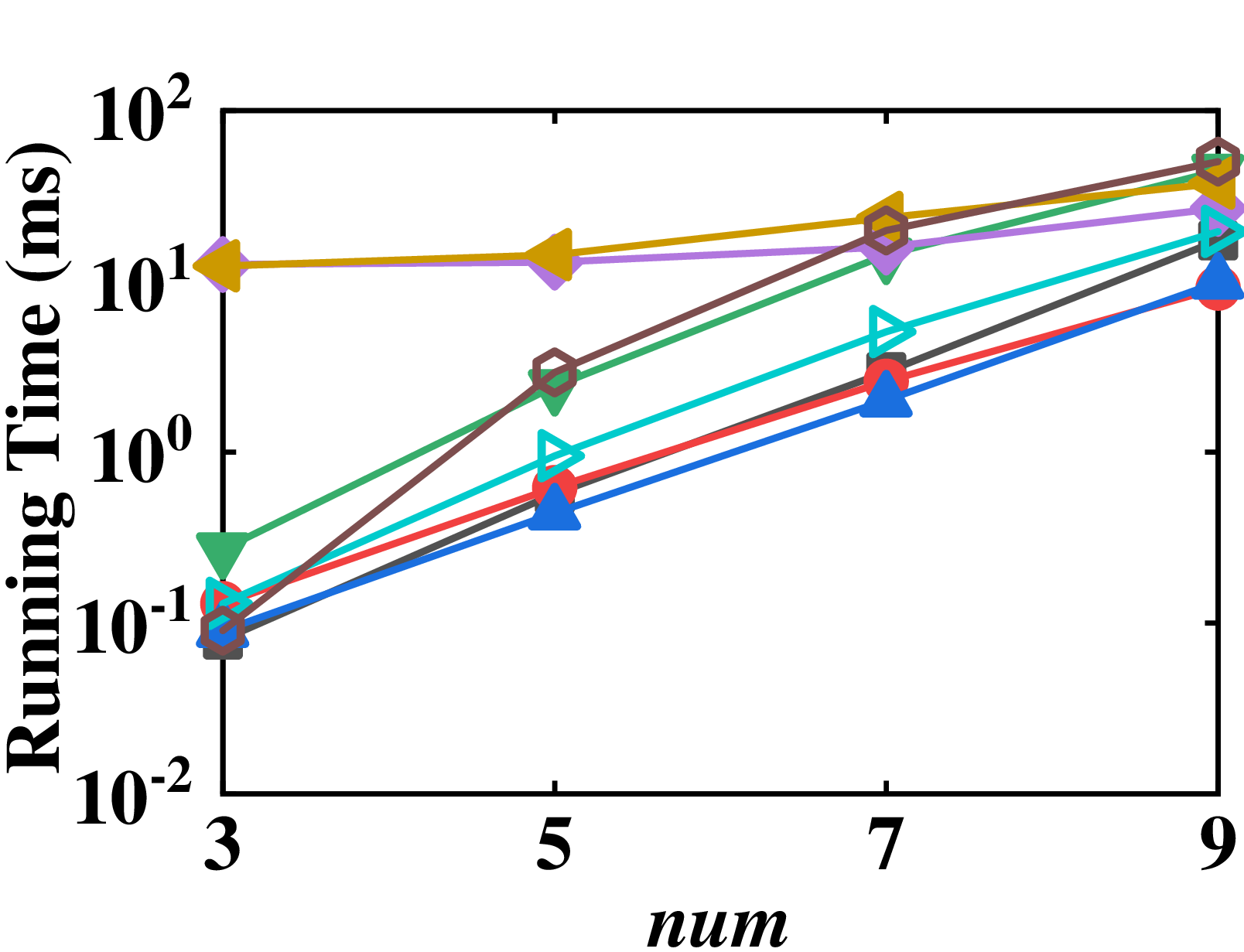}
		}
		\setlength{\abovecaptionskip}{0.1cm}
		\caption{{M\textit{k}NNQ Performance using Main-Memory based Metric Indexes vs \textit{num} on Synthetic Datasets}}
		\label{fig:ID Comparison Using MkNN queries_In_memory}
		\vspace{-0.3cm}
	\end{figure}

{Fig.~\ref{fig:ID Comparison Using MkNN queries_In_memory} plots the performance results of M$k$NN queries using main-memory metric indexes by varying \textit{num}, while all other parameters are set to their default values. As can be observed, the running times of LASEA and EPT* are less sensitive to \textit{num}. This is because, LASEA and EPT* are stored as tables, while the others are stored as trees. We need to scan two entire indexes (LASEA and EPT*) during search for any value of \textit{num}. However, the performance of the tree structures (BST, SAT, BKT, FQT, MVPT, and GNAT) degrades as \textit{num} increases (i.e., more sub-trees are visited).}
	
		\begin{figure}
	\centering
	\subfigtopskip=0cm
	\subfigbottomskip=0cm
	\subfigcapskip=0cm
	\includegraphics[width=\linewidth]{External_memory.eps}
	\\
	\subfigure[PA]{
		\includegraphics[width=0.3\linewidth]{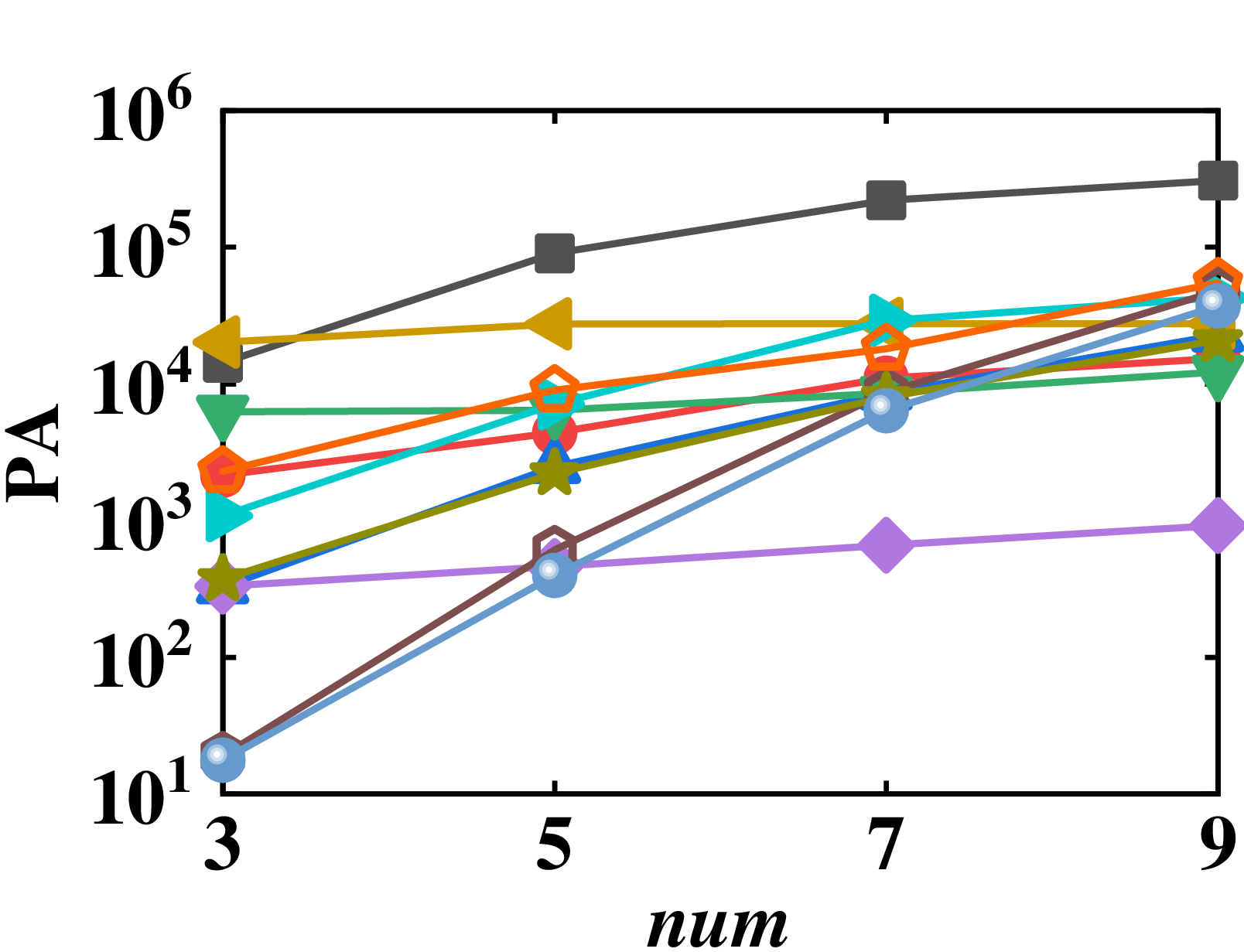}
	}
	\subfigure[Compdists]{
		\includegraphics[width=0.3\linewidth]{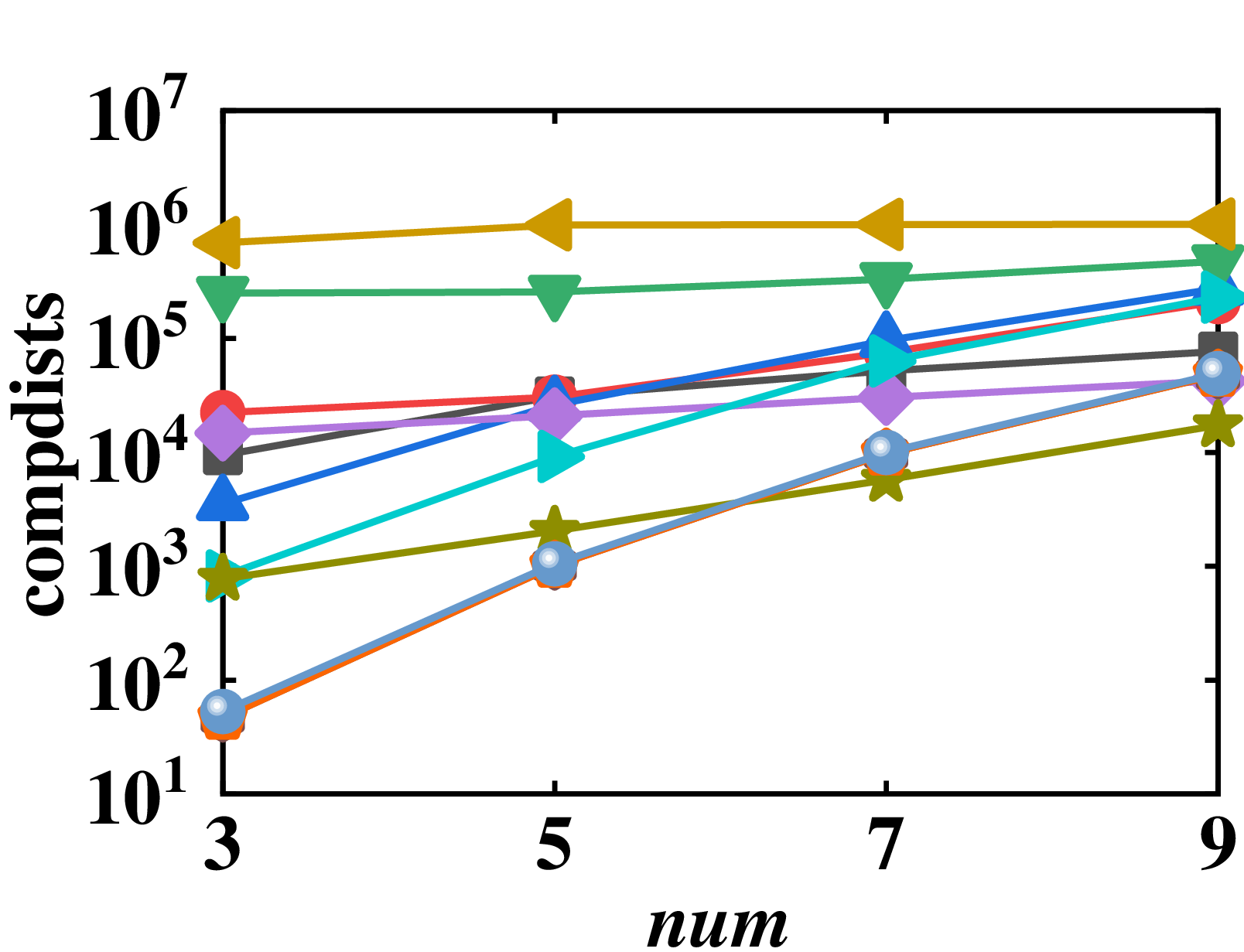}
	}
	\subfigure[Running Time]{
		\includegraphics[width=0.3\linewidth]{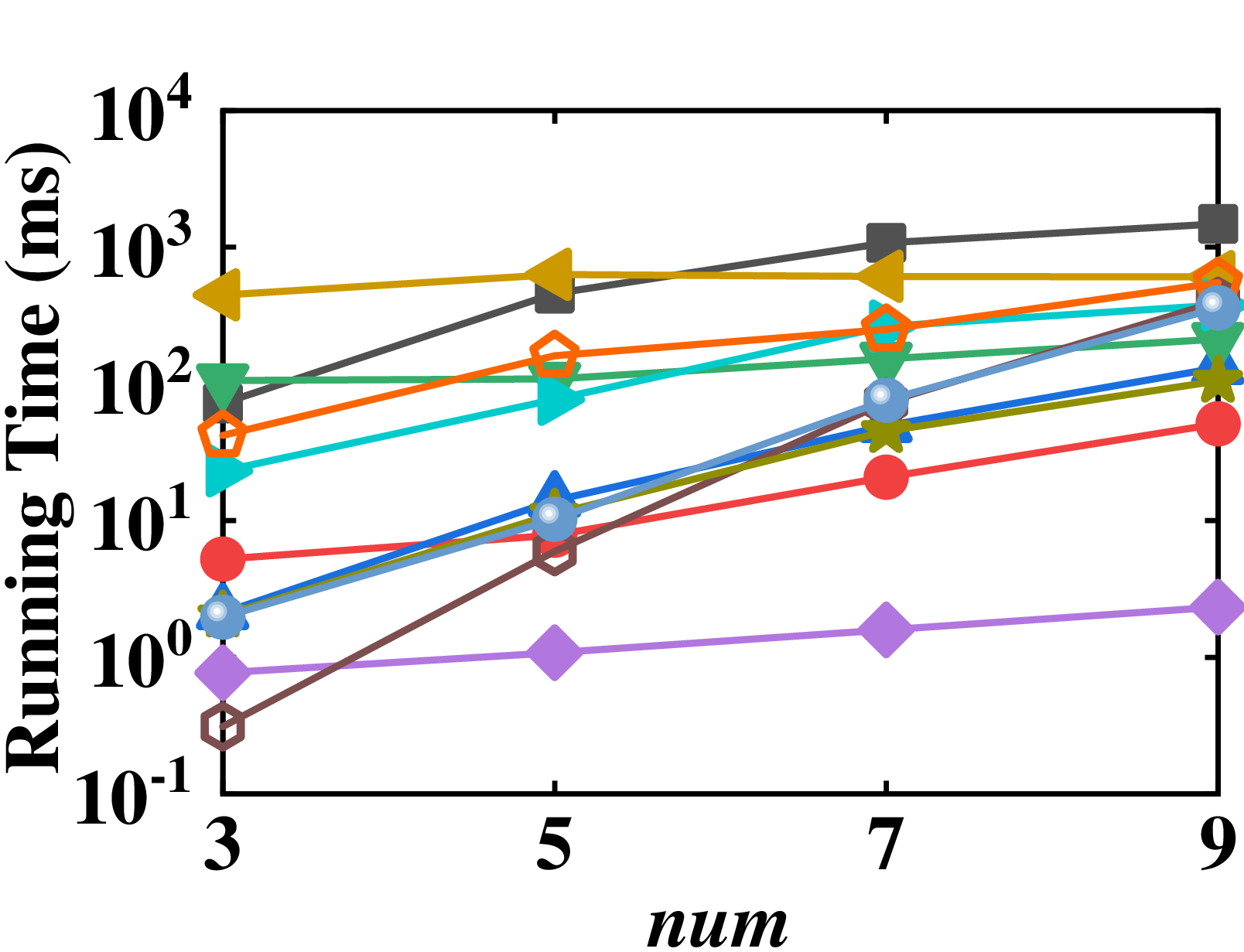}
	}
	\setlength{\abovecaptionskip}{0.1cm}
	\caption{{M\textit{k}NNQ Performance using Secondary-Memory based Metric Indexes vs \textit{num} on Synthetic Datasets}}
	\label{fig:ID Comparison Using MkNN queries}
	\vspace{-0.3cm}
\end{figure}

{Fig.~\ref{fig:ID Comparison Using MkNN queries} illustrates the performance results of M$k$NN queries using secondary-memory metric indexes. The first observation is that pivot-based indexes (i.e., the OmniR-tree, the SPB-tree, and the M-index*) are more sensitive to \textit{num} when compared with the compact-partitioning methods. The reason is that, the number of pivots is fixed in our experiments, which limits the pruning capabilities when the intrinsic dimensionality increases. More pivots are required with the growth of the intrinsic dimensionality. The second observation is that LC and the MB$^+$-tree are most stable across all performance metrics as \textit{num} grows. This is because, (i) LC is a list of clusters, where the query is performed by sequential scanning of clusters; and (ii) the query using the MB$^+$-tree needs to verify almost all data objects. The third observation is that the OmniR-tree and the SPB-tree are the indexes that are most sensitive to \textit{num}. The reason is that, the performance of the R-tree (used by the OmniR-tree) and the spatial locality of spacing-filling curve (used by the SPB-tree) degrades with the growth of the intrinsic dimensionality.}

	\subsection{Significance Evaluation}	
\begin{figure}
		\centering
		\subfigtopskip=0cm
		\subfigbottomskip=0cm
		\subfigcapskip=0cm
		\hspace{-2mm}\subfigure[\textit{Compdists} of MRQ]{
			\includegraphics[width=0.24\linewidth]{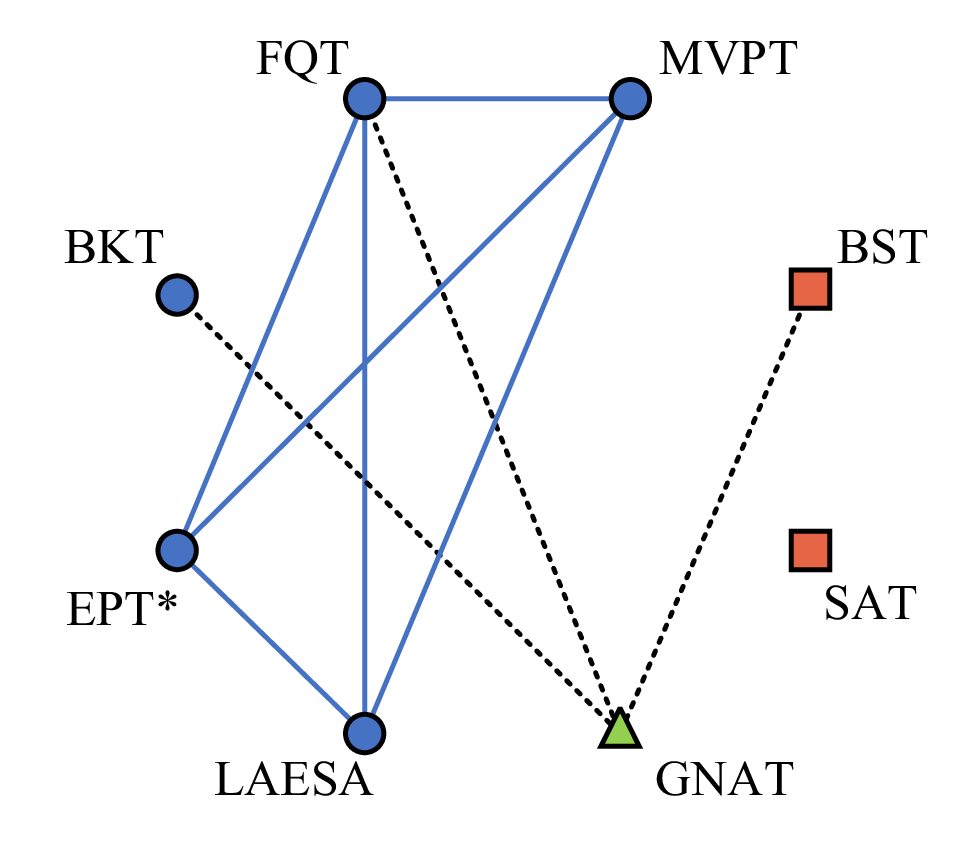}
		}
		\hspace{-1mm}\subfigure[Running time of MRQ]{
			\includegraphics[width=0.24\linewidth]{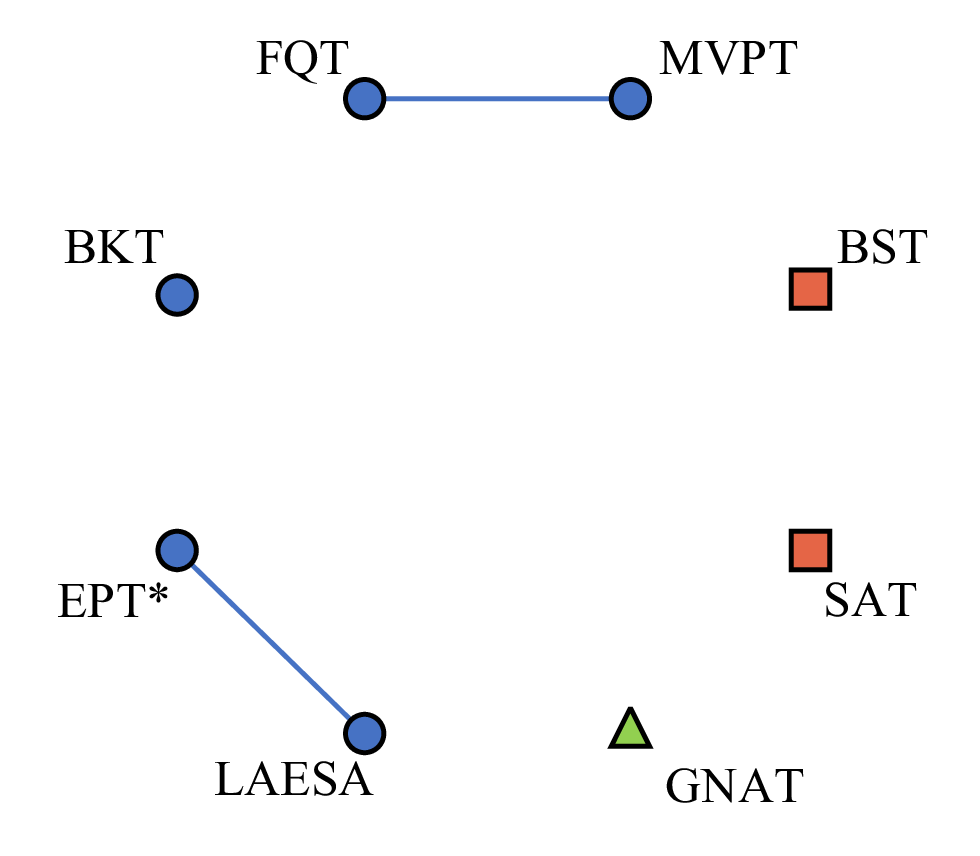}
		}
		\hspace{-1mm}\subfigure[\textit{Compdists} of M$k$NN]{
			\includegraphics[width=0.24\linewidth]{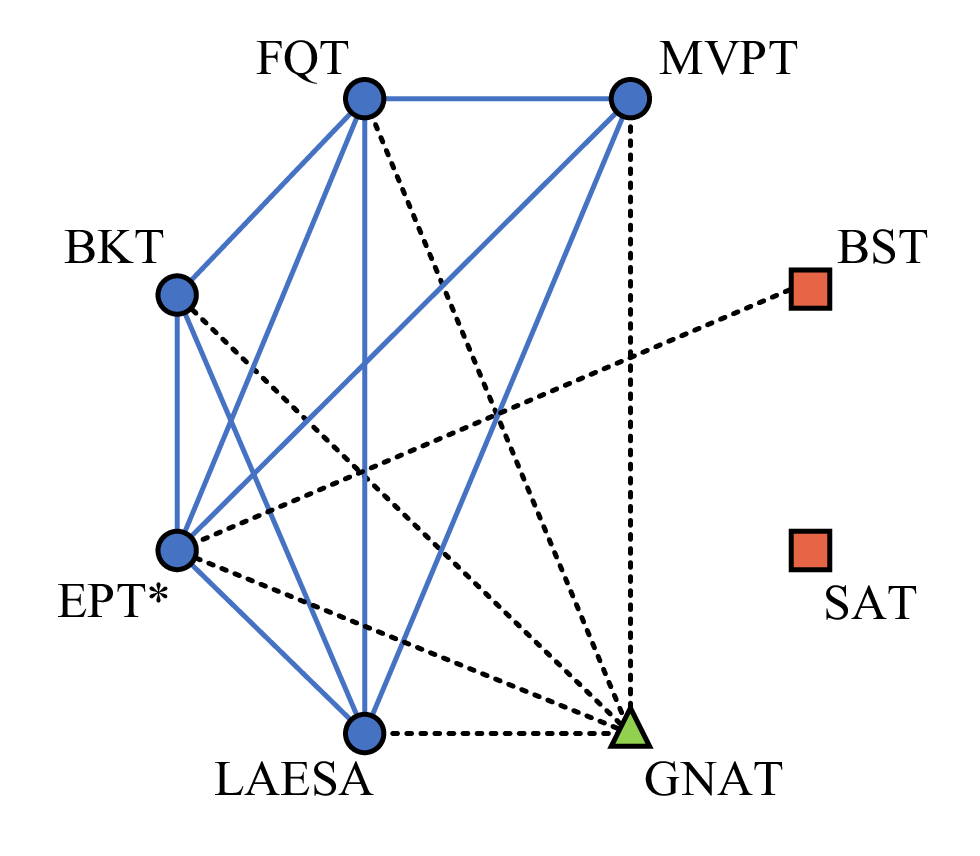}
		}
		\hspace{-1mm}\subfigure[Running time of M$k$NN]{
			\includegraphics[width=0.24\linewidth]{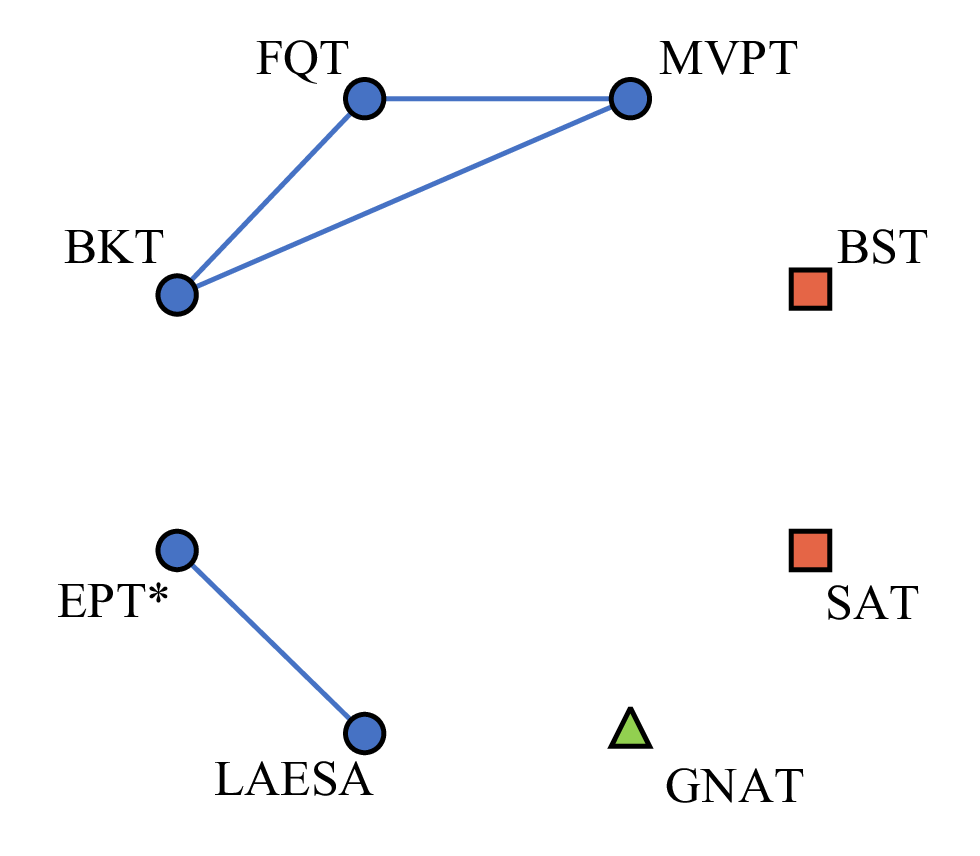}
		}	
		\setlength{\abovecaptionskip}{0.1cm}
		\caption{{Z-test on In-Memory Metric Indexes}}
		\label{fig:significanceI}
		\vspace{-0.3cm}
	\end{figure}

\begin{figure}
		\centering
		\subfigtopskip=0cm
		\subfigbottomskip=0cm
		\subfigcapskip=0cm
		\subfigure[PA of MRQ]{
			\includegraphics[width=0.25\linewidth]{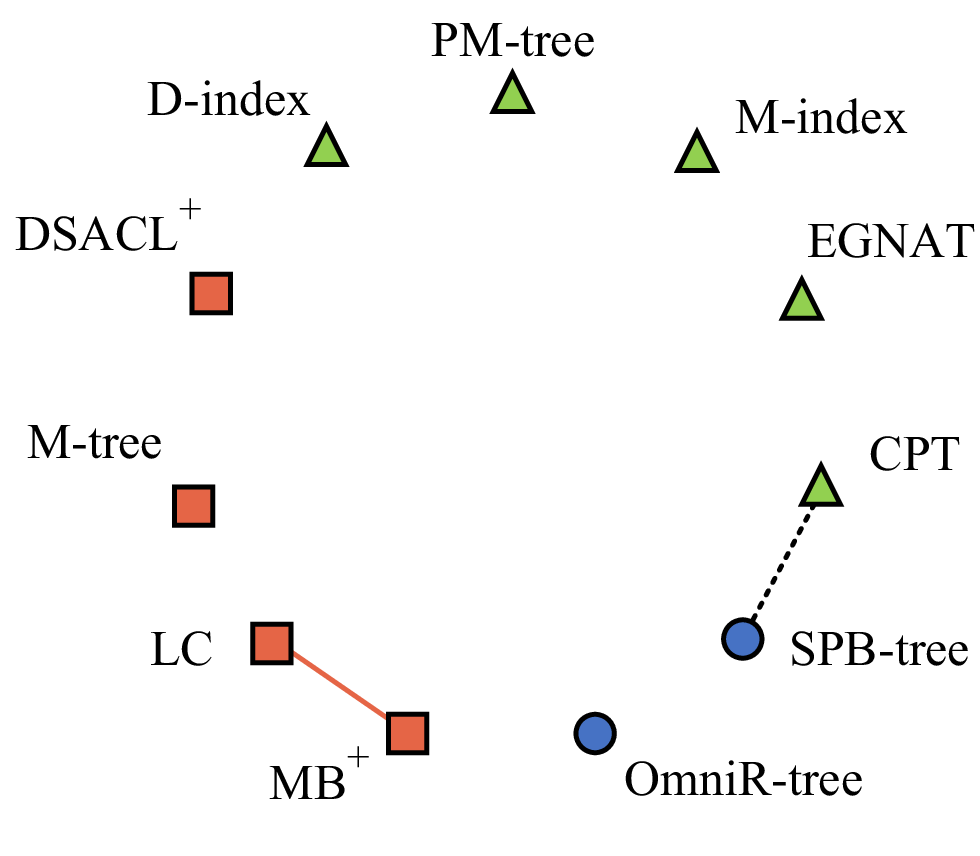}
		}\hspace{3mm}
		\subfigure[\textit{Compdists} of MRQ]{
			\includegraphics[width=0.25\linewidth]{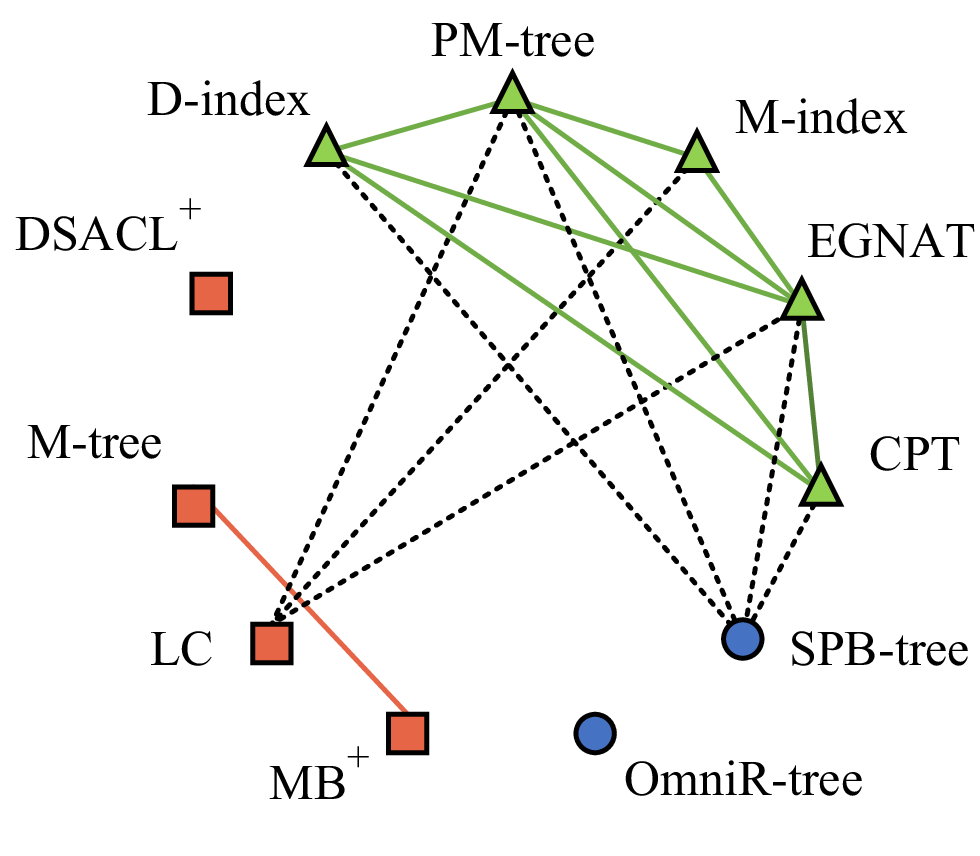}
		}\hspace{3mm}
		\subfigure[Running time of MRQ]{
			\includegraphics[width=0.25\linewidth]{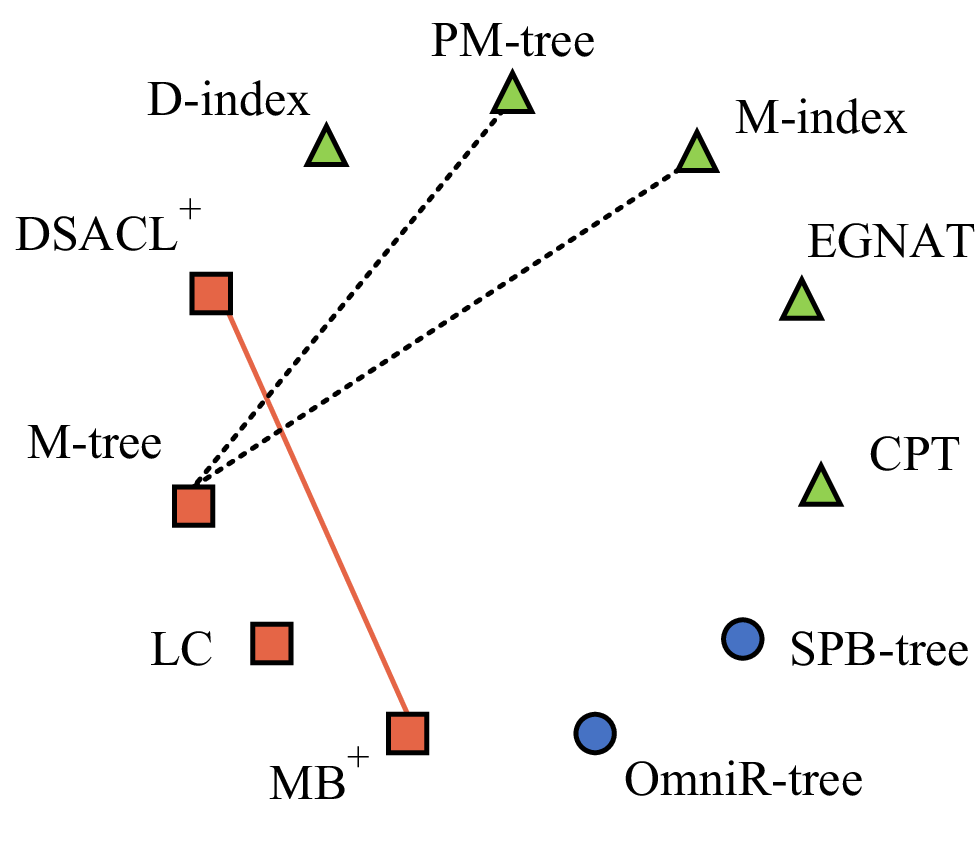}
		}
		\\		
		\subfigure[PA of M$k$NN]{
			\includegraphics[width=0.25\linewidth]{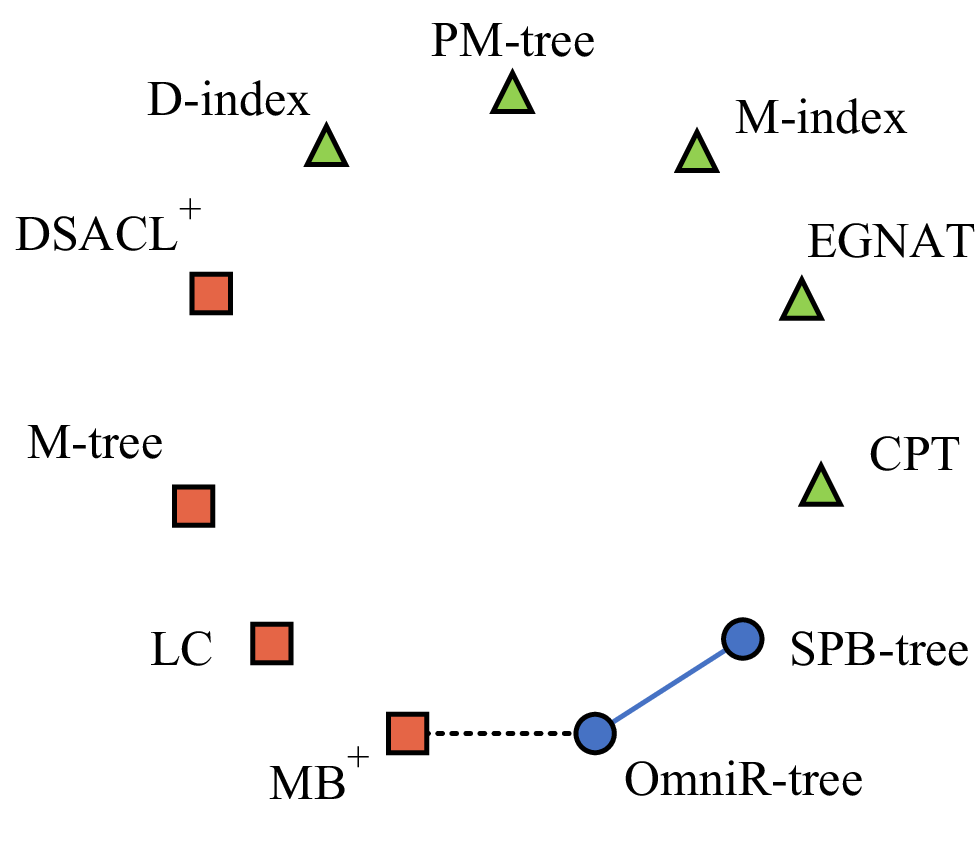}
		}\hspace{3mm}
		\subfigure[\textit{Compdists} of M$k$NN]{
			\includegraphics[width=0.25\linewidth]{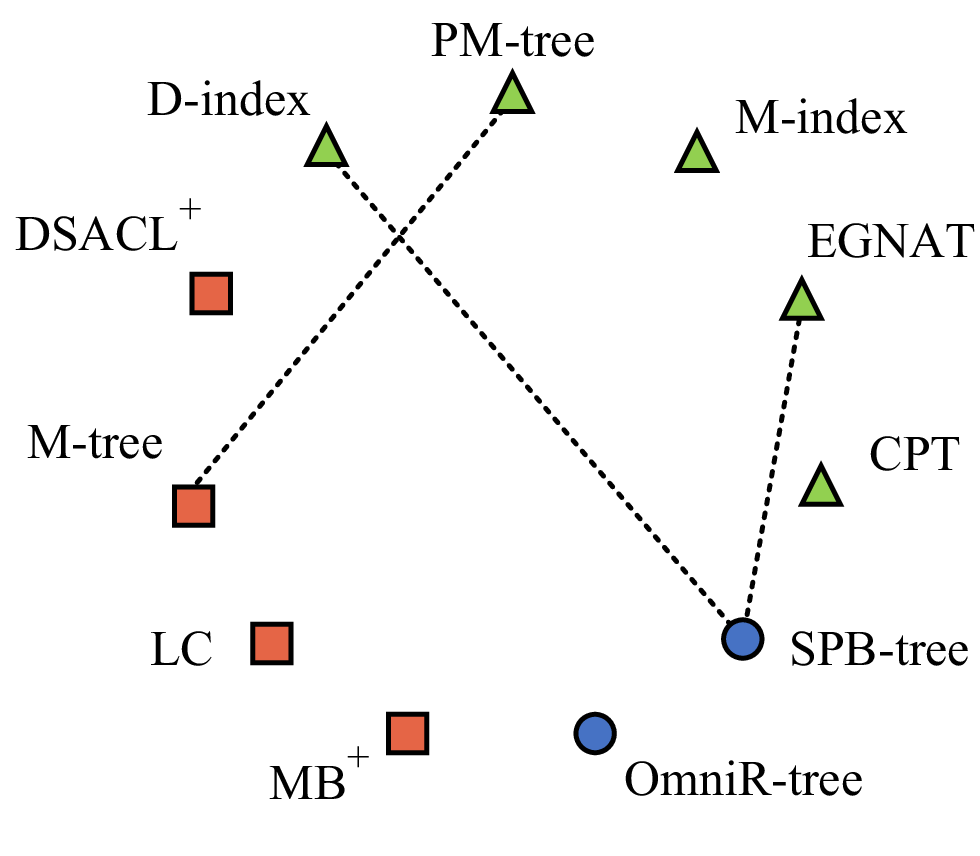}
		} \hspace{3mm}
		\subfigure[Running time of M$k$NN]{
			\includegraphics[width=0.25\linewidth]{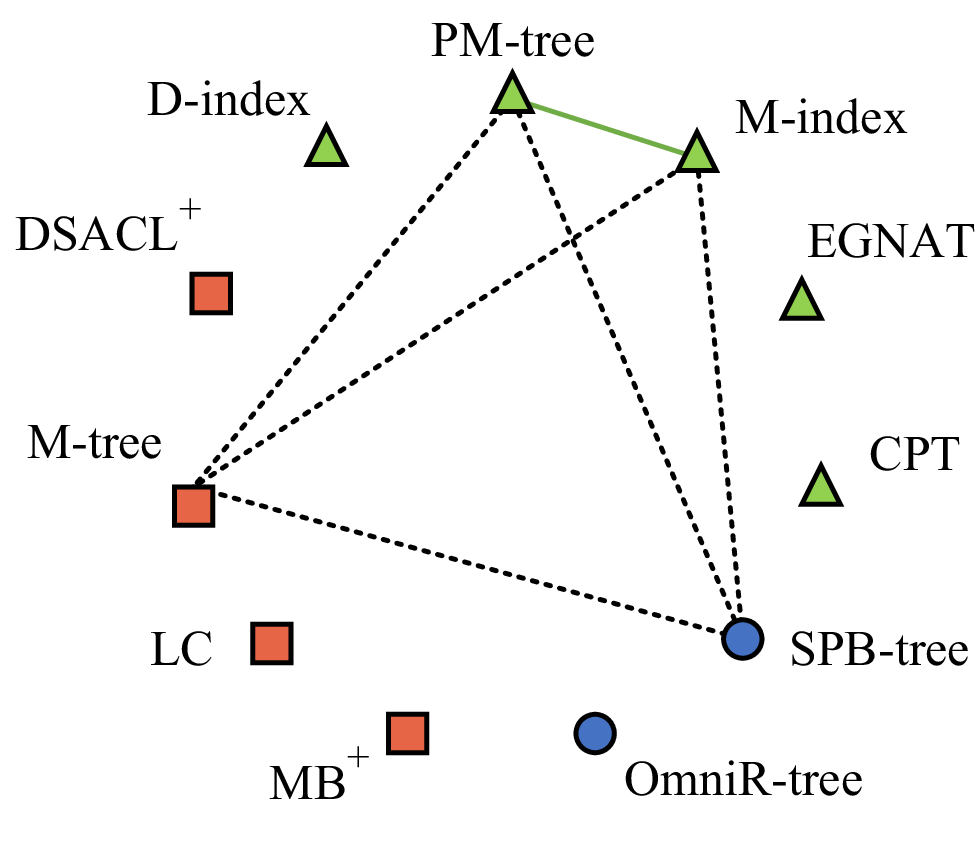}
		}
		
		\setlength{\abovecaptionskip}{0.1cm}
		\caption{{Z-test on Secondary-Memory Metric Indexes}}
		\label{fig:significanceD}
		\vspace{-0.3cm}
	\end{figure}

	{Here, we report on statistical significance tests of the differences in findings across metric indexes. Specifically, we run a Z-test with a $p$-value of 0.05 to test whether the performance findings (including \textit{compdists}, running time, and \textit{PA}) for different metric indexes when performing MRQ (with $r$ = 8\%, $l$ = 5) or M$k$NNQ (with $k$ = 20, $l$ = 5) are significantly different. Due to the space limitation, we only report the Z-test on the \textit{Words} dataset with all parameters being set to their default values. Fig.~\ref{fig:significanceI} shows the results on the in-memory indexes, while Fig.~\ref{fig:significanceD} shows the results on the secondary-memory indexes, where the indexes belonging to different categories are plotted as circles with different colors, i.e., compact-partitioning based indexes are orange squares, pivot-based indexes are blue circles, and hybrid indexes are green triangles. Two indexes with insignificantly different performance (i.e., $p$-value smaller than 0.05) are connected with a solid line segment if they belong to the same category, and they are connected with dotted line segments if they belong to different categories. The resulting sparse connections between metric indexes indicate that the performance of a pair of indexes is most often significantly different. In addition, the connections for \textit{compdists} are the densest, especially for the pivot-based and hybrid indexes. This is because, \textit{compdists} depends on the pivots, and because we use the same set of pivots for the pivot-based and hybrid indexes.}
	
	\section{SUMMARY AND RESEARCH DIRECTIONS}
	\label{sec:SUMMARY AND RESEARCH DIRECTIONS}
	We have surveyed {metric indexes} that aim to accelerate exact similarity search, which can be classified into three categories, i.e., pivot-based methods, compact-partitioning based methods, and hybrid methods. Pivot-based methods store pre-computed distances between well selected pivots and data objects, utilizing one of three types of data structures (table, tree, and {secondary-memory} based index). Thus, pivot-based indexes can be further divided into three types according to the data structure used to store pre-computed distances. Then, compact-partitioning based methods use one of four different types of partitioning techniques (ball partitioning, hyperplane partitioning, hash partitioning, and hybrid partitioning) to cluster the data. Thus, compact-partitioning based methods can be further divided into four types according to the partitioning technique used.
	
	Next, we have covered all pruning and validation techniques based on pivots or centers that are employed to accelerate similarity search using metric indexes. In addition, we have covered time and space complexities for metric index construction. Finally, we have reported on experimental analyses of similarity search using the metric indexes. The resulting findings, as below, enable users to select the indexes that best support the intended use cases:
	\begin{enumerate}
		\item[1)] For small datasets with little need for scalability, the main-memory based metric indexes can be chosen.
		\begin{enumerate}
			\item[a.] For complex distance functions (i.e., the distance computation cost is the dominance CPU cost), $\rm EPT^*$, SAT, and GNAT are the best choices, among which SAT has the lowest construction cost, followed by GNAT and $\rm EPT^*$.
			\item[b.] For simple distance functions, MVPT, GNAT, and $\rm EPT^*$ are the best choices, where the MVPT has lower construction cost than that of GNAT and $\rm EPT^*$.
		\end{enumerate}
		\item[2)] For large datasets that call for scalability, the disk-based metric indexes can be chosen.
		\begin{enumerate}
			\item[a.] For complex distance functions, the PM-tree, the SPB-tree and the OmniR-tree  are the best choices. They achieve good performance in terms of \textit{compdists} and have relatively good I/O performance. We not recommend LC due to its huge construction cost for large datasets.
			\item[b.] For simple distance functions, the M-tree, the PM-tree and D-index are the best choices. They achieve relatively good performance in terms of \emph{PA}  and incur little additional CPU cost for pruning.
		\end{enumerate}
	\end{enumerate}
	
	Although many metric indexes have been proposed, a number of open issues require further attention. Possible future directions of metric indexes are summarized below:
	\begin{enumerate}
		\item[1)] Search space pruning is based on pivots when using pivot-based methods while it is based on centers for compact-partitioning based methods. To improve pruning performance, it is of interest to study how to select high-quality pivots or centers, especially for compact-partitioning based methods.
		\item[2)] To further improve the performance of metric indexes, intelligent metric indexing can be considered that exploits machine learning techniques. In addition, distributed platforms can be leveraged, and new approximation optimizations can be pursued.
		\item[3)] As metric spaces only utilize the triangle inequality to accelerate search, possible directions for further work are how to improve the search efficiency by integrating specific characteristics of metric data and how to achieve high search efficiency if the triangle inequality is not fully satisfied by desirable distance functions.
		\item[4)] As expectations of privacy increase, privacy concerns exist wherever personally identifiable information is collected, stored, and used. Hence, the last but not the least direction for future work is to offer privacy protection in metric indexing.
	\end{enumerate}

	\section*{Acknowledgments}
This work was supported in part by the NSFC under Grants No. (62102351, 62025206 and 61972338), the Zhejiang Provincial Natural Science Foundation under Grant No. LR21F020005 and the DIREC center project. Yunjun Gao is the corresponding author of the work.
	
\bibliographystyle{ACM-Reference-Format}
\bibliography{IndexingSurvey}
	
\end{document}